\documentclass[11 pt, reqno]{article}

\usepackage{amsmath,amsfonts,amssymb,amsthm}
\usepackage[colorlinks=true,hyperindex=true]{hyperref}
\usepackage{cancel,bbm}
\usepackage{mathabx} 
\usepackage{comment}
\usepackage{enumitem}
\usepackage{cite}

\usepackage{chngcntr}
\counterwithin*{equation}{section}

\usepackage[a4paper, left=2cm, right=2cm, top=2 cm, bottom= 2 cm]{geometry}
\usepackage{color}
\usepackage{fancyhdr}
\usepackage{latexsym}

\allowdisplaybreaks

\newtheorem{ccounter}{ccounter}[section]
\newtheorem{thm}[ccounter]{Theorem}
\newtheorem{lem}[ccounter]{Lemma}
\newtheorem{cor}[ccounter]{Corollary}
\newtheorem{defn}[ccounter]{Definition}
\newtheorem{prop}[ccounter]{Proposition}
\newtheorem{ass}[ccounter]{Assumption}
\newtheorem{ex}[ccounter]{Example}
\newtheorem{definition}[ccounter]{Definition}

\def\sumai{\sum_{a}^{(i)}}

\def\bet{\begin{thm}}
\def\eet{\end{thm}}
\def\bel{\begin{lem}}
\def\eel{\end{lem}}
\def\bas{\begin{ass}}
\def\eas{\end{ass}}
\def\bec{\begin{cor}}
\def\eec{\end{cor}}
\def\bed{\begin{defn}}
\def\eed{\end{defn}}
\def\bep{\begin{prop}}
\def\eep{\end{prop}}
\def\beq{\begin{equation}}
\def\eeq{\end{equation}}
\def\proof{\noindent {\bf Proof.}\ \ }
\def\bea{\begin{equation*}}
\def\eea{\end{equation*}}
\def\tr{\operatorname{tr}}
\def\bex{\begin{ex}}
\def\eex{\end{ex}}
\def\remark{\noindent{\bf Remark. }}

\let\oldproofname=\proofname
\renewcommand{\proofname}{\rm\bf{\oldproofname}}

\def\rr{\mathbb{R}}

\def\cc{\mathbb{C}}
\def\hh{\mathbb{H}}
\def\1{\boldsymbol{1}}
\def\Im{\operatorname{Im}}
\def\Re{\operatorname{Re}}
\def\e{\mathrm{e}}
\def\i{\mathrm{i}}
\def\del{\partial}
\def\d{\mathrm{d}}
\def\eps{\varepsilon}
\renewcommand\leq\varleq
\renewcommand\geq\vargeq
\def\ee{\mathrm{E}}

\def\O{\mathcal{O}}

\def\ee{\mathbb{E}}

\def\pp{\mathbb{P}}

\def\mfa{\mathfrak{a}}

\def\bm{\mathfrak{m}}
\def\mb{\mathfrak{m}}
\def\hatmb{\hat{\mathfrak{m}}}
\def\hatbm{\hat{\mathfrak{m}}}
\def\hatS{\hat{S}}
\def\bg{\mathfrak{g}}
\def\bd{\mathfrak{d}}

\def\bs{\mathfrak{s}}

\def\Oma{\Omega_{\mathfrak{a}}}
\def\mff{\mathfrak{f}}
\newcommand{\GOE}{\mathrm{GOE}}

\def\rhosc{\rho_\mathrm{sc}}

\def\gamsc{\gamma^{(\mathrm{sc} ) } }

\def\tilx{\tilde{x}}

\def\Gi{G^{(i)}}
\def\sumi{\sum^{(i)}}
\def\E{\mathcal{E}}
\def\Gc{G^{(c)}}
\def\Gj{G^{(j)}}
\def\tilf{\tilde{f}}
\def\sth{s^{(3)}}
\def\sfo{s^{(4)}}
\def\Ep{E^{+}}
\def\Var{\mathrm{Var}}
\def\hatz{ \hat{z}}
\def\hatw{\hat{w}}
\def\mto{m_{t_0}}
\def\pto{\rho_{t_0}}
\def\hatp{\hat{\rho}}
\def\hatm{\hat{m}}
\def\tillam{\tilde{\lambda}}

\def\D{\mathcal{D}}
\def\dto{\downarrow}
\def\mfb{\mathfrak{b}}
\def\mfc{\mathfrak{c}}
\def\hatE{\hat{E}}
\def\N{\mathcal{N}}
\def\Ikap{I_{\kappa}}
\def\Dkap{\mathcal{D}_\kappa}
\def\Pp{P_\perp}
\def\hatx{\hat{x}}
\def\haty{\hat{y}}
\def\hatz{\hat{z}}
\def\Adel{\mathcal{A}_\delta}
\def\omst{\omega_*}
\def\msc{m_{\mathrm{sc}}}

\newcommand{\unn}[2]{[\![#1,#2]\!]}

\begin{document}
\title{Deformed GOE}


\begin{table}
\centering

\begin{tabular}{c}

\multicolumn{1}{c}{\parbox{12cm}{\begin{center}\Large{\bf Single eigenvalue fluctuations of general Wigner-type matrices }\end{center}}}\\
\\
\end{tabular}
\begin{tabular}{ c c c  }

 Benjamin Landon & Patrick Lopatto &  Philippe Sosoe
 \\
 & & \\  
  \footnotesize{University of Toronto} & \footnotesize{Brown University} & \footnotesize{Cornell University}  \\
  \footnotesize{Department of Mathematics}  & \footnotesize{Division of Applied Mathematics} & \footnotesize{Department of Mathematics}  \\
 \footnotesize{\texttt{blandon@math.toronto.edu}} & \footnotesize{\texttt{patrick\_lopatto@brown.edu}} &\footnotesize{\texttt{psosoe@math.cornell.edu}} \\
  & & \\
\end{tabular}
\\
\begin{tabular}{c}
\multicolumn{1}{c}{\today}\\
\\
\end{tabular}

\begin{tabular}{p{15 cm}}
\small{{\bf Abstract:} We consider the single eigenvalue fluctuations of random matrices of general Wigner-type, under a one-cut assumption on the density of states.  For eigenvalues in the bulk, we prove that the asymptotic fluctuations of a single eigenvalue around its classical location are Gaussian with a universal variance.

\hspace{5 pt} Our method is based on a  dynamical approach  to mesoscopic linear spectral statistics which reduces their behavior on short scales to that on larger scales.  We prove a central limit theorem for linear spectral statistics on larger scales via resolvent techniques and 
show that for certain classes of test functions, the leading-order contribution to the variance 
agrees with the GOE/GUE cases.}
\end{tabular}
\end{table}

\tableofcontents

\section{Introduction}

There\let\thefootnote\relax\footnote{The work of B.L. is partially supported by an NSERC Discovery grant. P.L. is supported by NSF grant DMS-1926686. The work of P.S. is partially supported by NSF grants DMS-1811093 and DMS-2154090.} has been substantial interest in 
random matrices ever since Wigner's 
proposal to use 
their eigenvalue statistics 
to model the energy levels of large 
quantum systems \cite{wig1,wig2}.     
Motivated by  Wigner's vision, there have been significant advances in understanding the universality of various spectral statistics 
 -- that is, whether their asymptotic behavior depends 
on the finer details 
of the matrix ensemble.  The general expectation is that the asymptotic distributions of 
statistics associated with the eigenvalues or eigenvectors do not depend on the choice of matrix entry distribution, 
and are identical to the Gaussian case.
In this paper we investigate a fundamental spectral statistic, the deviation of an eigenvalue from its ``classical location.'' 
We consider these \emph{single eigenvalue fluctuations} for a very broad class of matrices, greatly generalizing existing work on this observable.

 In Wigner's original work \cite{wig1,wig2}, he introduced the class of random matrices now known as \emph{Wigner ensembles} and proved that their empirical eigenvalue distribution converges the \emph{semicircle distribution},
\beq \label{eqn:int-semi}
\lim_{N \to \infty} \frac{1}{N} \sum_{i=1}^N \delta_{ \lambda_i } (E) \d E \stackrel{d}{=} \frac{1}{ 2 \pi } \sqrt{ (4-E^2)_+ } \, \d E =: \rhosc (E) \, \d E .
\eeq
Wigner ensembles consist of $N \times N$ self-adjoint random matrices whose entries are independent, centered random variables  (up to the self-adjointness constraint) with variance $1/N$.  
We denote eigenvalues by $\{ \lambda_i \}_{i=1}^N$ in increasing order.

 Wigner ensembles are divided into two \emph{symmetry classes}, real symmetric and complex Hermitian, depending on whether the entries are all real or the off-diagonal entries are complex.  Two special cases arise when one takes the matrix entries to be real or complex Gaussians, which are known as the Gaussian Orthogonal and Unitary Ensembles, respectively (GOE/GUE). 

The eigenvalue densities of the GOE and GUE admit exact algebraic formulas, which were used by Dyson, Gaudin, and Mehta to explicitly characterize the limits of various local statistics, such as the local $k$-point correlation functions \cite{gaudin1,gm1,dyson1,dyson2,mehta1}.   

Mehta formulated the  bulk universality conjecture for Wigner matrices in his 1961 treatise \cite{mehta}, conjecturing that the formulas proven for the local $k$-point correlation functions for the GOE and GUE should hold for all Wigner matrices.  This conjecture was proven for all symmetry classes in a recent series of works \cite{local-relax,erdos2010bulk,erdos2012rigidity,localrelaxation}.  Parallel results were established in certain cases in \cite{tao2010random,tao2011random} with the key result being a ``four moment comparison theorem.'' 


Since these works, there have been numerous developments extending universality to matrix ensembles 
beyond the Wigner class.  For a random self-adjoint matrix consisting of centered entries, denote the matrix of variances $S$ by
\beq
S_{ij} = \ee\left[ |H_{ij} |^2 \right].
\eeq
Wigner matrices correspond to the case that $S$ is a constant matrix, $S_{ij} = N^{-1}$.  Generalized Wigner matrices correspond to relaxing the constraint that the variances are all equal but demanding that the matrix remains doubly stochastic, 
\beq \label{eqn:int-ds}
\sum_j S_{ij} =1,
\eeq
 as well as assuming that the entries are all of the same order, $S_{ij} \asymp N^{-1}$.  The semicircle distribution \eqref{eqn:int-semi} still holds for generalized Wigner matrices \cite{guionnet,az}, and in fact   Erd\H{o}s, Yau, and Yin proved  universality of the local $k$-point correlation functions  
  \cite{erdos2012rigidity,erdos2010bulk,yaubernoulli}.

The ensembles studied in the present article arise when one drops the constraint \eqref{eqn:int-ds}.   These 
are called matrices of general Wigner-type and have been considered in great detail in the works \cite{qve,univ-wig-type,sing} of Ajanki, Erd\H{o}s, and Kr{\"u}ger.  One of the new phenomena that emerge when studying this class of matrices is that the semicircle law \eqref{eqn:int-semi} no longer holds; in general, the spectral measure no longer has an analytic form.  The work \cite{sing} provides a complete classification of the kinds of measures that arise, showing that in general they will have multiple intervals of support (at the edges of which the density decays like a square-root), with possible cusp and cusp-like singularities when two intervals are close or merge.

Local universality was proven for matrices of general Wigner-type in \cite{univ-wig-type} and required two main inputs: local ergodicity for Dyson Brownian motion (DBM) with general initial data \cite{convergence,fixed,es}; and local law and rigidity estimates for these general ensembles (high probability estimates for the eigenvalue locations).


For these general ensembles, the replacement of the semicircle law \eqref{eqn:int-semi} is defined in terms of a solution to the quadratic vector equation
\beq \label{eqn:int-qve}
- \frac{1}{ \mb_i (z) } = z + ( S \mb(z))_i,
\eeq
which is an equation for $\mb (z): \hh_+ \to (\hh_+)^N$.  The spectral measure is then defined via boundary values of $\mb(z)$.  The works \cite{qve,univ-wig-type,sing} 
give a comprehensive study of the stability and solution of \eqref{eqn:int-qve}, as mentioned above. 
Moreover, these works establish the local law and rigidity estimates for matrices of general Wigner-type and obtain universality for the local eigenvalue statistics.

 So far, we have discussed the universality of the local $k$-point correlation functions. Parallel results also hold for the gap between consecutive eigenvalues -- this is another example of a local statistic, and its universality is similarly established for broad classes of random matrices.  Given these results, it is natural to investigate where the boundary between universal and model-dependent asymptotic distributions lies.   In this context, we will now discuss the single eigenvalue fluctuations and the related eigenvalue counting function, and central limit theorems for linear spectral statistics.

Gustavsson \cite{gust} considered the eigenvalues $\{ \lambda_i \}_{i=1}^N$ of the GUE and proved that as $N \to \infty$, the statistic
\beq \label{eqn:int-sev}
\frac{N}{ \sqrt{ \log(N) }} ( \lambda_i - \gamsc_i )
\eeq
converges to a centered normal random variable with explicit variance.  Here, the $\gamsc_i$ are the $N$-quantiles of the semicircle distribution, and the index $i$ refers to a sequence of indices satisfying $\kappa N \leq i \leq (1- \kappa)N$ for some $\kappa>0$.  O'Rourke \cite{orourke} used Gustavsson's result to show that this holds also for the GOE by applying a coupling of Forrester--Rains \cite{FR}. 

Due to the duality
$
\{ \lambda_k > E  \} = \{ | \{ i: \lambda_i  \leq E \} | < k\} ,
$
 Gustavsson's result is equivalent to the fact that the eigenvalue counting function
$
n (E) := | \{ i : \lambda_i  \leq E \} |  
$
has asymptotic Gaussian fluctuations on the scale $\sqrt{ \log(N)}$.  The eigenvalue counting function can be thought of as a \emph{linear spectral statistic},
\beq
\operatorname{LSS}( \varphi ) := \sum_i \varphi ( \lambda_i ),
\eeq
with $\varphi$ an indicator function. Linear spectral statistics are well-studied objects in random matrix theory, and we omit a comprehensive discussion.  The results of Shcherbina and Lytova--Pastur \cite{lytovapastur,shcherbina} show that for the eigenvalues of a Wigner matrix, one has that for sufficiently smooth $\varphi$,
\beq \label{eqn:int-clt}
\lim_{N \to \infty} \sum_i \varphi ( \lambda_i ) - \ee[ \varphi ( \lambda_i ) ]  \stackrel{d}{=} \N (0, V ( \varphi) ).
\eeq
Here $\N(0, V( \varphi))$ is a centered normal random variable with variance
\beq \label{eqn:int-var}
V( \varphi) = \frac{1}{ 2 \pi^2} \int_{-2}^2 \int_{-2}^2 \frac{ ( \varphi(x ) - \varphi(y) )^2}{ ( x-y)^2} \frac{ 4 - xy}{ \sqrt{4-x^2}\sqrt{4-y^2}}\, \d x\, \d y + \frac{s_4}{2 \pi^2} \sigma(\varphi)^2,
\eeq
where $s_4$ denotes the fourth cumulant of the matrix entries, and 
\beq
\sigma ( \varphi) := \int_{-2}^2 \varphi (x) \frac{2 - x^2}{ \sqrt{4-x^2}}\, \d x.
\eeq
In particular, the asymptotic fluctuations of linear spectral statistics are in general not universal within the class of Wigner matrices and instead depend on the fourth cumulant.  It is therefore not clear at first whether one expects universality of the eigenvalue counting function. Note that the indicator function is sufficiently irregular to lie outside the domain of the results \cite{shcherbina,lytovapastur}.  Note also that  up to the weighting by semicircle-like factors, the first term in \eqref{eqn:int-var} is similar to the integral norm of the homogeneous Sobolev space $\dot{H}^{1/2} ( \rr)$ (with the indicator function being only in $\dot{H}^{1/2+}$). 

Nonetheless, the works \cite{meso,bourgade2019gaussian} show that the eigenvalue counting function and single eigenvalue fluctuations are in fact universal for Wigner matrices; they have asymptotic Gaussian fluctuations with variance identical to the GOE/GUE.  
The apparent tension with the non-universal results on LSS is resolved by noting that the correct rescaling factor for these observables is $\sqrt{ \log(N)}$, as can be seen by an explicit computation in the Gaussian case \cite[Section 5.4]{potters2020first}.
Renormalizing by this factor suppresses the non-universal contributions of \eqref{eqn:int-var} in the limit $N \to \infty$, which are of sub-leading order.

This suggests that the single eigenvalue fluctuations are essentially on the boundary of non-universal fluctuations by a $\sqrt{ \log(N)}$ factor.  In fact, in \cite{meso}, an expansion for the expectation $\ee[ \lambda_i]$ down to $o (N^{-1})$ was derived, resolving a conjecture of Tao and Vu \cite{taovu-conj}.  The $1/N$ term in this expansion depends on the fourth cumulant of the matrix entries; in terms of the single eigenvalue fluctuations, this is not seen as it is lower order by the  $\sqrt{ \log(N)}$ factor.

We are therefore interested in whether  this universality persists when we move beyond the Wigner class and encounter matrices whose global spectral measure is no longer semicircular.  We note that before the present work, single eigenvalue fluctuations have not been considered even for the class of generalized Wigner matrices.  

Our main result is that if one rescales the observable \eqref{eqn:int-sev} by the local particle density given by the spectral measure associated with the variance matrix $S$ (through \eqref{eqn:int-qve}), then one recovers the universal Gaussian fluctuations proven for the GUE and GOE by Gustavsson and O'Rourke.  That is, the single eigenvalue fluctuations remain universal, up to a rescaling by the local particle density.  As a consequence, the eigenvalue counting fluctuations are seen to be universal.  

\subsection{Proof strategy}

We now explain the methodology of our paper. 
We begin by reviewing the approach of \cite{meso}, which studied single eigenvalue fluctuations in the Wigner case. 
Then, we discuss the contributions of this work.
First, we mention that we rely heavily on the local laws, rigidity estimates, and properties of the quadratic vector equation and its solution  proven in the works of Ajanki, Erd\H{o}s, and Kr{\"u}ger \cite{sing,univ-wig-type,qve}, some of which are reviewed in Section \ref{sec:main}. 

 In \cite{meso} it was shown that single eigenvalue fluctuations for Wigner matrices could be deduced from two separate results in random matrix theory.  The first is the homogenization theory for Dyson Brownian motion, developed for Wigner matrices in the work \cite{fixed-wig}. The main point of this result is that, through the use of DBM, local fluctuation results on the scale $N^{-1}$ can be reduced to studying \emph{mesoscopic linear spectral statistics}.  These are LSS of the form,
\beq
\varphi (x) := f (N^{\omst} (x-E) )
\eeq
where $0 < \omst <1$.  Note that since $0 < \omst <1$, such a statistic lives in between the macroscopic and microscopic scales.  Therefore, the second ingredient of the proof in \cite{meso} is a central limit theorem for mesoscopic LSS.

Mesoscopic central limit theorems for functions $f$ of compact support are well studied \cite{heknowles,lodhiasimm}, and results analogous to \eqref{eqn:int-clt} hold.  Note that when $f$ is of compact support, the non-universal contributions to $V( \varphi)$ vanish, while the $\dot{H}^{1/2}$ norm is invariant under rescaling and survives the limit. 

However, special attention must be paid to the type of function $f$ that arises from the homogenization theory; up to lower order corrections it is a smoothed out step function.  In particular, it is not of compact support, and lies outside the results of works such as \cite{heknowles,lodhiasimm}, and so a contribution of \cite{meso} is to prove a mesoscopic CLT for test functions of non-compact support. 

At a high level, the present work follows a similar strategy to \cite{meso}.  Instead of the homogenization theory of \cite{fixed-wig}, which holds only for semicircular initial data, we apply \cite{fixed}, which establishes homogenization results in a more general setting.  We therefore arrive at a similar reduction to analyzing mesoscopic linear spectral statistics for non-compactly supported test functions.

The main technical contribution of this work is then proving a central limit theorem for mesoscopic linear spectral statistics for matrices of general Wigner-type, and moreover showing that the form of the variance is universal.  We begin similar to \cite{meso} and  use Stein's method and resolvent expansions to analyze the characteristic function of a linear spectral statistic.

However, there are significant complications that arise once one leaves the setting of Wigner matrices; we enumerate two here.  The first main obstacle is that one immediately encounters quantities of the form
\beq
T_{xy} (z, w) := \sum_j S_{xj} G_{yj} (z) G_{jy} (w), \qquad G(z) := (W-z)^{-1}.
\eeq
In the Wigner case,  the sum over $j$  becomes matrix multiplication and this quantity is easily seen to be related to $(G(z) - G(w) ) / (w-z)$ which may be analyzed via local laws.  Even in the case of generalized Wigner matrices this quantity is difficult to handle.  For $z=w$, it was encountered in the work \cite{diffusion} in the context of band matrices, and significant technical effort was expended in obtaining precise estimates.  For $z \neq w$, the generalized Wigner case was handled in \cite{li-xu}.  

For the general class of matrices considered here, there are no existing results for the quantity $T_{xy}(z, w)$.  In general, one has to analyze $z$ and $w$ whose imaginary parts are on the order of $N^{-\omst}$, where $\omst$ comes from the mesoscopic linear spectral statistic.  Error estimates in resolvent expansion techniques degenerate as $\Im[z]$ and $\Im[w]$ become smaller.  Unfortunately, the $\omst$ coming from the homogenization result of \cite{fixed} is very close to $1$, so the mesoscopic scale considered is just barely above the microscopic scale $1/N$.  The error estimates in the resolvent expansion therefore degenerate rapidly unless carefully controlled.

In order to develop our main result, one could try to control the quantity $T_{xy}$ via involved resolvent methods, carefully tracking error estimates.  Another approach could be to develop further the homogenization theory of \cite{fixed}; the reason for the short scales appearing in the test function is that the homogenization theory (which analyzes a parabolic difference equation) was proven only to hold for a certain range of times $t \sim N^{-1}$ where the time $t$ corresponds to the scale of the mesoscopic test function, $t = N^{-\omst}$.  However, the proof of \cite{fixed} is quite involved, and the restriction on the time range is in part due to certain drift terms that arise, which are difficult to control directly.

A new idea here is that one may use Dyson Brownian motion to give a \emph{decomposition} of a short-scale mesoscopic LSS as the independent sum of an explicit Gaussian random variable and a large-scale mesoscopic LSS.  This is based on the characteristic approach to DBM developed in the context of linear spectral statistics in \cite{huang2019rigidity}. 
In that work, the authors directly calculated the trajectory of the Stieltjes transform along a characteristic, deducing a CLT for compactly supported test functions. Clearly, this does not suffice for our purposes.  For non-compactly supported functions, we further develop and refine this approach by using the Helffer--Sj{\"o}strand representation, which represents a general test function as an integral of Stieltjes transforms and allows us to arrive at our decomposition. 
 We are able to show that the leading-order contribution of the variance of the Gaussian  is in fact universal, leaving us with the task of establishing a CLT for mesoscopic linear statistics on large scales.

For mesoscopic linear statistics on large scales, we then proceed as above via resolvent techniques.  The main simplification is that now we do not need precise error estimates due to the fact that we are working on larger scales; we do not have to track the dependence on $\Im[z]$ and $\Im[w]$ in error terms carefully.

After proving a CLT for these classes of test functions, we are left with a second obstacle: the formula for the variance we find involves a complicated double integral involving the matrix $S$; in particular it is difficult to find any term that resembles the $\dot{H}^{1/2}$ or $(x-y)^{-2}$ contribution in \eqref{eqn:int-var} that is a prerequisite for the universality. These mesoscopic non-compactly supported linear spectral statistics in fact fluctuate on the same scale as the single eigenvalue fluctuations and so the variance must be computed to leading order and must be shown to have a universal character.

Let $\mb(z)$ be as in \eqref{eqn:int-qve}. 
  Our formula for the variance involves the  
  operator
\beq \label{eqn:int-stab}
\frac{1}{ 1 - S \mb(z) \mb (w) },
\eeq
where now the vectors $\mb(z)$ are identified with diagonal matrices.  When $z= x + \i 0$ and $w = y - \i 0$, this operator is singular when $x=y$, hinting at the appearance of the kernel $(x-y)^{-1}$ (whose derivative is $(x-y)^{-2}$).  We apply perturbation theory to a generalization of the ``stability operator'' of \cite{qve,univ-wig-type,sing}, $F(z, w) = |\mb(z) \mb(w) |^{1/2} S | \mb(z) \mb(w) |^{1/2}$.  The singularity in \eqref{eqn:int-stab} stems from the contribution of the largest eigenvalue--eigenvector pair of this operator.  In order to isolate this contribution we use the Sherman--Morrison formula to expand the resulting quantity around $x=y$ using perturbation theory.  The coefficient of the leading-order contribution in the expansion in powers of $(x-y)^{-1}$ turns out be universal and independent of the matrix $S$.  We are therefore able to obtain the desired 
 universal behavior of the kernel and obtain universality of the mesoscopic LSS fluctuations, and consequently the universality of the single eigenvalue fluctuations through our application of the homogenization theory of \cite{fixed}. 


\subsection{Relation to prior work}

We now review some additional literature related to our contributions.  As stated above, single eigenvalue fluctuations were first obtained for the GUE by Gustavsson \cite{gust}, using as input a central limit theorem for determinantal point processes of Costin--Lebowitz \cite{costin1995gaussian}.  O'Rourke extended this result to the GOE \cite{orourke} by relying on a coupling of the GOE and GUE of Forrester--Rains \cite{FR}.  
Tao and Vu \cite{tao2011random} applied their four-moment theorem to show that this holds for any matrix ensemble whose moments match the GOE or GUE to fourth order.

  Universality for the single eigenvalue fluctuations for general Wigner matrices was obtained in \cite{meso} by two of the authors of the present article and by Bourgade--Mody in \cite{bourgade2019gaussian} in parallel.  The work \cite{bourgade2019gaussian} also studied determinant fluctuations for Wigner matrices and relies on the stochastic advection equation introduced and analyzed in \cite{bourgade2018extreme}.

Linear spectral statistics are well-studied in random matrix theory and we list only the articles most relevant to our discussion, referring the reader to the references within.  Central limit theorems were established by Shcherbina \cite{shcherbina} and Lytova--Pastur \cite{lytovapastur}. Our approach to linear spectral statistics via resolvent methods is most closely related to the work \cite{meso}, which in turn was inspired by the approach using Stein's method of Shcherbina \cite{shcherbina}.  Fluctuations for mesoscopic linear spectral statistics of Wigner matrices have been obtained by Lodhia--Simm \cite{lodhiasimm} on scales $\Im[z] \gg N^{-1/3}$ and by He--Knowles \cite{heknowles} on scales $\Im[z] \gg N^{-1}$.  Further results in the Wigner case have been obtained by Bao--He \cite{bao2021quantitative}, as well as Cipolloni--Erd{\H{o}}s--Schr{\"o}der \cite{erdos2020functional}. Fluctuations for generalized Wigner matrices were studied by Li--Xu \cite{li-xu} and for some general classes of random matrices by Anderson--Zeitouni \cite{az} and by Guionnet \cite{guionnet}.  Fluctuations for some polynomial test functions for matrices with a variance profile were obtained by Adhikari--Jana--Saha \cite{adhikari2019linear}. 
Linear statistics of heavy-tailed and half-heavy-tailed matrices have been considered by Benaych-Georges and Maltsev \cite{benaych2016fluctuations}, Benaych-Georges, Guionnet and Male \cite{benaych2014central}, and Lodhia and Maltsev \cite{lodhia2020covariance}; sparse matrices were considered by He in \cite{he2020fluctuations}.

Gustavsson's result for the GUE extends also to eigenvalues near the edge as well as calculates correlations between eigenvalues even at macroscopically separated distances. It is likely that one could obtain Gustavsson's result near the edge for our model, at least for some range of eigenvalue indices satisfying $1 \ll i \leq N^{c}$, some $c>0$ by using the results of \cite{landonedge} and a Green function comparison theorem. Near the edge, the eigenvalue fluctuation scale is above the fluctuations of the linear spectral statistics (as can be seen by the main result of \cite{landonedge} - there is no LSS coming from a homogenization - it is lower order) and so no CLT would be necessary. There is, however, an intermediate regime of eigenvalue indices $N^c \leq i \leq N^{1-c}$  not covered by \cite{landonedge} or \cite{fixed} which would require further development on the DBM side. Concerning correlations, the homogenization theory of \cite{fixed} is local. However, it is likely that the result of \cite{fixed} could nonetheless reduce the question of correlations of two distant eigenvalues to a question of correlations of LSS. At least in the bulk, the latter question is likely within reach of the methods developed in this paper.

\medskip

\noindent{\bf Acknowledgements.} P.L. thanks J. Marcinek for helpful conversations.  The authors thank L. Erd\H{o}s for helpful comments on a draft of this work. 

\subsection{Organization of remainder of paper}

In Section \ref{sec:main} we introduce our model and state our main result, Theorem \ref{thm:main}.  In Section \ref{sec:prelim} we recall various local laws for general Wigner-type matrices.  Section \ref{sec:notation} collects notations that we will use throughout the paper.  In Section \ref{sec:technical} we state various technical results from which our main result follows.  Section \ref{sec:stab} contains stability properties of the quadratic vector equation and estimates for various operators associated with it, as well as estimates on the solution $\bm$ of the quadratic vector equation.  

In Sections \ref{sec:clt} and \ref{sec:var-calc} we compute the characteristic function of linear spectral statistics on relatively large mesoscopic scales. In Section \ref{sec:meso} we state a result on DBM that allows to calculate linear spectral statistics on short mesoscopic scales in terms of linear spectral statistics on larger mesoscopic scales. We use all of these components to prove universality for Gaussian divisible ensembles in Section \ref{sec:rest}.

In Section \ref{sec:dbm-proofs} we prove the results of Section \ref{sec:meso}. In Section \ref{sec:homo} we  apply the homogenization theory of \cite{fixed} to deduce a technical input that is stated in Section \ref{sec:technical}. We prove Theorem \ref{thm:main} in Section \ref{sec:proof-of-main}.  In Section \ref{sec:expect-calc} we calculate the leading-order corrections to the expectation of linear spectral statistics as well as to the expectation of a single eigenvalue.  Various appendices collect some elementary properties of the free convolution and other auxilliary results.


\section{Model and main results} \label{sec:main}

In this section we will first introduce our model and then state our main results.  Following \cite{qve,univ-wig-type} we will consider the following class of $N \times N$ self-adjoint random matrices, known as \emph{matrices of general Wigner-type}.

\bed\label{d:wignertype}
An $N \times N$ real symmetric matrix $W$ is of general Wigner-type if the following hold. For all $i,j \in \unn{1}{N}$, we have
\beq\label{sijdef}
\ee[ W_{ij} ] = 0, \qquad \ee[ (W_{ij} )^2 ] = S_{ij},
\eeq
where
\beq \label{eqn:S-bd}
c \leq N S_{ij} \leq C
\eeq
for some constants $C, c >0$. Further for every integer $p \geq 2$ there is a constant $C_p$ such that
\beq \label{eqn:moment-bd}
\ee[ |W_{ij} |^p ] \leq \frac{C_p}{ N^{p/2}}.
\eeq 
\eed
\remark In this work we consider only real symmetric Wigner matrices for notational simplicity.  Our methods do not depend on the choice of symmetry class and the case of complex Hermitian matrices can be handled by similar methods.\qed

\vspace{5 pt}

Our definition of matrices of general Wigner-type makes a standard assumption on the variance matrix $S$ in \eqref{eqn:S-bd}.  We will make two additional assumptions on $W$ through  
$S$.  
To state these assumptions, we need to define the associated self-consistent density of states.  Define  the upper half-plane
\beq
\hh_+ := \{ z \in \cc : \Im [z] >0 \}.
\eeq
  Following \cite{qve} we consider a quadratic vector equation (QVE) for $z \in \hh_+$ and a vector $\bm_i (z) \in \cc^N$,
\beq \label{eqn:qve-def}
- \frac{1}{ \bm_i (z)} = z + \sum_j S_{ij} \bm_j (z).
\eeq
It was shown in \cite[Theorem 2.1]{qve} that this has a unique solution $\bm : \hh_+ \to \hh_+^N$.   As in \cite{univ-wig-type}, we assume the following uniform (in $N$) bound on the vector $\mb(z)$. 
\begin{enumerate}[label=(A)]
\item \label{it:ass-1} There is a constant $C>0$ such that 
\beq
| \bm_i (z) | \leq C
\eeq
for all $i\in\unn{1}{N}$, $z \in \hh_+$.
\end{enumerate}
Sufficient conditions for this assumption to hold are known \cite[Theorem 2.10]{qve}.  The density of states associated with the matrix $S$ is denoted by $\rho (E)$ and defined by the limit
\beq
\rho (E) := \lim_{ \eta \to 0 }  \frac{1}{ \pi N} \sum_{i=1}^N \Im [ \bm_i (E + \i \eta ) ].
\eeq
We remark that $\rho$ and $\mb$ in fact depend on $N$ through the variance matrix $S$ but we omit this from the notation.

Given Assumption \ref{it:ass-1}, the possible behavior of $\rho (E)$ is summarized in Theorem 4.1 of \cite{univ-wig-type}.     In particular, it is a continuous density of compact support which consists of finitely many intervals.  At the extremal edges, $\rho (E)$ vanishes as a square root.  We will assume that $\rho (E)$ has only one interval of support and is bounded below away from the spectral edges; in particular, we rule out the possibility of small local minima or cusps.  This assumption is similar to the standard one-cut regularity conditions typically made in studying the local statistics of $\beta$-ensembles; see e.g. \cite{KM}. 
\begin{enumerate}[label=(B)]
\item \label{it:ass-2} There exist (sequences of real numbers) $\alpha < \beta$ and constants $C, c >0$ such that $\beta - \alpha >c$, $| \alpha | + | \beta | \leq C$, and 
\beq
\rho (E) = \1_{ \{ \alpha < x < \beta\} } \sqrt{ (E -\alpha )( \beta - E) } F(E),
\eeq
where $c \leq F(E) \leq C$.
\end{enumerate}
Before stating our main result we require the following concept of $N$-quantiles.  We define $\gamma_i$ by,
\beq \label{eqn:quant-def}
\frac{i}{N} = \int_{ \alpha}^{\gamma_i} \rho (E) \d E.
\eeq
Recall the semicircle distribution,
\beq
\rhosc(E) := \frac{1}{ 2 \pi} \sqrt{ (4-E^2)_+},
\eeq
and denote its $N$-quantiles by $\gamsc_i$, defined similarly to \eqref{eqn:quant-def}.   
The following is our main result on the single eigenvalue fluctuations of matrices of general Wigner-type.  
\bet \label{thm:main}
Let $W$ be a matrix of general Wigner-type obeying Assumptions \ref{it:ass-1} and \ref{it:ass-2}.  Let $c >0$ and let $i$ be an index satisfying $c N \leq i \leq (1-c)N$.  Let $F$ be a Schwarz function, and denote the eigenvalues of $W$ by $\{ \lambda_k \}_{k=1}^N$, labelled in increasing order.  There exists a constant $C>0$, depending on $F$, such that the estimate,
\beq
\left| \ee \left[ F \left( \frac{N \rho ( \gamma_i )}{\sqrt{ \log(N) }} ( \lambda_i - \gamma_i ) \right) \right] - \ee^{(\GOE)} \left[ F \left( \frac{N \rhosc ( \gamsc_i )}{\sqrt{ \log(N) }} ( \mu_i - \gamsc_i ) \right) \right] \right| \leq \frac{C}{ \log(N)^{1/10}}
\eeq
holds, where the $\{ \mu_i \}_{i=1}^N$ are the eigenvalues of a GOE matrix in increasing order.  As a consequence, the random variable $\frac{ N \rho ( \gamma_i )}{ \sqrt{ \log(N) }} ( \lambda_i - \gamma_i )$ converges to a normal random variable with a universal variance.
\eet

\remark We expect that the above theorem is true in the absence of Assumption \ref{it:ass-2}, as long as one restricts, for example, to eigenvalues lying in intervals of length order $1$ on which the density of states $\rho$ is bounded below.  Moreover, our proof should apply to this case with some additional attention paid to the degeneration of error estimates near cusps and small local minima in the density of states.  We do not pursue this  primarily in the interest of notational simplicity, as even under Assumption \ref{it:ass-2} we have a rich class of models with non-semicircular spectral measures. \qed 

\vspace{5 pt}

We have the following theorem that calculates corrections to the expectation of a single eigenvalue around its classical location.  For its statement, we introduce the notation $\sfo_{ab}$, which denotes the fourth cumulant of the random variable $\sqrt{N} W_{ab}$.  It generalizes Theorem 1.4 of \cite{meso}. 
\bet \label{thm:sev-expect}
Let $W$ be a matrix of general Wigner-type satisfying Assumptions \ref{it:ass-1} and \ref{it:ass-2}. Let $c>0$ and assume that the index $i_0$ satisfies $c N \leq i_0 \leq (1-c)N $. Then there is a $c_1>0$ such that
\begin{align}
   2 \pi N \ee[ \rho ( \gamma_{i_0} ) ( \lambda_{i_0} - \gamma_{i_0} ) ] &= \Im \left[ \log \det ( 1 - S \mb^2 ( \gamma_{i_0} ) ) \right] + \Im \left[ \tr S \mb^2 ( \gamma_{i_0} ) \right] \nonumber\\
   &- \frac{1}{2} \frac{1}{N^2} \sum_{ij} \sfo_{ij} \Im\left[ ( \mb_i ( \gamma_{i_0} ) \mb_j ( \gamma_{i_0} ) )^2 \right] -\pi+ \O (N^{-c_1} ),
\end{align}
where we have denoted the boundary values by $\mb(E) := \mb(E + \i 0)$. The first three terms on the right side of the above estimate are all $\O(1)$. 
\eet
The above theorem is proven in Section \ref{sec:expect-calc}. Theorem \ref{thm:main} implies the following corollary for the fluctuations of the eigenvalue counting function.  The proof is the same as in \cite{gust,dallaporta2011note} and is omitted.
\bec
Let $W$ be a matrix of general Wigner-type satisfying Assumptions \ref{it:ass-1} and \ref{it:ass-2}.  Fix $\kappa >0$ and let $E \in (\alpha +\kappa, \beta - \kappa)$ be given.  Then the random variable
\beq
\frac{\left| \{  i : \lambda_i \leq E \} \right| - N \int_{-\infty}^E \rho (x) \d x }{\pi^{-1} \sqrt{ \log(N) }}
\eeq
converges in distribution to a standard Gaussian random variable.
\eec

\subsection{Results on linear spectral statistics of matrices of general Wigner-type}

While our main focus is on single eigenvalue fluctuations, our intermediate results may be used to deduce a few results on linear spectral statistics of matrices of general Wigner-type.   These results are essentially reformulations of other results proved elsewhere in our paper, and so the proofs are deferred to Appendix \ref{a:gfct-lss}. 
For mesoscopic linear spectral statistics we have the following.
\bet \label{thm:meso-clt-gen}
Let $W$ be a matrix of general Wigner-type satisfying Assumptions \ref{it:ass-1} and \ref{it:ass-2}.  Let $E_0 \in (\alpha, \beta)$.  Let $g $ be a smooth function of compact support, and let $0 < \omst  < 1$.  Let
\beq
f_N(x) := g (N^{\omst}(x-E_0 ) ).
\eeq
The random variable
\beq
\tr f_N (W) - N \int f_N (x) \rho (x) \, \d x 
\eeq
converges to centered Gaussian random variable with variance
\beq
\frac{1}{ 2 \pi^2} \int \frac{ (g(x) - g(y) )^2}{ (x-y)^2} \,\d x \, \d y.
\eeq
Further, let $h(x)$ be a smooth function such that $h'(x) = 0$ for $|x| > 1$, $h(-1) =0$ and $h(1) = 1$.  Let
\beq
f_N(x) := h (N^{\omst} (x -E_0 ) ).
\eeq
Then, the random variable 
\beq
\frac{1}{ \sqrt{ \log (N) } } \left( \tr f_N (W) - N \int f_N(x) \rho (x) \, \d x \right)
\eeq
converges to a centered Gaussian random variance with variance $ \omega_* / \pi^2$. 
\eet
For global linear spectral statistics the following holds.
\bet \label{thm:global-clt-gen} 
 Let $W$ be a matrix of general Wigner-type satisfying Assumptions \ref{it:ass-1} and \ref{it:ass-2}.  
Let $f$ be a smooth function such that $f'(x) = 0$ if $|x- \alpha | < \kappa$ or $|x - \beta | < \kappa$, for some $\kappa >0$.  Let $\hat{V} (f)$ be as in \eqref{eqn:V-def}.  Then $\hat{V} (f) \leq C$.  If there is a constant $c >0$ such that $\hat{V} (f) \geq c$, then
\beq
\frac{1}{ \hat{V} (f) } ( \tr f (W) - \ee[ \tr f (W) ] )
\eeq
converges to a standard normal random variable.
\eet
\remark We did not obtain a general condition for $\hat{V} (f) \geq c$, as it is not necessary for our main result on single eigenvalue fluctuations.  However, one does have the following.  
Let $g$ be a smooth function of compact support and $E_0 \in ( \alpha, \beta)$.  Let $f (x) = g (r^{-1} (x-E_0 ))$.  Using our methods it is possible to show that
\beq
\lim_{r \dto 0 } \limsup_{N \to \infty} \left| \hat{V} (f) - \frac{1}{ 2 \pi^2} \int \frac{ (g (x) - g(y))^2}{ (x-y)^2} \, \d x\, \d y \right| = 0.
\eeq
so the condition $\hat{V} (f) \geq c$ is seen to hold for some classes of functions. \qed

\subsection{Preliminary results on matrices of general Wigner-type} \label{sec:prelim}

In this section we collect some preliminary results  
on matrices of general Wigner-type that we will use frequently in our work.  In order to state them, introduce the following notion of \emph{overwhelming probability}.
\bed \label{def:op}
We say that a set of events $\{\mathcal A(u)\}_{u \in U^{(N)}}$, where  $U^{(N)}$ is a parameter set which may depend on $N$, holds with \emph{overwhelming probability} if, for any $D>0$, there exists $N\big(D, U^{(N)}\big)$ such that for $ N \geq N\big(D, U^{(N)}\big)$,
\begin{equation}
 \inf_{u \in U^{(N)}} \pp \left( \mathcal A(u) \right) \geq 1 - N^{-D}.
\end{equation}
\eed

For an $N \times N$ symmetric matrix $W$ of general Wigner-type and $z \in \cc \backslash \rr$, we define the Green's function (or resolvent) and empirical Stieltjes transform by
\begin{equation}
G(z) = ( W - z)^{-1}, \qquad m_N (z) = \frac{1}{N} \tr \left( \frac{1}{ W - z } \right).
\end{equation}
We define the function $m (z)$ by
\beq
m (z) = \frac{1}{ N} \sum_{i=1}^N \bm_i (z),
\eeq
where $\bm (z) = ( \bm_i (z))_i$ is the solution to the QVE in \eqref{eqn:qve-def}.  

We will require the following result, which is a specialization  of \cite[Theorem 1.7]{univ-wig-type} to our setting.  In particular, under our assumptions, the function $\kappa(z)$ defined in (1.23) of \cite{univ-wig-type} is bounded.

\bet \label{thm:local-law}
Let $W$ be a matrix of general Wigner-type obeying Assumptions \ref{it:ass-1} and \ref{it:ass-2}.  Let $\gamma >0$ and $\eps >0$.  Then with overwhelming probability, for any $z = E + \i \eta$ satisfying
$
\frac{N^{\gamma}}{N} \leq \eta \leq \gamma^{-1}$ and $ |E| \leq \gamma^{-1},
$
we have the estimates
\beq \label{eqn:entry-ll}
\max_{i, j} |G_{ij} (z) - \bm_i (z) \delta_{ij} | \leq N^{\eps} \left( \sqrt{ \frac{ \Im m (z) }{ N \eta } } + \frac{1}{N \eta} \right),
\eeq
and
\beq \label{eqn:ll}
\left| m_N (z) - m (z) \right| \leq N^{\eps} \frac{1}{ N \eta}.
\eeq
\eet
We have also the following, from (1.21) of \cite{univ-wig-type} and \cite[Theorem 1.13]{univ-wig-type}.
\bet
Let $u, w, v \in \rr^N$ satisfy $\|u_i\|_\infty \leq 1$ and $\|w\|_2 = \|v\|_2 = 1$.  Let $\eps>0$. Under the assumptions of the previous theorem we have with overwhelming probability that
\beq \label{eqn:tr-ll}
\left| \frac{1}{N} \sum_{i=1}^N u_i (G_{ii} (z) - \bm_i (z) ) \right| \leq \frac{N^\eps}{N \eta}
\eeq
and
\beq \label{eqn:iso}
\left| \sum_{i, j=1}^N w_i G_{ij} (z) v_j - \sum_{i=1}^N \bm_i (z) w_i  v_i \right| \leq N^{\eps}  \left( \sqrt{ \frac{ \Im \mb (z) }{ N \eta } } + \frac{1}{N \eta} \right).
\eeq
\eet
Finally, the following rigidity result holds \cite[Corollary 1.11]{univ-wig-type}.
\bet \label{thm:rigidity}
Let $\{ \lambda_i \}_{i=1}^N$ denote the eigenvalues of a matrix of general Wigner-type, and assume that Assumptions \ref{it:ass-1} and \ref{it:ass-2} hold.  Let $\eps >0$.  Then with overwhelming probability we have for any $i$ that
\beq \label{eqn:rig}
| \lambda_i - \gamma_i | \leq \frac{ N^{\eps}}{N^{2/3} \min\{ i^{1/3}, (N+1-i)^{1/3} \}}.
\eeq
\eet

\subsection{Notation} \label{sec:notation}

In this section we collect some notation that we will use in the remainder of the paper, as well as list some notation already defined, in order to organize all of the notation of the paper in a single subsection.  We have already defined our notion of high probability events in Definition \ref{def:op} (\emph{overwhelming probability.})  If $S$ is the matrix of variances associated with a matrix of general Wigner-type, we recall the solution $\mb(z)$ to the QVE \eqref{eqn:qve-def}, and denote
$
m(z) = \frac{1}{N} \sum_i \mb_i (z)$ and $ \rho(E) = \frac{1}{ \pi} \lim_{\eta \dto 0 } \Im[m(z) ].
$
Note that the functions $\mb_i (z)$ satisfy $\mb_i (z) = \bar{ \mb}_i ( \bar{z} )$ and by \cite{univ-wig-type}[Corollary 1.3], admit continuous extensions to the closed upper or lower half-planes. We will denote their boundary values by,
\beq
\mb_i (E \pm \i 0) = \lim_{ \eta \dto 0} \mb_i (E \pm \i \eta ).
\eeq
Throughout this paper we will consider various quantities that depend on $\mb(z)$, (such as $F(z, w)$ defined below) and their boundary values will also be denoted using the notation $E \pm \i 0$. Note that since $|\mb_i (E + \i 0) | = | \mb_i (E- \i 0)|$, if a quantity depends only on the absolute value of $\mb(z)$ then we will sometimes omit the $\pm \i 0$ from the argument.  A similar statement holds for the real part of $\mb (z)$.

We will often identify the vector $\mb(z)$ with the diagonal matrix with $i$th entry equal to $\mb_i(z)$.  For example, the matrix $(1- \mb^2 S)^{-1}$ plays an important role in the study of matrices of general Wigner-type.  We also define the $N \times N$ matrix
\beq \label{eqn:F-def}
F(z, w) := | \mb(z) \mb(w) |^{1/2} S | \mb(z) \mb(w) |^{1/2},
\eeq
where $| \mb(z) \mb(w) |^{1/2}$ denotes the diagonal matrix with $i$th entry $| \mb_i(z) \mb_i(w) |^{1/2}$.

We will use $c$ and $C$ to denote arbitrary small and large positive constants, respectively, whose specific values may change from line-to-line (finitely many times).  These constants will often depend on $S$, but only through the model parameters in \eqref{eqn:S-bd} and Assumptions \ref{it:ass-1} and \ref{it:ass-2}, and possibly on the moment bounds \eqref{eqn:moment-bd}.  For two positive (and possibly $N$-dependent) parameters $a_N$ and $b_N$, the notation 
\beq
a_N \asymp b_N
\eeq
means that there are constants $c, C >0$ such that $c a_N \leq b_N \leq C a_N$. 


For vectors $v \in \cc^N$ we denote the $\ell^p$ norms by
\beq
\| v \|_p := \left( \sum_i |v_i|^p \right)^{1/p}, \qquad \|v\|_\infty = \max_i |v_i|.
\eeq
The operator norm of an $N \times N$ matrix $A :\ell^p \to \ell^q$ is denoted,
\beq
\| A\|_{\ell^p \to \ell^q} := \sup_{ v : \|v\|_p=1} \|Av\|_q.
\eeq 
We denote the inner product on $\cc^N$ and  sum of a vector $v \in \cc^N$ 
\beq
\langle v, w \rangle := \sum_i \bar{v}_i w_i, \qquad \langle v \rangle := \sum_i v_i.
\eeq

We introduce the following notion of bulk intervals.
\begin{definition}
Let $W$ be a matrix of general Wigner-type with variance matrix satisfying Assumptions \ref{it:ass-1} and \ref{it:ass-2}.  For $\kappa>0$, we define the interval
\beq \label{eqn:Ikap-def}
\Ikap := [ \alpha + \kappa, \beta -\kappa ].
\eeq
\end{definition}
We will make frequent use of the following notion of test function.
\begin{definition} \label{def:half-f}
A smooth real-valued function $f$ is said to be a \emph{half-regular bump function} with data $(t, M, E_0, E_1, C', c')$ if the following holds.  
(The parameters $t, M, C'$ and $c'$ are all positive, and $E_0, E_1 \in \rr$.)
First, there is a $\kappa>0$ such that $E_0, E_1 \in \Ikap$, and the parameters $c'$ and $tM$ satisfy $tM, c' < \kappa/10$.  Secondly, $f (x) \geq 0$ and $f$ is compactly supported.  Furthermore, $f'(x) \neq 0$ only for $|x-E_0| \leq tM$ or $|x-E_1 | \leq c'/2$, and we have
\beq
0 \leq f^{(k)}(x) \leq \frac{1}{t^{k-1}} \frac{C' t}{ (x-E_0)^2 + t^2}, \quad |x-E_0| \leq t M, \qquad k=1, 2
\eeq
and
\beq
|f^{(k)}(x) |\leq C', \quad |x-E_1| \leq c'/2, \qquad k=1, 2.
\eeq
Furthermore, for $E_0 + t M < x < E_1 - c'/2$ we have $f(x) =1$.  Finally,
\beq
\|f\|_1 + \|f'\|_1 \leq C', \qquad \|f''\|_1 \leq \frac{C'}{t}.
\eeq
We also demand that $E_1 -E_0 \geq c'$. 
\end{definition}
We will only consider half-regular bump functions with $t = o(1)$ and $M \to \infty$ sufficiently slowly so that $Mt \to 0$ (they will depend on $N$ as $t = N^{-\omega}$ and $M = N^{\delta_M}$ with some $\delta_M < \omega$), and $E_0$ and $E_1$ being close, but an order $1$ distance from each other.  

Finally, we need a certain spectral domain on which the operators we introduce have good properties.  Let $U_*>0$ be a constant such that
\beq
\|S\|_{\ell^2 \to \ell^2} + \|S\|_{\ell^\infty \to \ell^\infty}  + \sup_{z\in \mathbb C} \|\mb(z) \|_\infty +1 \leq \frac{U_*}{10},
\eeq
and 
\beq
\| \bm (z) \|_\infty \leq \frac{U_*} {10 |z|}.
\eeq
We will see that such a constant exists by the results of Section \ref{sec:stab}.  Hence, there is an $L_*>1+ \max\{ |\alpha|, |\beta|\}$ (where $\alpha$ and $\beta$ are as in Assumption \ref{it:ass-2}) such that, if either the real part of $| \Re[z] | \geq L_*$ or $| \Im[z]| \geq L_*$ then,
\beq
\sup_{w\in \mathbb C} \| |\mb(z) \mb(w)| S \|_{\ell^2 \to \ell^2} + \sup_{w\in \mathbb C} \| |\mb (z) \mb (w) | S \|_{\ell^\infty \to \ell^\infty} \leq \frac{1}{2}.
\eeq
We will need the following domain over which we will integrate various quasi-analytic extensions of test functions.
\begin{definition} \label{def:dc}
For any $\kappa>0$ the domain $\Dkap$ is defined by
\begin{align}\label{d:dkap}
\Dkap := \{ z : \Re[z] \in \Ikap \} \cup \{ 2L_* < |\Re[z] | \leq 4L_* \} \cup \{ 2 L_* < | \Im[z] | < 3 L_* \}.
\end{align}
\end{definition}

Finally, we require another class of test functions.
\begin{definition} \label{def:reg-f}
A smooth test function $f$ is called regular with data $(t, c', C')$ if $f$ is of compact support and,
\beq
\| f \|_1 + \|f'\|_1 \leq C', \qquad \| f''\|_1 \leq \frac{C'}{t}
\eeq
and $f'(x)$ is non-zero only if either $x \in I_{c'}$ or $2L_* < |x| < 2 L_* +1$.
\end{definition}
Note that half-regular bump functions are in fact regular.  We will be able to calculate the characteristic function of the linear spectral statistics associated with regular test functions (for $t$ not too small).  However, it is only for functions whose support is restricted to  a small order $1$ interval for which we will be able to explicitly calculate leading-order contributions to the variance arising in the CLT.  Half-regular bump functions arise naturally from our application of the homogenization theory of \cite{fixed} and so we single them out as test functions with variance of the order $\log(N)$ with a leading order contribution that we are able to calculate.

\subsection{Definition of free convolution} \label{sec:fc-def}

Given a probability measure $\mu$ with Stieltjes transform $m_\mu (z)$, the free convolution of $\mu$ with the semicircle distribution at time $t$ is the following construction from free probability (we refer the reader to \cite{biane} for more details). Consider the following functional equation for $m_{\mu, t} (z): \cc_+ \to \cc_+$,
\beq
m_{\mu, t} (z) = m_\mu (z + t m_{\mu, t} (z) ).
\eeq
There is a unique solution obeying the condition $m_{\mu, t} (z) \sim |z|^{-1}$ at $|z| \sim \infty$.  The measure $\mu_t$ is defined via the boundary value,
\beq
\mu_t (E) \d E = \frac{1}{ \pi} \lim_{ \eta \dto 0 } \Im [ m_{\mu, t} (E + \i \eta ) ] \d E.
\eeq
For $ t>0$, the measure $\mu_{t}$ has a density, is of bounded support, and is analytic on the interior of its support \cite{biane}.

Given a density $\rho(E)$, which in our applications will always be the spectral measure associated with a matrix of variances $S$ satisfying Assumptions \ref{it:ass-1} and \ref{it:ass-2}, we will denote its free convolution with the semicircle distribution at time $t$ by $\rho_t (E)$.  Standard calculations give the following. A detailed proof can be found in Appendix \ref{a:free-1}. 
\bel \label{lem:free-conv-aa1}
If $\rho (E)$ comes from a matrix of general Wigner-type satisfying Assumptions \ref{it:ass-1} and \ref{it:ass-2}, then there is an $\eps >0$ such that for $t < \eps_0$, $\rho_t$ comes from a variance matrix $S_t$ that satisfies the same assumptions, after adjusting constants appropriately. Moreover, the spectral edges are within $\O ( t)$ of each other. 
\eel

We remark here that the role of the free convolution in the present work is that it gives the spectral measure of Dyson Brownian motion,  a stochastic eigenvalue flow at time $t$.  Equivalently, it is the spectral measure for the matrix
$
W_t := W +\sqrt{t} G
$
where $W$ is the matrix of general Wigner-type with matrix of variances $S$ and $G$ is a GOE matrix independent of $W$.

\section{Main technical inputs to proof of Theorem \ref{thm:main}} \label{sec:technical}

In this section we collect statements of several technical results which will be used to prove our main result, Theorem \ref{thm:main}.  



The first result we give here reduces single eigenvalue fluctuations for Gaussian divisible ensembles to studying the mesoscopic fluctuations of linear spectral statistics.  It is a straightforward application of the homogenization results of \cite{fixed} and is proven in Section \ref{sec:homo}.

\bet \label{thm:homog}
Let $W$ be a matrix of general Wigner-type satisfying Assumptions \ref{it:ass-1} and \ref{it:ass-2}, with spectral measure $\rho$.  Recall that $\rho_t$ denotes the free convolution of $\rho$ with the semicircle distribution at time $t$, with quantiles $\gamma_{i, t}$.  Fix two times $t_0 = N^{\tau_0-1}$ and $t_1 = N^{\tau_1-1}$, and an index $i_0$ satisfying $c' N \leq i_0 \leq (1-c') N'$.  Assume that $0 < \tau_1 < 1 /100$ and $0< 1-\tau_0 < 1/10$.  Fix $\omega >0$. 

There is a function $f(x)$ such that the following hold.
\begin{enumerate}
\item We have that $f( - t_1 N^{\omega} ) = 0$, $f( t_1 N^{\omega} ) = 1$.
\item $f'(x) \neq 0$ only for $|x| \leq t_1 N^{\omega}$. 
\item The estimate
\beq
0 \leq f'(x) \leq \frac{C t_1}{x^2 +(t_1)^2 }
\eeq
holds.
\item For higher $k$, 
\beq
|f^{(k)} (x) | \leq \frac{C_k}{ t_1^{k-1}} \frac{t_1}{x^2 + t_1^2}.
\eeq
\end{enumerate}
There are stochastic processes $\{ x_i (t) \}_{i=1}^N$, $\{ y_i (t) \}_{i=1}^N$ and $ \{ z_i (t) \}_{i=1}^N$ such that the following hold. 
\begin{enumerate}
\item The particles $\{x_i (0) \}_{i=1}^N$ are independent from $\{ y_i (t), z_i (t) \}_{i, t}$. Analogous statements hold  for $\{y_i (0) \}_{i=1}^N$ and $\{ z_i (0) \}_{i=1}^N$.  
\item For each $t$, the marginal distribution of $\{ x_i (t) \}_{i=1}^N$ is that of the eigenvalues of $W + \sqrt{t_0+t} G$.  
\item Let $a = \rho_t ( \gamma_{i_0, t} )/ \rhosc (0)$.  For each fixed $t$, the marginal distribution of $\{y_i (t) \}_{i=1}^N$ is that of the eigenvalues of $\sqrt{1+a^2 t } G$.  The same statement holds for $\{ z_i (t) \}_{i=1}^N$ for any fixed $t$.
\end{enumerate}
The estimates
\begin{align}
& \rho_{ t_0} ( \gamma_{i_0, t_0 } ) ( x_{i_0} (t_1) - \gamma_{i_0, t_0+t_1} ) - \rhosc (0) y_{N/2} (t_1) \\
& =\frac{1}{N}  \left( \sum_j f (   \rho_{ t_0} ( \gamma_{i_0, t_0 } )  ( x_j(0)  - \gamma_{i_0, t_0} ) ) - \int ( f ( \rho_{ t_0} ( \gamma_{i_0, t_0 } )  ( s - \gamma_{i_0, t_0} ) )  \rho_{ t_0} (s)\d s \right) \\
&-  \frac{1}{N} \left( \sum_j f ( \rhosc (0) y_j (0)) - \int f( \rhosc(0) s) \rhosc ( s) \d s\right) + \O (N^{-1-\tau_1/100} + N^{-1-\omega/3} )
\end{align}
and a similar estimate with $\rhosc(0) z_{N/2} (t)$ replacing $\rho_{ t_0} ( \gamma_{t_0, i_0 } ) ( x_{i_0} (t) - \gamma_{t_0+t, i_0 } ) $ hold.
\eet

The following result computes to leading order the characteristic function for mesoscopic linear spectral statistics of the form coming from Theorem \ref{thm:homog}.  It is proven in Section \ref{sec:final-clt-homog}.

\bet \label{thm:final-clt-homog}
Let $W$ be a matrix of general Wigner-type satisfying Assumptions \ref{it:ass-1} and \ref{it:ass-2}. Let $\rho(x)$ be the associated spectral measure with free convolution denoted by $\rho_t$.  Let $t_0 = N^{-\tau_0}$ satisfy $0 < \tau_0 < 1/10$.   Let $t_1 = N^{\omega_1-1}$ be a scale with $1 > \omega_1 > 0$ and also $\tau_0 < 1 - \omega_1$.  Let $M = N^{\delta_M}$ with $0 < \delta_M < \omega_1 /100$.  Let $E_0 \in I_c$ for some fixed $c>0$ and $p(x)$ be a function satisfying
\beq
0 \leq p'(x) \leq C \frac{t_1}{ (x-E_0)^2 + t_1^2 }, \qquad |p''(x) | \leq \frac{C}{ (x-E_0)^2 + t_1^2}
\eeq
and $p'(x) = 0$ for $|x-E_0 | > t_1 M$, as well as $p (-\infty) = 0$ and $p ( \infty ) = 1$.  Let $G$ be an independent GOE matrix and denote
\beq
W_t := W + \sqrt{t} G.
\eeq
Then, for $| \lambda| \leq \log(N)^{1/4}$, we have
\begin{align}
\ee\left[ \exp\left[ \i  ( \log(N) )^{-1/2} \lambda \left( \tr \,p ( W_{t_0}) - N \int p(x) \rho_{t_0} (x) \d x \right) \right] \right] =& \exp \left[ - \frac{\lambda^2}{2} \frac{ |\log(t_1) |}{ \pi^2 \log(N) }\right] \\
+& \O ( ( \log(N))^{-1/4} ).
\end{align}
\eet

The previous two results are used to prove the following.  It is proven in Section \ref{sec:gde-univ} and proves Theorem \ref{thm:main} for Gaussian divisible ensembles.


\bet \label{thm:gde-univ}
Let $W$ be a matrix of general Wigner-type satisfying Assumptions \ref{it:ass-1} and \ref{it:ass-2}.  Let $t_0 = N^{- \tau_0}$ and $t_1 = N^{\omega_1 -1}$ with $0 < \tau_0 < 1/10$ and $0 < \omega_1 < 1/1000$.  Let $i_0$ be an index satisfying $c_1 N \leq i_0 \leq (1-c_1) N$ for some $c_1 >0$.  Let $F$ be a Schwartz function.  Let $G$ be a GOE matrix, independent of $W$.  Denote the spectral measure associated with $W$ by $\rho$  and the free convolution at time $t$ by $\rho_t$, with $N$-quantiles denoted by $\gamma_{i, t}$.  Denote
\beq
W_t := W + \sqrt{t} G.
\eeq
 Then
\begin{align}
& \left| \ee\left[ F \left( \frac{ N \rho_{t_0+t_1} ( \gamma_{i_0, t_0+t_1} )}{ \sqrt{ \log(N) }} ( \lambda_{i_0} (W_{t_0+t_1} ) - \gamma_{i_0, t_0+t_1}  )\right) \right] - \ee \left[ F \left( \frac{N \rhosc (0) }{ \sqrt{ \log(N) } }\lambda_{N/2} (G) \right)\right] \right|\nonumber \\
 \leq & C ( \log(N) )^{-1/10}
\end{align}
\eet

\section{Estimates on stability operator and related quantities} \label{sec:stab}

In this section we collect properties of the solution to the quadratic vector equation \eqref{eqn:qve-def} and associated operators, as well as properties of QVE solutions corresponding to variance matrices that are small perturbations of a matrix $S$ satisfying Assumptions \ref{it:ass-1} and \ref{it:ass-2}.

The methods in this section follow closely those of \cite{qve}. See also \cite{mde-notes} for a pedagogical treatment.  One of the main differences is that our stability operator $F(z, w)$ defined in \eqref{eqn:F-def} involves two complex parameters where \cite{qve} handles only the diagonal $F(z, z)$. However, a basic convexity result, Lemma \ref{l:convexity} below, shows that the behavior of $F(z, w)$ is no worse than the diagonal operator.

\subsection{Properties of \texorpdfstring{$\bm$}{m} and stability operator}

In this section we summarize a few properties of the solution $\bm(z)$ of the quadratic vector equation.  The following is a consequence of Theorem 4.1 of \cite{univ-wig-type}.
\bep Let $S$ be the matrix of variances of a matrix of general Wigner-type satisfying Assumptions \ref{it:ass-1} and \ref{it:ass-2}.  Then for every $i \in \unn{1}{N}$ there is a probability density $\rho_i (E)$ supported on $[\alpha, \beta]$ such that
\beq \label{eqn:rho-i}
\mb_i (z) = \int \frac{ \rho_i (x) }{ x- z } \d x
\eeq
and moreover there are constants $c, C$ such that
\beq
c \rho_i (E) \leq \rho (E) \leq C \rho_i (E)
\eeq
for every $i$.
\eep

The following collects some elementary estimates for $\mb(z)$.  It is a consequence of Theorem 7.2.2 and Lemma 7.3.2 of \cite{mde-notes} and the representation \eqref{eqn:rho-i}.  Note that \eqref{eqn:im-mb} follows from the representation \eqref{eqn:rho-i} and Assumption \ref{it:ass-2}. 
\bep \label{prop:basic-m}
Let $S$ satisfy Assumptions \ref{it:ass-1} and \ref{it:ass-2}.  Every component of $\mb(z)$ satisfies
\beq \label{eqn:m-size}
\frac{c}{1 + |z| } \leq | \mb_i (z) | \leq \frac{C}{1 + |z|}.
\eeq
In particular each component satisfies $|\mb_i (z) | \asymp 1$ in any compact region of $\cc$. 

For any $C'>0$ we have in the domain $|z| \leq C'$ that
\beq
\| \del^k_z \mb (z) \|_\infty \leq C_k \left( \frac{1}{N} \sum_i \Im [ \mb_i (z) ]^2 \right)^{-j_k}
\eeq
for some $C_k >0$ and exponent $j_k >0$ (with $j_1 =1$).  

For any $c'>0$ and $z$ satisfying $| \Re[z] | \geq \max \{ |\alpha|, |\beta|\} + c'$ or $|\Im[z]| \geq c'$ we have,
\beq
\| \partial^{k}_z \mb_i(z)\|_\infty \leq C_k.
\eeq

For any $\kappa >0$ and $C' >0$ such that $\Re[z] \in \Ikap$, and $\Im[z] \geq 0$ and $|z| \leq C'$ we have
\beq \label{eqn:im-mb}
c \leq \Im [ \mb_i (z) ] \leq C.
\eeq
\eep

During our proofs we will require estimates for operators like 
\beq
\frac{1}{ 1 - \bm (z) \bm (w) S }, \qquad \frac{1}{1 - S \mb(z) \mb(w) } ,
\eeq
where we have identified $\mb$ with the diagonal matrix whose $i$th entry is $\mb_i$.   We first prove the following.
\bep \label{prop:F-est}
Let $S$ be the variance matrix of a matrix of general Wigner-type matrix that satisfies Assumptions \ref{it:ass-1} and \ref{it:ass-2}.  Let
\beq\label{Fdef}
F(z, w) := | \mb(z) \mb(w)|^{1/2} S | \mb(z) \mb(w)|^{1/2} .
\eeq
This is a symmetric matrix with positive entries and its eigenvalue that is largest in magnitude is positive and simple, with an $\ell^2$-normalized eigenvector $v(z, w)$ that has strictly positive entries.  We have the estimate
\begin{align} \label{eqn:F-2-2}
\| F(z, w) \|_{\ell^2 \to \ell^2} &\leq 1 -\frac{1}{2} \left( | \Im[w]| \frac{ \langle v (w, w) | \mb (w) | \rangle }{ \langle v(w, w)  \frac{ | \Im[\mb(w) ]|}{ | \mb(w) |} \rangle } +| \Im[z]| \frac{ \langle v (z, z) | \mb (z) | \rangle }{ \langle v(z, z)  \frac{ | \Im[\mb(z) ]|}{ | \mb(z) |} \rangle }  \right).
\end{align}
In particular $\|F(z, w) \|_{\ell^2 \to \ell^2} \leq 1$.  

If there is a $C' > 0$ such that $|z|, |w| \leq C'$ then the following estimates hold.  If $\kappa >0$ and $\Re[z], \Re[w] \in \Ikap$ then
\beq \label{eqn:F-norm-bulk}
\|F(z, w) \|_{\ell^2 \to \ell^2} \leq 1 - c( | \Im[z]| + | \Im[w]| ).
\eeq
If there is a $c'>0$ such that $\Re[z] \geq \beta + c'$ or $\Re[z] \leq \alpha - c'$ or $| \Im[z] | \geq c'$ then
\beq \label{eqn:F-out}
\|F(z, w) \|_{\ell^2 \to \ell^2} \leq 1 - c,
\eeq
and by symmetry the estimate holds if $w$ satisfies any of these constraints.  Finally, for $|z|, |w| \leq C'$, the entries of $v(z, w)$ are all comparable in magnitude,
\beq \label{eqn:v-size}
c \leq \sqrt{N} v(z, w) \leq C,
\eeq
and moreover there exists a $c_g>0$ such that
\beq \label{eqn:F-gap}
\inf_{j \neq 1} \left| \left( \lambda_1 (F (z, w) ) - | \lambda_j (F(z, w) )| \right) \right| \geq c_g.
\eeq
Finally, there is a $C'' >0$ such that if either $|z|$ or $|w| > C''$, then,
\beq \label{eqn:F-large-z}
\|F(z, w)\|_{\ell^2 \to \ell^2 } \leq \frac{1}{2}.
\eeq
\eep

Before proving the above proposition we require the following elementary lemma.

\bel\label{l:convexity}
Let $A$ and $B$ be diagonal matrices with positive entries and let $S$ be a symmetric matrix with positive entries.  For $0 < s <1$ we have
\beq
\| A^s B^{1-s} S B^{1-s} A^s \|_{\ell^2 \to \ell^2 } \leq s \| A S A\|_{\ell^2 \to \ell^2 } + (1-s) \|B S B \|_{\ell^2 \to \ell^2}
\eeq
\eel
\proof  Since the matrix $A^s B^{1-s} S B^{1-s} A^s$ is symmetric and has positive entries, the Perron--Frobenius theorem asserts that there is a positive eigenvalue $\lambda$ that is the largest in magnitude of all of its eigenvalues and whose $\ell^2$-normalized eigenvector $f$ has positive entries.  Since the matrix is symmetric, necessarily the $\| \cdot\dot \|_{\ell^2 \to \ell^2}$ norm equals $\lambda$. Denote the diagonal entries of $A$ and $B$ by $a_i$ and $b_i$, respectively. Then
\beq      \label{eqn:fsum}
\lambda = \sum_{i, j} f_i a_i^{s} b_i^{1-s} S_{ij} b_j^{1-s} a_j^s f_j = \sum_{i, j} (f_i f_j S_{ij} ) ( a_i a_j)^s (b_i b_j)^{1-s}.
\eeq
The function $s \to a^s b^{1-s}$ is convex, and \eqref{eqn:fsum} is a positive linear combination of convex functions, so it too is convex as a function of $s$.  Therefore,
\begin{align} 
\lambda \leq s \langle f, A SA f \rangle + (1-s) \langle f, B S B f \rangle \leq s \|ASA\|_{2 \to 2} + (1-s) \|B S B\|_{2 \to 2},
\end{align}
and the claim follows.
\qed 

\noindent{\bf Proof of Proposition \ref{prop:F-est}.} 
Applying Lemma \ref{l:convexity} we see that,
\beq
\|F(z, w) \|_{\ell^2 \to \ell^2} \leq \frac{1}{2} \left( \|F(z, z) \|_{\ell^2 \to \ell^2} + \| F(w, w) \|_{\ell^2 \to \ell^2} \right)
\eeq
and so \eqref{eqn:F-2-2} follows from Proposition 7.2.9 of \cite{mde-notes}.  Since $F$ is symmetric and has positive entries, by the Perron-Frobenius theorem, its eigenvalue of largest magnitude is positive and simple, and the corresponding eigenvector can be taken to have positive entries. 

If $|z|, |w| \leq C'$ then
\beq
\frac{c}{N} \leq F_{ij} \leq \frac{C}{N}
\eeq
by \eqref{eqn:m-size}.  Hence, $\|F(z, w) \|_{\ell^2 \to \ell^2} \geq c$ and moreover \eqref{eqn:v-size} holds.  The claim \eqref{eqn:F-gap} then follows from Lemma 7.5 of \cite{qve}.  The estimate \eqref{eqn:F-norm-bulk} follows from \eqref{eqn:F-2-2}, \eqref{eqn:im-mb}, \eqref{eqn:v-size} and \eqref{eqn:m-size}.  The estimate \eqref{eqn:F-out} follows from the fact that 
\beq
c | \Im[z] | \leq | \Im[ \mb_i (z) ] | \leq C |\Im[z]|,
\eeq
for such $z$.  Finally, \eqref{eqn:F-large-z} follows from \eqref{eqn:m-size} applied to the definition \eqref{Fdef} of $F(z,w)$. 
\qed

From Proposition \ref{prop:F-est} we obtain various bounds on the stability operators that we encounter in our proofs, summarized in the following proposition.

\bep \label{prop:stab-3}  Let $C'>0$ such that $|z|, |w| \leq C'$.  Fix $\kappa >0$.
For any $z, w $ with $ \Re [z], \Re [w] \in \Ikap$ we have,
\beq
\left\| (1 - \bm(z) \bm (w) S )^{-1} \right\|_{\ell^2 \to \ell^2 } + \left\| ( 1 - \bm(z) \bm (w) S )^{-1} \right\|_{\ell^\infty \to \ell^\infty } \leq C  (| \Im [z]| + | \Im [w] |)^{-1} ,
\eeq
and
\beq
\left\| ( 1 -  S \bm(z) \bm (w) )^{-1} \right\|_{\ell^2 \to \ell^2 } + \left\| ( 1 -S   \bm(z) \bm (w) )^{-1} \right\|_{\ell^\infty \to \ell^\infty } \leq C ( \Im [z]| + | \Im [w] | )^{-1}.
\eeq
Let $c' >0$.  If either $\Re[z] \geq \beta + c'$ or $\Re[z] \leq \alpha - c'$ or $| \Im[z] | \geq c'$ then,
\beq \label{eqn:stab-out-1}
\left\|  ( 1 - \bm(z) \bm (w) S )^{-1} \right\|_{\ell^2 \to \ell^2 } + \left\| ( 1 - \bm(z) \bm (w) S )^{-1} \right\|_{\ell^\infty \to \ell^\infty } \leq C
\eeq
and
\beq \label{eqn:stab-out-2}
\left\| ( 1 -  S \bm(z) \bm (w) )^{-1} \right\|_{\ell^2 \to \ell^2 } + \left\| ( 1 -S   \bm(z) \bm (w) )^{-1} \right\|_{\ell^\infty \to \ell^\infty } \leq C
\eeq
Moreover, there is a $C''>0$ such that if either $|z|$ or $|w| > C''$ then \eqref{eqn:stab-out-1} and \eqref{eqn:stab-out-2} hold.
\eep
\proof  Let $U$ be the unitary operator $U = \bm(z) \bm (w) / | \bm(z) \bm(w) |$.  Then,
\begin{align}
\frac{1}{ 1 - \bm (z) \bm (w) S } &= \frac{1}{ 1 - U | \bm(z) \bm (w) |S} = \frac{1}{U^*-| \bm(z) \bm (w) |S} U^* \nonumber \\
&= | \bm(z) \bm (w) |^{-1/2} \frac{1}{ U^* - | \bm(z) \bm (w) |^{1/2}S | \bm(z) \bm (w) |^{1/2} } | \bm(z) \bm (w) |^{1/2} U^*.
\end{align}
The $\ell^2 \to \ell^2$ bounds for $(1- \mb(z) \mb(w) S)^{-1}$ then follow from 
\eqref{eqn:m-size} and \eqref{eqn:F-norm-bulk}, and the $\ell^2 \to \ell^2$ bounds for $(1- S \mb(z) \mb(w) )^{-1}$ follow from a similar factorization.

  The $\ell^\infty $ bounds follow in the same manner as the proof of (7.3.4) via (7.4.10) in \cite{mde-notes}.  That is, given a matrix $R$, we have
\beq
\frac{1}{1-R} = 1 + R + R \frac{1}{1 -R} R.
\eeq
If $|R_{ij}| \leq C /N$ then, 
\beq
\|R\|_{\ell^\infty \to \ell^\infty} \leq C, \qquad \|R\|_{\ell^2 \to \ell^\infty} \leq CN^{-1/2}, \qquad \|R\|_{\ell^\infty \to \ell^2} \leq CN^{1/2}.
\eeq
Applying this with $R = \mb(z) \mb(w) S$ or $R = S \mb(z) \mb(w)$ yields the claim.
 \qed

We also note that the above estimates can be used to derive the following estimates on the matrix elements of the stability operator.

\bep \label{prop:stab-elements}
Let $C'>0$ such that $|z|, |w| \leq C'$.  Fix $\kappa >0$.
For any $z, w $ with $ \Re [z], \Re [w] \in \Ikap$ we have
\beq\label{e:stab-elements}
\left| \left[ ( 1 - \bm (z) \bm (w) S )^{-1} \right]_{xy} \right| \leq \delta_{xy}+ C N^{-1} ( | \Im[z] | + | \Im[w] | )^{-1}
\eeq
Let $c' >0$.  If either $\Re[z] \geq \beta + c'$ or $\Re[z] \leq \alpha - c'$ or $| \Im[z] | \geq c'$ then
\beq
\left| \left[ (1 - \bm (z) \bm (w) S )^{-1} \right]_{xy} \right| \leq \delta_{xy} + C N^{-1}.
\eeq
There is a $C'' >0$ such that if either $|z|$ or $|w| > C''$ then the above estimate also holds.  All of the above estimates also hold for the matrix elements of $(1- S \mb(z) \mb(w) )^{-1}$.
\eep
\proof 
From the expansion
\beq
\frac{1}{1 - R} = 1 + R + R \frac{1}{1-R} R
\eeq
and the fact that
\beq
\left[ R \frac{1}{1-R} R\right]_{xy} \leq ( \sup_{1 \leq a,b\leq N} |R_{ab}|^2 ) N \left\|(1- R)^{-1} \right\|_{\ell^\infty \to \ell^\infty},
\eeq
we obtain the conclusion using  Proposition \ref{prop:stab-3} and \eqref{eqn:m-size}. 
\qed

If $z$ and $w$ are in the same half-space, then we have a better estimate.
\bel \label{lem:stab-bulk}
Let $z,w \in \cc$, and $\kappa >0$. If $(\Im z) (\Im w) >0$  and $\Re [z], \Re [w] \in \Ikap$, we have
\beq \label{eqn:stab-a1}
\left\| (1 - \mb(z) \mb (w) S )^{-1} \right\|_{\ell^\infty \to \ell^\infty} + \left\| (1 - S  \mb(z) \mb (w) )^{-1} \right\|_{\ell^\infty \to \ell^\infty}  \leq C,
\eeq
as well as
\beq\label{claim555}
\left| \left[ ( 1 - \mb(z) \mb(w) S )^{-1} \right]_{xy}\right| + \left| \left[ (1 -  S \mb(z) \mb(w) )^{-1} \right]_{xy}\right|  \leq \delta_{xy} + C N^{-1}.
\eeq
\eel
\proof Define $U$ as in the proof of Proposition \ref{prop:stab-3}. By following the proof of the $\ell^\infty$ bounds in that proposition, we see that to show \eqref{eqn:stab-a1}, it suffices to prove
\beq\label{suffices554}
\| (U^* - F)^{-1} \|_{\ell^2 \to \ell^2 } \leq C.
\eeq
This is proven similarly to Lemma 7.3.2 of \cite{mde-notes} which uses Lemma 7.4.7 of \cite{mde-notes}.  To be more precise, 
 let $v(z, w)$ be the Perron--Frobenius eigenvector for $F(z, w)$.  Similarly to the calculation after Lemma 7.4.7 of \cite{mde-notes},  we just need to find a lower bound for the quantity
\beq
\Re[1 - v^T \mb(z) \mb(w) | \mb(z) \mb(w) |^{-1} v ].
\eeq
We have
\begin{align}
\Re[ v^T \mb(z) \mb(w) | \mb(z) \mb(w) |^{-1} v ] =& -2 v^T \Im [ \mb (w) ] \Im [ \mb (z) ] | \mb (z) \mb (w)|^{-1} v \nonumber \\ &+ v^T \Re[ \mb(z) \bar{\mb} (w) ] | \mb(z) \mb(w)|^{-1} v \nonumber \\
\leq& -2 v^T \Im [ \mb (w) ] \Im [ \mb (z) ] | \mb (z) \mb (w)|^{-1} v  + 1,
\end{align}
and so the claim \eqref{suffices554} follows from \eqref{eqn:im-mb} and \eqref{eqn:m-size}. Given \eqref{eqn:stab-a1}, the claim \eqref{claim555} follows as in the proof of the previous proposition.
\qed

\section{Stein's method for general Wigner-type matrices} \label{sec:clt}

We introduce some notation for cumulants.  For a matrix $W$ of general Wigner-type, denote the $k$th cumulant of $N^{1/2} W_{ij}$ by $s_{ij}^{(k)}$ with the shorthand $s^{(2)}_{ij} = s_{ij} = N S_{ij}$.  The main result of this section on characteristic functions of linear spectral statistics is Proposition \ref{prop:char-func}, which will be used  later in Theorem \ref{thm:clt-homog}.  Proposition \ref{prop:char-func} is a general result on the characteristic functions of regular test functions on sufficiently large scales as in Definition \ref{def:reg-f}, and Theorem \ref{thm:clt-homog} specializes this to the case of half-regular bump functions, as in Definition \ref{def:half-f}. The latter are the functions arising from the homogenization theory, for which we can compute the leading-order contribution to the variance.

\subsection{Expansion for \texorpdfstring{$T_{xy}$}{Txy}}
Before proving our main results on linear spectral statistics, we will determine the leading-order contribution of a quantity we denote by $T_{xy} (z, w)$ which arises naturally in the calculation of the characteristic functions of linear spectral statistics.  Define,
\beq
T_{xy} (z, w) := \frac{1}{N} \sum_{i} s_{ix} G_{iy} (z) G_{yi} (w).
\eeq
To give the reader a sense of the scale of this quantity, we remark that in the Wigner case when $s_{ix} =1$ and $z=w$ we have that $T_{xy}$ equals $N^{-1} \del_z G_{yy}$ and so is order $\O(N^{-1})$ for large $\Im [z]$.  Since $W$ is real symmetric, the resolvent is symmetric, $G_{ij}(z) = G_{ji} (z)$, a fact we will use repeatedly without comment.  

We first recall some notation.  For the matrix $W$ we let $W^{(i)}$ be the $i$th minor.  We denote its resolvent by $\Gi (z) := (W^{(i)} -z )^{-1}$.   We have the identity, valid for $i \neq j$ (see, e.g., Lemma 3.5 of \cite{semi-lec}),
\beq \label{eqn:gij-minor}
G_{ij}(z)  = - G_{ii} \sumi_k W_{ik} \Gi_{kj}.
\eeq
Here the notation $\sumi$ means
\beq
\sumi_k := \sum_{ k : k \neq i }.  
\eeq
For random variables $X$ we denote the partial expectation operators by
\beq
P_i [X] := \ee[ X | W^{(i)} ], \qquad Q_i [X] := X - P_i [X].
\eeq
We first derive the following self-consistent equation for $T_{xy}$.  The method is very similar to Section 4 of \cite{diffusion} except that we allow two spectral parameters $z$ and $w$; we also make no attempt to derive optimal error estimates. 
\bep \label{prop:T-self}
Define $T_{xy} (z, w)$ by
\beq
T_{xy} (z, w) := \frac{1}{N} \sum_{i} s_{ix} G_{iy} (z) G_{yi} (w).
\eeq
Let $C'>0$ and assume $|z|, |w| \leq C'$.  
Fix $\gamma >0$, let $\eta = \min \{ | \Im[z]|, | \Im[w] | \}$, and assume $ \eta \geq N^{\gamma-1}$.   We have
\beq
T_{xy} (z, w) = [ S \bm (z) \bm (w) ]_{xy} + (S \bm (z) \bm (w) T )_{xy} + \E_{xy} (z, w),
\eeq
where $\E_{xy} (z, w)$ is a matrix that satisfies, for any $\eps >0$,
\beq
\ee[ | \E_{xy} (z, w) |^2]\leq C \frac{N^{\eps}}{ (N \eta)^3},
\eeq
for some $C(\eps,\gamma)>0$.
\eep
\proof  We write, for $i \neq y$, using \eqref{eqn:gij-minor},
\beq
G_{iy} (z) G_{iy} (w) = G_{ii} (z) G_{ii} (w) \left( \sumi_k W_{ik} \Gi_{ky} (z) \right) \left( \sumi_l  W_{il} \Gi_{ly} (w) \right).
\eeq

We have from the local laws \eqref{eqn:entry-ll} that with overwhelming probability,
\beq
G_{ii} (z) = \bm_i (z) + \O (N^{\eps} (N \eta)^{-1/2}),
\eeq
as well as
\beq
\left( \sumi_k  W_{ik}\Gi_{ky} (z) \right) =\frac{G_{iy}(z)}{ G_{ii}(z)}= \O (N^{\eps } (N \eta)^{-1/2} ),
\eeq
since $\mb_i(z)$ satisfies \eqref{eqn:m-size}. 
We have therefore arrived at the preliminary identity, valid for $i \neq y$,
\begin{align}
G_{iy} (z) G_{iy} (w) = \bm_i (z) \bm_i (w) \left( \sumi_k W_{ik} \Gi_{ky} (z) \right) \left( \sumi_l  W_{il} \Gi_{ly} (w) \right) + \O (N^{\eps} (N \eta)^{-3/2} ),
\end{align}
with overwhelming probability, using again \eqref{eqn:m-size}.  
We decompose $T_{xy}(z, w)$ as
\begin{align}
T_{xy}(z, w) &= \frac{1}{N} s_{xy} G_{yy} (z) G_{yy} (w) \notag  \\
&+ \frac{1}{N} \sum_{i \neq y} s_{xi} P_i \left[ \bm_i (z) \bm_i (w) \left( \sumi_k W_{ik} \Gi_{ky} (z) \right) \left( \sumi_l  W_{il} \Gi_{ly} (w) \right) \right] \label{eqn:Pi}\\
&+ \frac{1}{N} \sum_{i \neq y } s_{xi} Q_i [ G_{iy} (z) G_{iy} (w) ] \notag \\
&+ \O ( (N\eta)^{-3/2} N^{\eps} )
\end{align}
with overwhelming probability.  By the local law \eqref{eqn:entry-ll},
\beq
G_{yy} (z) G_{yy} (w) = \mb_y(z) \mb_y (w) + \O ( N^{\eps} ( N \eta)^{-1/2} )
\eeq
with overwhelming probability.  We calculate the term  \eqref{eqn:Pi}.  We have,
\beq
P_i \sumi_{k, l} W_{ik} W_{il} \Gi_{ky} (z) \Gi_{ly} (w) = \frac{1}{N} \sumi_k s_{ik} \Gi_{ky} (z) \Gi_{ky} (w).
\eeq
For $k, y$ satisfying $k \neq i$ and $y \neq i$ we have the resolvent identity (again see, e.g., Lemma 3.5 of \cite{semi-lec}),
\beq
G_{ky} = \Gi_{ky} + \frac{G_{ki} G_{iy}}{G_{ii} } = \Gi_{ky} + \O (N^{\eps} (N \eta)^{-1} ),
\eeq
with overwhelming probability; here we again used the local law \eqref{eqn:entry-ll}.  Hence, we have for $i \neq y$ that, with overwhelming probability
\beq
P_i \sumi_{k, l} W_{ik} W_{il} \Gi_{ky} (z) \Gi_{ly} (w)  = T_{iy} (z, w) + \O ( N^{\eps} (N \eta)^{-3/2} ).
\eeq
From this, we have for \eqref{eqn:Pi}, with overwhelming probability,
\begin{align}
&\frac{1}{N} \sum_{i \neq y} s_{xi} P_i \left[ \bm_i (z) \bm_i (w) \left( \sumi_k W_{ik} \Gi_{ky} (z) \right) \left( \sumi_l  W_{il} \Gi_{ly} (w) \right) \right]  \nonumber \\
=& \frac{1}{N} \sum_{i \neq y } s_{xi} \mb_i (z) \mb_i (w) T_{iy} (z, w) + \O (N^{\eps} (N \eta)^{-3/2} ) \nonumber\\
= &(S \mb (z) \mb (w) T)_{xy} + \O (N^{\eps} (N \eta)^{-3/2} ),
\end{align}
where in the last line we used that $T_{yy} (z, w) = \O (N^{\eps} (N \eta)^{-1} )$ with overwhelming probability (by \eqref{eqn:entry-ll}).  
Therefore, with overwhelming probability,
\begin{align}
T_{xy} (z, w) =& (S \mb (z) \mb (w) )_{xy} + (S \mb(z) \mb (w)  T)_{xy} \nonumber \\
+ &\O (N^{\eps} ( N \eta)^{-3/2} ) + \frac{1}{N} \sum_{i \neq y } s_{xi} Q_i [ G_{iy} (z) G_{iy} (w) ].
\end{align}
We define $\E_{xy} (z, w)$ as the terms on the last line of the above.  The second moment of the second term (the sum) is estimated in Lemma~\ref{52} below. \qed

 The proof of the following is straightforward and based on elementary resolvent identities and so is deferred to Appendix \ref{sec:Txy-error-2m}. 

\bel\label{52} Let $\eta = \min \{ | \Im[z]|, | \Im[w] | \}$, and assume $\gamma^{-1} \geq \eta \geq N^{\gamma-1}$, for some $\gamma >0$.  For every $\eps>0$, we have
\beq
\ee \left| \frac{1}{N} \sum_{i \neq y } s_{xi} Q_i [ G_{iy} (z) G_{iy} (w) ] \right|^2 \leq C \frac{N^{\eps}}{ (N \eta)^3}
\eeq
for some $C(\eps,\gamma)>0$.
\eel

Our expansion for $T_{xy} (z, w)$ is then the following theorem, which follows immediately from Proposition~\ref{prop:T-self} and \eqref{e:stab-elements}. We recall that $\Dkap$ was defined in \eqref{d:dkap}.
\bet \label{thm:Txy}
Let $\eta = \min \{ | \Im[z]|, | \Im[w] | \}$, and assume $\gamma^{-1} \geq \eta \geq N^{\gamma-1}$, for some $\gamma >0$. 
Define $A_{xy}$ by
\beq
T_{xy} (z, w) = [ (1 - S \bm (z) \bm (w) )^{-1} S \bm(z) \bm (w) ]_{xy} + A_{xy}.
\eeq
For any $\eps, \kappa >0$, we have for $z, w \in \Dkap$ that
\beq
\ee[ |A_{xy} |^2 ] \leq \frac{C N^{\eps}}{N^3 \eta^3( | \Im[w]| + | \Im[z]| )^2}
\eeq
for some $C(\eps,\gamma)>0$.
\eet

\subsection{Calculation of characteristic function} \label{sec:char-calc}

Throughout the present section, we will let $f$ be a regular test function as in Definition \ref{def:reg-f}, with data $(t, c', C')$.  The big $\O$ error notation hides dependence on $c', C'$ but not on $t$.  
The method in this section is similar to \cite{meso}.  At times, we will refer to \cite{meso} for estimates that are similar.

We will need to differentiate various quantities wrt matrix elements. Our notation here and throughout the rest of the paper is as follows. We parameterize the space of symmetric $N \times N$ matrices by their upper triangular part $\{ H_{ij} \}_{i \leq j }$ and for $i \leq j$ denote,
$$
\del_{ij} F (H) = \frac{ \del F(H)}{ \del H_{ij} }  = \lim_{ h \to 0 } \frac{ F ( H + h \Delta_{ij} ) - F (H) }{h}
$$
where $F$ is any differentiable function on the space of $N \times N$ symmetric matrices. Above, $\Delta_{ij}$ is the symmetric matrix whose entries are $0$ except for the $(i, j)$th and $(j, i)$th entries which are $1$ (when $i=j$ everything is $0$ except that the $(i, i)$th entry is $1$). For $i > j$ we put $\del_{ij} = \del_{ji}$.

 We denote the quasi-analytic extension of $f$ by
\beq
\tilf (x + \i y ) = (f (x) + \i y f' (x) ) \chi (y),
\eeq
where $\chi (y)$ is a smooth cut-off function such that $\chi (y) = 1 $ for $|y| <2 L_*$ and $\chi (y) = 0$ for $|y| > 2L_* +1$.  Here, the constant $L_*$ is as in Definition \ref{def:dc}.  We also choose $\chi$ to be an even function of $y$.  We observe that
\beq\label{ftilde}
\partial_z \tilf (z) = \frac{ \i y \chi (y) f'' (x) + \i ( f (x) + \i f' (x) y) \chi' (y) }{2}.
\eeq

Now let $\mfa >0$ and define
\beq \label{eqn:oma-def}
\Oma := \{ (x, y) \in \rr^2 : |y| > N^{\mfa-1} \}.
\eeq
We have the following estimate, proven identically to (4.12) of \cite{meso} using the Helffer--Sj{\"o}strand representation of $f(x)$ as an integral of $\partial_z \tilf (z) $:
\begin{align}\label{eqn:oma-error}
\tr f(W) - \ee[ \tr f (W) ]= & \frac{1}{ 2 \pi} \int_{ \Oma} ( \i y \chi (y) f'' (x) + \i ( f (x) + \i f' (x) y) \chi' (y) ) N (m_N (z) - \ee[ m_N (z) ] ) \d x \d y \nonumber \\
+ & \O (N^{\mfa+\eps-1} \|f''\|_1 ),
\end{align}
which holds for any $\eps >0$ with overwhelming probability.
 Note that the integrand in \eqref{eqn:oma-error} is non-zero only in $\D_{c'}$, where $\D_{c'}$ is defined in Definition \ref{def:dc}.   Note furthermore that since $\chi$ is even, the integral above over $\Oma$ is real, so that we have $| \e_\mfa ( \lambda) | \leq 1$, where
\beq
\e_\mfa ( \lambda) := \exp \left[ \i \lambda \frac{1}{ 2 \pi} \int_{ \Oma} ( \i y \chi (y) f'' (x) + \i ( f (x) + \i f' (x) y) \chi' (y) ) N (m_N (z) - \ee[ m_N (z) ] ) \d x \d y\right] .
\eeq
From \eqref{eqn:oma-error} we have that for any $\eps >0$, that with overwhelming probability
\beq
| \e_\mfa ( \lambda) - \e ( \lambda ) | \leq C |\lambda| N^{\eps+\mfa-1} \|f''\|_1,
\eeq
where
\beq
\e ( \lambda) := \exp \left[ \i \lambda ( \tr f(W) - \ee[ f (W) ]) \right].
\eeq
Define
\beq
E_\mfa (z) := \sum_i \ee[ \e_\mfa ( G_{ii}(z) - \ee[G_{ii}(z) ]) ], \qquad \psi_\mfa (\lambda ) := \ee[ \e_\mfa (\lambda ) ],
\eeq
so that
\begin{align} \label{eqn:d-dlamb}
\frac{ \d }{ \d \lambda} \psi_\mfa ( \lambda) = \frac{ \i}{ 2 \pi } \int_{ \Oma } ( \i y \chi (y) f''(x) + \i ( f (x)  + \i y f' (x) ) \chi' (y) ) E_\mfa (z) \d x \d y.
\end{align}
We have the following, which is the same as Lemma 4.4 of \cite{meso}.
\bel \label{lem:H}
Let $H(z)$ be a function holomorphic on $\cc \backslash \rr$.  Assume that,
\beq
|H(z) | \leq \frac{K}{ | \Im[z] |^s}
\eeq
holds for some $1 \leq s \leq 2$.  There is a C such that
\begin{align}
\left| \int_{\Oma} \left( \i y \chi (y) f''(x) \right) H (x+ \i y ) \d x \d y \right| \leq C K \log(N) \left( 1 + \|f''\|_1 \right)^{s-1}.
\end{align}
\eel
Usually, $H(z)$ is the difference between some random quantity like $G_{ii}(z)$ and a deterministic quantity such as $\mb_i(z)$; the size of this error depends on $\Im[z]$, and Lemma~\ref{lem:H} is used to integrate such an error term using integration by parts in the region where $\Im[z] \geq \|f''\|^{-1}_1$ to obtain an optimal estimate.

We also will use the following elementary consequence of the Cauchy integral formula.
\bel \label{lem:cauchy-int}
Let $H(z)$ be a function holomorphic in the upper half-plane and let $z_0 = E_0 + \i \eta_0 \in \hh_+$.  Then, for $k \geq 1$,
\beq
\left| \partial^k_z H(z_0) \right| \leq C_k \frac{1}{\eta_0^k} \sup_{ z : |z-z_0| < \eta_0/2} |H(z) |.
\eeq
\eel
Next we give a lemma that collects some estimates on derivatives of the characteristic function with respect to matrix entries.
\bel
For any $\eps >0$ the following estimates hold with overwhelming probability. For any $i, a$ we have for $k \geq 1$ that
\beq \label{eqn:delia-ea-bd}
| \partial_{ia}^k \e_\mfa | \leq (1 + | \lambda | )^{k} N^{\eps}.
\eeq
We also have
\begin{align} \label{eqn:deleia}
(1 + \delta_{ia} ) \partial_{ia} \e_{\mfa} = - \frac{2 \i \lambda}{ \pi} \e_{\mfa} \int_{\Oma} \bar{\del}_w \tilf (w) \del_w G_{ia} (w) \d w \d \bar{w} ,
\end{align}
and for $i \neq a$,
\beq \label{eqn:ea-der-1}
\left|   \partial_{ia} \e_{\mfa}  \right| \leq N^{\eps} (1 + | \lambda | ) (N^{-1/2} (1 + \|f''\|_1 )^{1/2} ).
\eeq
For the second derivative, for $i \neq a$,
\begin{align} \label{eqn:ea-der-2}
\del_{ia}^2 \e_\mfa &= \e_{\mfa} \frac{ \i \lambda}{ \pi} \int_{\Oma} \bar{\del}_w \tilf (w) 2\del_w( \mb_a (w) \mb_i (w)) \d w \d \bar{w}  + \O (N^{\eps} (1  + | \lambda| )^2 N^{-1/2} ( 1+ \|f''\|_1 )^{1/2} ). 
\end{align}
\eel
\proof Equation \eqref{eqn:delia-ea-bd} is proven similarly to (4.23) of \cite{meso}. The equality \eqref{eqn:deleia} is an identity based on the fact that $(1+\delta_{ia} ) \del_{ia} G_{xx} = - 2 G_{xi} G_{xa}$. The estimate \eqref{eqn:ea-der-1} is proven similarly to (4.38) of \cite{meso} (or just integrating \eqref{eqn:deleia} using Lemma \ref{lem:H}).  The last estimate is proven in the same way as (4.42) of \cite{meso}.  \qed

We turn to the calculation of $E_\mfa (z)$.   The following is the analogue of  Lemma 4.5 of \cite{meso}.
\bel \label{lem:char-calc-1}
For any $\eps,\gamma >0$, we have for $\gamma^{-1} \geq |\Im[z] | \geq N^{\gamma-1}$ that
\begin{align}
\frac{-1}{ \mb_i} \ee[ \e_\mfa (G_{ii} (z) - \ee[ G_{ii} (z) ] )] =& - \frac{1}{N} \sum_{a} \mb_i s_{ia} \ee[ \e_\mfa (G_{aa} - \ee[ G_{aa} ] )] \label{eqn:char-a2}  + \frac{1}{N} \sum_a s_{ia} \ee[ ( \partial_{ia} \e_\mfa ) G_{ia} ]\nonumber \\
& - \ee[ \e_\mfa (T_{ii}(z,z)- \ee[ T_{ii} (z,z)])] - \frac{1}{2 N^2} \sum_a \sfo_{ia} \mb_a \mb_i \ee[ \partial_{ia}^2 \e_{\mfa} ] \nonumber \\
&+ N^{\eps-1} (1 + | \lambda |^4) \O ( (N \eta)^{-1/2} \eta^{-1} + (N \eta)^{-1/2} (1 + \|f''\|_1)^{1/2} ),
\end{align}
where $\eta = | \Im[z]|$.  
\eel 

\proof Beginning as in the proof of Lemma 4.5 of \cite{meso} we have via a cumulant expansion that
\begin{align}
&z \ee[ \e_\mfa  (G_{ii} (z)- \ee[ G_{ii}(z)] ) ]= \sum_{a} \ee[ \e_\mfa ( G_{ia} W_{ai} - \ee[ G_{ia} W_{ai} ] ) ] \label{eqn:char-a1} \\
= & \frac{1}{N}\sum_a s_{ia}\ee[(\partial_{ia}\e_\mfa(\lambda))G_{ia}] \\
- & \frac{1}{N} \sum_a s_{ia}  \ee[ \e_\mfa (G_{ii} G_{aa} - \ee[ G_{ii} G_{aa} ] ) ] \label{eqn:term-1} \\
- & \frac{1}{N} \sum_a s_{ia} \ee[ \e_\mfa (G_{ia}^2 - \ee[ G_{ia}^2] ) ] \label{eqn:term-2} \\
+ & \frac{1}{N} s_{ii} \ee[ \e_\mfa (G_{ii}^2 - \ee[ G_{ii}^2 ] )] \label{eqn:term-3} \\
+&\frac{1 }{2 N^{3/2}} \sum_a \sth_{ia}\left\{ \ee[ ( \partial_{ia}^2 \e_\mfa ) G_{ai} ] + 2 \ee[ \partial_{ai} \e_\mfa \partial_{ai} G_{ai} ] + \ee[ \e_\mfa ( \partial_{ai}^2 G_{ai} - \ee[ \partial_{ai}^2 G_{ai} ] )] \right\}\label{eqn:term-4} \\
+& \frac{ 1}{6 N^2} \sum_a \sfo_{ia} \left\{ \ee[ ( \partial_{ia}^3 \e_\mfa ) G_{ai} ] + \dots + \ee[ \e_\mfa ( \partial_{ai}^3 G_{ai} - \ee[ \partial_{ai}^3 G_{ai} ] ) ] \right\} \label{eqn:term-5}\\
+& \O (N^{\eps-3/2} (1+ |\lambda|)^4 ).
\end{align}
Note that we suppressed the argument and wrote $G_{xy} = G_{xy} (z)$, etc., above. 
We now go through each term one by one.  Starting with \eqref{eqn:term-1}, we have
\begin{align}
- &\frac{1}{N} \sum_a s_{ia}  \ee[( \e_\mfa )(G_{ii} G_{aa} - \ee[ G_{ii} G_{aa} ] ) ]  =- \frac{1}{N} \sum_a s_{ia} \ee[ ( \e_\mfa- \ee[ \e_\mfa ] ) ( G_{ii} - \ee[ G_{ii} ] )(G_{aa} - \ee[ G_{aa} ] )] \nonumber \\
-& \frac{1}{N} \sum_a  s_{ia} \ee[ \e_\mfa (G_{ii} - \ee[ G_{ii} ] ) ]\ee[ G_{aa} ]-\frac{1}{N} \sum_a   s_{ia} \ee[ \e_\mfa (G_{aa} - \ee[ G_{aa} ])] \ee[ G_{ii} ] 
\end{align}
For any $\eps >0$ we have from \eqref{eqn:tr-ll} that
\beq
\frac{1}{N} \sum_a s_{ia} G_{aa} = \frac{1}{N} \sum_a s_{ia} \mb_{a} + \O (N^{\eps} (N \eta)^{-1} )
\eeq
with overwhelming probability.  Combining this with
 $G_{ii} = \mb_i + \O (N^{\eps} (N \eta)^{-1/2} )$ 
with overwhelming probability from \eqref{eqn:entry-ll}, we  see that
\begin{align}
 &- \frac{1}{N} \sum_a s_{ia}  \ee[ \e_\mfa (G_{ii} G_{aa} - \ee[ G_{ii} G_{aa} ] ) ]  \notag \\
=& -\ee[ \e_\mfa ( G_{ii} - \ee[ G_{ii} ] ) ]   \frac{1}{N} \sum_a s_{ia} \mb_a  -\frac{1}{N} \sum_{a}  \mb_i s_{ia}  \ee[ \e_\mfa (G_{aa} - \ee[G_{aa} ] )] + \O (N^{\eps} (N \eta)^{-3/2} ). \nonumber \\
=& \ee[ \e_\mfa (G_{ii} - \ee[ G_{ii} ] ) ] (z + \mb_i^{-1} ) -\frac{1}{N} \sum_{a}  \mb_i s_{ia}  \ee[ \e_\mfa (G_{aa} - \ee[G_{aa} ] )] + \O (N^{\eps} (N \eta)^{-3/2} ).
\end{align}
In the second equality we used
\beq
-\frac{1}{N} \sum_a s_{ia} \mb_a = z + \mb_i^{-1}
\eeq
from \eqref{eqn:qve-def}.
The factor with $z$ cancels with the left side of \eqref{eqn:char-a1} and the $\mb_i^{-1}$ contributes the term on the left side of \eqref{eqn:char-a2}, after moving it to the other side of the equality in \eqref{eqn:char-a1}.  
The term \eqref{eqn:term-2} is, by definition,
\beq
-\frac{1}{N} \sum_a s_{ia} \ee[ \e_\mfa (G_{ia}^2 - \ee[ G_{ia}^2] ) ]  = - \ee[ \e_\mfa (T_{ii}(z,z)- \ee[ T_{ii} (z,z)])]
\eeq
and the term \eqref{eqn:term-3} is, by \eqref{eqn:entry-ll},
\beq
 \frac{1}{N} s_{ii} \ee[ \e_\mfa (G_{ii}^2 - \ee[ G_{ii}^2 ] )] = \O (N^{\eps} N^{-1} (N \eta)^{-1/2} ).
\eeq

We now estimate the various contributions to \eqref{eqn:term-4}.  First, from \eqref{eqn:ea-der-2}, we have
\begin{align}
\frac{1 }{2 N^{3/2}} \sum_a \sth_{ia} \ee[ ( \partial_{ia}^2 \e_\mfa ) G_{ai}(z) ] &= \frac{1}{ 2 N^{3/2}} \sumai \sth_{ia} \frac{ \i \lambda}{ \pi} \int_{\Oma} \bar{\del}_w \tilf (w) 2\del_w( \mb_a (w) \mb_i (w)) \ee[ \e_{\mfa} G_{ai} (z) ] \d w \d \bar{w} \nonumber \\
&+ N^{\eps-1}(1+ |\lambda|^2) \O\left( (N \eta)^{-1/2} (1 + \|f''\|_1 )^{1/2} \right),
\end{align}
where we used $|G_{ia}(z) | = \O ( N^{\eps} ( N \eta)^{-1/2})$ with overwhelming probability (from \eqref{eqn:entry-ll}).
We claim that by the isotropic local law \eqref{eqn:iso} and the fact that the derivatives of $\mb(w)$ are bounded for $w \in \Dkap$ by Proposition~\ref{prop:basic-m} that
\beq \label{eqn:revision-1}
\frac{1}{N^{1/2} } \sumai \sth_{ia} \del_w ( \mb_a (w) \mb_i (w) ) G_{ai} (z) = \O(N^{\eps} (N \eta)^{-1/2} ). 
\eeq
Here, in order to apply \eqref{eqn:iso}, we write
\beq \label{eqn:revision-2}
\frac{1}{N^{1/2} } \sumai \sth_{ia} \del_w ( \mb_a (w) \mb_i (w) ) G_{ai} (z) = \boldsymbol{v} G(z) \boldsymbol{w} ,
\eeq
where $\boldsymbol{w}$ is the vector that is $1$ only at coordinate $i$ and $0$ elsewhere, and $\boldsymbol{v}$ has $j$th entry equal to $ \boldsymbol{v}_j = N^{-1/2} \sth_{ij} \del_w (\mb_j (w) \mb_i (w) )$ except for the $i$th entry which is $0$. Then we have $\| \boldsymbol{w} \|_2 + \| \boldsymbol{v} \|_2 \leq C$ as well as $\sum_{j=1}^N \mb_j (z) \boldsymbol{w}_j \boldsymbol{v}_j = 0$ as they have disjoint support, and so the estimate \eqref{eqn:revision-1} follows from an immediate application of \eqref{eqn:iso} to the RHS of \eqref{eqn:revision-2}.

Therefore, applying Lemma \ref{lem:H} (with, e.g.,  $s=1$) we see that,
\beq
 \frac{1}{ 2 N^{3/2}} \sumai \sth_{ia} \frac{ \i \lambda}{ \pi} \int_{\Oma} \bar{\del}_w \tilf (w) 2\del_w( \mb_a (w) \mb_i (w)) \ee[ \e_{\mfa} G_{ai} (z) ] \d w \d \bar{w} = N^{\eps-1} \O ( (N \eta)^{-1/2} ).
\eeq
In summary, for the first term of \eqref{eqn:term-4} we have,
\beq
\frac{1 }{2 N^{3/2}} \sum_a\sth_{ia} \ee[ ( \partial_{ia}^2 \e_\mfa ) G_{ai} ] = N^{\eps-1} (1 + | \lambda |^2 ) \O( (N \eta)^{-1/2} (1 + \|f''\|_1)^{1/2} ).
\eeq

From  \eqref{eqn:ea-der-1} and the local law \eqref{eqn:entry-ll} we have 
\begin{align}
\frac{1}{N^{3/2} }\sum_a \sth_{ia} \ee[ \partial_{ai} \e_\mfa \partial_{ai} G_{ai} ] &= - \frac{1}{N^{3/2}} \sumai \sth_{ia} \mb_i \mb_a \ee[ \partial_{ai} \e_\mfa ] \nonumber \\
&+ N^{\eps-1}(1+ |\lambda | ) \O ( N^{-1/2} \eta^{-1} + N^{-1/2} \eta^{-1/2} (1 + \|f''\|_1 )^{1/2}  ).
\end{align}
Moreover, using \eqref{eqn:deleia},
\begin{align}
 \frac{1}{N^{3/2}} \sumai \sth_{ia} \mb_i(z) \mb_a (z)\ee[ \partial_{ai} \e_\mfa ] &= \int ( \bar{\partial}_w \tilf ) \frac{- \i 2 \lambda}{N^{3/2}\pi } \partial_w \sumai \mb_i (z)\ee[ \e_\mfa  \sth_{ia} \mb_a(z) G_{ia} (w) ] \d w  \d \bar{w}\nonumber \\
&=\O (N^{\eps-1} (1+ | \lambda | )(N^{-1/2} (1+ \|f''\|_1 )^{1/2} ),
\end{align}
where we used the isotropic local law \eqref{eqn:iso} (similarly to the argument in \eqref{eqn:revision-1}-\eqref{eqn:revision-2}) to conclude,
\beq
\frac{1}{N^{1/2} } \partial_w \sumai \sth_{ia} \mb_a (z) G_{ia} (w) = \O \left( N^{\eps} N^{-1/2} | \Im[w]|^{-3/2} \right)
\eeq
with overwhelming probability, 
 and then Lemma \ref{lem:H}.  For the last term of \eqref{eqn:term-4}, we have for $i \neq a$ that
$
\partial_{ia}^2 (G_{ai} ) = 6 G_{ia} \mb_i \mb_a + \O( N^{\eps} (N \eta)^{-3/2} ),
$
and so
\begin{align}
\frac{1}{N^{3/2}} \sum_a \sth_{ia} \ee[ \e_\mfa ( \partial_{ai}^2 G_{ai} - \ee[ \partial_{ai}^2 G_{ai} ] )]  = N^{\eps-1} \O( N^{-1/2} \eta^{-1}),
\end{align}
by the isotropic local law \eqref{eqn:iso} (again using  a similar argument to \eqref{eqn:revision-1}, \eqref{eqn:revision-2}).

For the terms \eqref{eqn:term-5}, we see that the term $i = a$ contributes $\O(N^{-2+\eps})$.  Any other term that contains at least one off-diagonal resolvent entry contributes $\O (N^{\eps-1} ((N \eta)^{-1/2} )$ after the sum over $a$.  Similarly,
$
\partial_{ia}^3 G_{ai} - \ee[ \partial_{ia}^3 G_{ai} ] = \O (N^{\eps} (N \eta )^{-1/2} ).
$
Hence, 
\begin{align}
& \frac{ 1}{6 N^2} \sum_a \sfo_{ia} \ee[ ( \partial_{ia}^3 \e_\mfa ) G_{ai} ] + \dots \ee[ \e_\mfa ( \partial_{ai}^3 G_{ai} - \ee[ \partial_{ai}^3 G_{ai} ] ) ] \nonumber \\
= &- \frac{1}{2 N^2} \sum_a \sfo_{ia} \mb_a \mb_i \ee[ \partial_{ia}^2 \e_{\mfa} ] + \O (N^{\eps-1}( 1 + | \lambda |^3) (N \eta)^{-1/2} ).
\end{align}
This yields the claim. \qed

We make some further calculations. The next lemma is the analogue of Lemma 4.6 of \cite{meso}.
\bel \label{lem:char-calc-2}
We have,
\beq
\ee[ \e_\mfa (T_{ii}(z,z)- \ee[ T_{ii} (z,z)])]  = N^{\eps-1} \O ( N^{-1/2} \eta^{-5/2} ),
\eeq
as well as
\begin{align}
 & \frac{1}{N}  \sum_a s_{ia} \ee[ ( \partial_{ia} \e_\mfa ) G_{ia} ] 
=    \frac{ \i \lambda}{ \pi} \frac{s_{ii}}{N} \int_{\Oma} ( \bar{\partial}_w \tilf (w) )  \mb_i' (w) \mb_i (z) \ee[ \e_{\mfa} ] \d w \d \bar{w} \nonumber \\
 - & \frac{  \i 2 \lambda}{ \pi} \int_{\Oma} ( \bar{\partial}_w \tilf (w) )  \ee[ \e_{\mfa} ] \partial_{w}  [ (1 - S \bm (z) \bm (w) )^{-1} S \bm(z) \bm (w) ]_{ii} \d w \d \bar{w} \nonumber \\
+ & N^{\eps-1}(1 + | \lambda | ) \O ( N^{-1/2} \|f''\|_1 ( N^{1-\mfa})^{3/2}  +(N \eta)^{-1/2} (1 + \|f''\|_1 )^{1/2} )
\end{align}
and
\begin{align}
 -\frac{1}{2 N^2} \sum_a \sfo_{ia} \mb_a \mb_i \ee[ \partial_{ia}^2 \e_{\mfa} ] &= - \frac{ \i \lambda}{ \pi} \frac{1}{ N^2} \sum_a \sfo_{ia} \mb_a(z) \mb_i (z) \int_{\Oma} ( \bar{ \partial}_w \tilf (w) ) \partial_w ( \bm_a (w) \bm_i (w) ) \ee[ \e_\mfa ] \d w \d \bar{w} \nonumber \\
+& N^{-1+\eps} ( 1 + | \lambda |^2) \O (N^{-1/2} (1 + \|f''\|_1 )^{1/2} ).
\end{align}
\eel

\proof The first estimate follows from Theorem \ref{thm:Txy} and the bound $| \e_\mfa ( \lambda) | \leq 1$; the latter follows from the definition of $\e_\mfa ( \lambda)$.  We now turn to proving the second estimate of the lemma. We apply \eqref{eqn:deleia} to obtain
\begin{align}
\frac{1}{N} \sum_a s_{ia} \ee[ ( \partial_{ia} \e_\mfa ) G_{ia} ] &= \frac{ - \i 2 \lambda}{ \pi} \frac{1}{N} \sum_{a} s_{ia} \int_{\Oma} ( \bar{\partial}_w \tilf (w) ) \ee[ \e_{\mfa} G_{ia} (z) \partial_w G_{ia} (w) ] \d w \d \bar{w} \nonumber \\
&+ \frac{ \i \lambda}{  \pi} \frac{s_{ii}}{N} \int_{\Oma} ( \bar{\partial}_w \tilf (w) ) \ee[ \e_{\mfa} G_{ii} (z) \partial_w G_{ii} (w) ] \d w \d \bar{w} .
\end{align}
First, we claim that
\beq \label{eqn:revision-3}
\partial_w G_{ii} (w) = \mb'_i (w) + \O (N^{\eps}N^{-1/2} | \Im[w]|^{-3/2} )
\eeq
with overwhelming probability. Indeed, this follows from expressing $\del_w G_{ii} (w)$ as a contour integral, and a direct application of the entry-wise local law \eqref{eqn:entry-ll}: 
\beq
| \del_w G_{ii} (w) - \mb'_i (w) | = \left| \frac{1}{2 \pi } \int_{ |\xi - w| = \frac{1}{2} |\Im[w]| } \frac{ G_{ii} (\xi ) - \mb_i (\xi ) }{ (w - \xi )^2 } \d \xi \right| \leq C N^{\eps} \frac{1}{(N |\Im[w] |)^{1/2} } \frac{ | \Im[w] | }{ | \Im[w] |^2}.
\eeq
The last inequality is due to the fact that $\Im[\xi] \asymp \Im[w]$ (so that the application of \eqref{eqn:entry-ll} picks up a factor $(N | \Im[w] | )^{-1/2}$), that $|\xi - w | = \frac{1}{2} | \Im[w] |$ and that the length of the contour integral is order $| \Im[w]|$. 

Using Lemma \ref{lem:H} and \eqref{eqn:revision-3} we have,
\begin{align}
 &\frac{ \i \lambda}{ \pi} \frac{s_{ii}}{N} \int_{\Oma} ( \bar{\partial}_w \tilf (w) ) \ee[ \e_{\mfa} G_{ii} (z) \partial_w G_{ii} (w) ] \d w \d \bar{w} \nonumber \\
 =&  \frac{ \i \lambda}{  \pi} \frac{s_{ii}}{N} \int_{\Oma} ( \bar{\partial}_w \tilf (w) )  \mb_i' (w) \mb_i (z) \ee[ \e_{\mfa} ] \d w \d \bar{w}+ N^{\eps-1} (1+ | \lambda | )\O( (N \eta)^{-1/2} (1 + \|f''\|_1 )^{1/2} ).
\end{align}
For the other term, using Theorem \ref{thm:Txy} and the Cauchy--Schwarz inequality (as well as the Cauchy integral formula as in Lemma \ref{lem:cauchy-int} to estimate the derivative in $w$ below in terms of the original function), we conclude for $z, w$ satisfying $| \Im[z]|, |\Im[w] | \geq N^{\mfa-1}$ that
\begin{align}
& \left| \partial_w \ee[ \e_{\mfa} ( T_{ii} (z, w) - [(1-S \mb(z) \mb(w))^{-1} S \mb(z) \mb(w) ]_{ii} )] \right| \nonumber\\
\leq & \frac{N^{\eps}}{N^{3/2} \min\{ | \Im[z]|, | \Im[w]| \}^{3/2}} \frac{1}{ | \Im[w] | ( | \Im[w]| + | \Im[z] | )} 
\leq  \frac{C N^{\eps}}{ N^{3/2} | \Im[w]|^2}  (N^{1-\mfa})^{3/2} .
\end{align}
Integrating this estimate in $w$ using Lemma~\ref{lem:H}, we find
\begin{align}
& \frac{ - \i 2\lambda}{ \pi} \frac{1}{N} \sum_{a} s_{ia} \int_{\Oma} ( \bar{\partial}_w \tilf (w) ) \ee[ \e_{\mfa} G_{ia} (z) \partial_w G_{ia} (w) ]  \d w \d \bar{w} \nonumber \\
= &\frac{ - \i 2\lambda}{ \pi} \int_{\Oma} ( \bar{\partial}_w \tilf (w) )  \ee[ \e_{\mfa} ( \partial_w T_{ii} )(z, w) ] \d w \d \bar{w} \nonumber \\
=&\frac{ - \i 2 \lambda}{ \pi} \int_{\Oma} ( \bar{\partial}_w \tilf (w) )  \ee[ \e_{\mfa} ] \partial_{w}  [ (1 - S \bm (z) \bm (w) )^{-1} S \bm(z) \bm (w) ]_{ii}  \d w \d \bar{w} 
+N^{\eps-1} \O ( N^{-1/2} \|f''\|_1 ( N^{1-\mfa})^{3/2} )
\end{align}
This proves the second estimate of the lemma.  The last estimate follows directly from \eqref{eqn:ea-der-2}.  \qed

For the moment, define the vector $v \in \mathbb{C}^N$ by its coordinates,
\beq
v_i = \ee[ \e_{\mfa} ( \lambda) ( G_{ii} (z) - \ee[ G_{ii} (z) ])].
\eeq
By combining Lemmas \ref{lem:char-calc-1} and \ref{lem:char-calc-2}, we derive the expansion
\begin{align}
[(1 - \mb(z)^2 S) v]_i &= - \frac{ \i \lambda}{ \pi} \ee[ \e_\mfa ] \int_{\Oma} ( \bar{ \partial}_w \tilf (w) ) [S \bm' (w) \bm^2(z)]_{ii} \d w \d \bar{w} \nonumber \\
&+ \frac{2 \i \lambda}{ \pi } \ee[ \e_{\mfa} ]\int_{\Oma} ( \bar{ \partial}_w \tilf (w) ) \partial_w [ \mb (z) (1 - S \bm (z) \bm (w) )^{-1} S \bm(z) \bm (w) ]_{ii} \d w \d \bar{w} \nonumber \\
&+\frac{ \i \lambda}{ \pi} \frac{1}{ N^2} \sum_a \sfo_{ia} \mb_a(z) \mb^2_i (z) \int_{\Oma} ( \bar{ \partial}_w \tilf (w) ) \partial_w ( \bm_a (w) \bm_i (w) ) \ee[ \e_\mfa ]  \d w \d \bar{w} \nonumber \\
&+N^{\eps-1} ( 1 + | \lambda|^4) \O( ( N^{-1/2} \|f''\|_1 ( N^{1-\mfa})^{3/2}  +(N \eta)^{-1/2} (1 + \|f''\|_1 )^{1/2}+ N^{-1/2} \eta^{-5/2} ). \label{eqn:lop-a1}
\end{align}
By Proposition \ref{prop:stab-3} and Lemma \ref{lem:stab-bulk}, we have $\|(1 - \mb(z)^2 S)^{-1}\|_{\ell^\infty \to \ell^\infty} \leq C$  for $z \in \Oma$ and wherever $\bar{\partial}_z \tilf (z) \neq 0$. 
 Here we use that the set of such $z$ lies in $\mathcal{D}_{c'}$ by Definition~\ref{def:reg-f}.  If we temporarily denote the the expression on the first three lines on the RHS of \eqref{eqn:lop-a1} by $\tilde{q}_i$ we find the estimate,
\beq
v_i = \sum_{j=1}^N (1 - \mb(z)^2 S)^{-1}_{ij} \tilde{q}_i + N^{\eps-1} (1+ | \lambda|^4) \O \left(\eta^{-1} N^{-1/2} ( 1 + \| f'' \|_1 ) ( N^{1-\mfa} )^{3/2} \right)
\eeq
where we simplified some error terms using $\eta \geq N^{\mfa-1}$. 
Therefore, plugging in this estimate for $v_i$ into \eqref{eqn:d-dlamb}, we derive the following preliminary statement, after applying again Lemma \ref{lem:H}, 
\begin{align}
\frac{ \d}{ \d \lambda} \ee[ \e_\mfa ( \lambda ) ] = - \lambda \ee[ \e_{\mfa} ( \lambda ) ] \hat{V} (f) + N^{\eps}(1 +|\lambda|^4) \O ( N^{-1/2}(1+ \|f''\|_1 )(N^{1-\mfa})^{3/2} )
\end{align}
where
\begin{align}
\hat{V} (f) &:= \label{eqn:V-def}  \frac{1}{  \pi^2} \int_{\Oma \times \Oma } ( \bar{ \partial}_w \tilf (w) ) ( \bar{ \partial}_z \tilf (z) ) \nonumber \\
& \times\bigg\{ 2 \sum_{i, j} (1 - \bm^2(z) S)^{-1}_{ij} \partial_w [ \mb (z) (1 - S \bm (z) \bm (w) )^{-1} S \bm(z) \bm (w) ]_{jj} \nonumber  \\
&- \sum_{i, j} (1- \bm^2 (z) S)^{-1}_{ij} [S \bm' (w) \bm^2(z) ]_{jj}\nonumber \\
&+ \frac{1}{N^2} \sum_{i, j, a} \sfo_{ja} (1- \bm^2 (z) S)^{-1}_{ij}  \mb_a (z) \mb^2_j (z) \partial_w ( \mb_a (w) \mb_j (w) ) \bigg\} \d z \d \bar{z} \d w \d \bar{w}.
\end{align}
It is important to note that the functional $\hat{V} (f)$ depends on the parameter $\mfa >0$. The dependence will later be removed (at the cost of an additional error term) for certain classes of test functions.

We summarize the above calculations in the first equation of the following proposition. The second equation comes from \eqref{eqn:oma-error}.
\bep \label{prop:der-char}
Let $f$ be a regular test function with data $(\|f''\|_1^{-1}, c', C')$.  Assume that there is a $\delta >0$ such that
\beq
\|f''\|_1 \leq N^{1/5-\delta}.
\eeq
Let $c_1 < \delta$ and choose $\mfa >0$ satisfying $4/5 \leq \mfa \leq 4/5 +\delta -c_1$.  Then
\beq
\frac{ \d}{ \d \lambda} \ee[ \e_\mfa ( \lambda ) ] = -\lambda \ee[ \e_{\mfa} ( \lambda ) ] \hat{V} (f) +(1+ | \lambda|)^4 \O (N^{\eps-\delta})
\eeq
and
\begin{align}
\tr f (W) - \ee[ \tr f (W) ] &= \frac{1}{ 2 \pi} \int_{ \Oma} ( \i y \chi (y) f'' (x) + \i ( f (x) + \i f' (x) y) \chi' (y) ) N (m_N (z) - \ee[ m_N (z) ] ) \d x \d y \nonumber \\
&+ \O (N^{\eps-c_1} ).
\end{align}
\eep

Note that our calculations are sufficient to also use $\hat{V} (f)$ to bound the variance of the linear spectral statistic.
\bel\label{l:59}
Let $f$ be a regular test function with data $(\| f''\|_1, c', C')$.  Assume that there is a $\delta >0$ such that
\beq
\|f''\|_1 \leq N^{1/5-\delta}.
\eeq
Let $c_1 < \delta$ and choose $\mfa >0$ satisfying $4/5 \leq \mfa \leq 4/5 +\delta -c_1$.  Then
\beq
\mathrm{Var} ( \tr f (W) ) =  \hat{V} (f) + \O\left( N^{\eps} (N^{-c_1} + N^{-\delta/2})\right).
\eeq
We have also,
\beq \label{eqn:lop-a3}
\hat{V} (f) \geq - C N^{\eps-\delta / 2} .
\eeq
\eel
\proof Let $X$ denote
\beq
X := \frac{1}{ 2 \pi} \int_{ \Oma} ( \i y \chi (y) f'' (x) + \i ( f (x) + \i f' (x) y) \chi' (y) ) N (m_N (z) - \ee[ m_N (z) ] ) \d x \d y.
\eeq
By \eqref{eqn:oma-error},
\beq \label{eqn:lop-a2}
\mathrm{Var} ( \tr f(W) ) = \mathrm{Var} (X) + \O (N^{\eps-c_1} ),
\eeq
so it suffices to compute $\mathrm{Var} (X)$.

 By the local law \eqref{eqn:ll}, the fact that $f$ is a regular test function, and the definitions of $\Oma$ and $\chi(y)$,
\beq\label{Xbound}
|X| \leq N^{\eps}
\eeq
with overwhelming probability for any $\eps >0$, where we also used Lemma \ref{lem:H}.

Take $ \lambda = N^{-\delta/2}$.   We have, by Taylor expansion of $\e^{ \i \lambda X}$ and \eqref{Xbound},
\beq
\ee[ \e^{ \i \lambda X} \i X ] = - \lambda \ee[ X^2] + \O ( \lambda^2 N^{\eps} ) = - \lambda \ee[ \e^{ \i \lambda X } ] \ee[ X^2] + \O ( \lambda^2 N^{\eps} ),
\eeq
where the last equality follows from $\ee[ \e^{ \i \lambda X } ] = 1 + \O (N^{\eps} | \lambda |)$.  On the other hand, by Proposition \ref{prop:der-char},
\beq
\ee[  \e^{ \i \lambda X} \i X ] = \frac{ \d }{ \d \lambda} \ee[ \e^{ \i \lambda X} ] =  -\lambda \ee[ \e^{ \i \lambda X } ] \hat{V} (f) + \O(N^{\eps-\delta} ),
\eeq
where we used $\ee[ \e_{\mfa} ( \lambda ) ] =\ee[ \e^{ \i \lambda X } ]$, by the definition of $\e_{\mfa}$.
Subtracting the previous two equations yields,
\beq
\hat{V} (f) = \Var(X) + \O (N^{\eps-\delta/2} ),
\eeq
which gives the second estimate of the lemma. The first estimate follows from the above and \eqref{eqn:lop-a2}. 
 \qed

The estimate \eqref{eqn:lop-a3} ensures that $| \e^{ - \lambda^2 \hat{V}(f)/2} | \leq C$ for $\lambda < N^{\delta/5}$. 
Hence, we can integrate the derivative in Proposition \ref{prop:der-char} and obtain the following.
\bep \label{prop:char-func}  Let $f$ be a regular test function with data $(\|f''\|_1^{-1}, c', C')$.  
Assume 
 that there is a $\delta >0$ such that 
\beq
\|f''\|_1 \leq N^{1/5-\delta}.
\eeq
Let $c_1 < \delta$ and choose $\mfa >0$ satisfying $4/5 \leq \mfa \leq 4/5 +\delta -c_1$.  Then for $|\lambda| \leq N^{\delta/10}$ we have
\beq
\ee[ \exp[ \i \lambda ( \tr(f) - \ee[ \tr (f) ] ) ] ] = \exp[ - \lambda^2 \hat{V} (f) /2] + \O(N^{-\delta/10} + N^{\eps-c_1} ).
\eeq
\eep

\section{Calculation of variance and CLT}\label{sec:var-calc}

In this section, we will analyze the variance functional $\hat{V} (f)$, which was defined in \eqref{eqn:V-def}, for a few different types of test functions $f$.  Note that in the previous section we assumed only that $f$ was a regular test function as in Definition \ref{def:reg-f}; in this section we will impose more assumptions where necessary.  Then we will use this to conclude Theorem \ref{thm:clt-homog}, our main result on CLTs for linear spectral statistics. 

We first have the following purely algebraic calculation.
\bep\label{p:definegh}
Let $f$ be a regular test function.  Then
\begin{align}
\hat{V} (f) &= \frac{1}{ \pi^2}  \int_{ \Oma \times \Oma} ( \bar{ \del}_w \tilf (w) ) ( \bar{\del}_z \tilf (z) ) \nonumber \\
&\times \bigg\{ 2 g(z, w)  - \tr ( \bm'(z) S \bm' (w) ) +  \frac{1}{2 N^2} \sum_{ja} \sfo_{aj} \partial_z \partial_w (\bm_a (z) \bm_j (z) \bm_a (w) \bm_j (w) ) \bigg\},
\end{align}
where
\begin{align}
g(z, w) &=  \tr( \bm'(z) \bm(z)^{-1} (1 - S \bm(z) \bm(w))^{-1} S \bm(z) \bm'(w) (1-S \bm(z) \bm (w) )^{-1} ) \nonumber \\
&= \partial_w \tr \left( \bm'(z) \bm(z)^{-1} (1 - S \bm(z) \bm(w) )^{-1} \right) =: \partial_w h(z, w).
\end{align}
The function $g(z, w)$ is symmetric in $z$ and $w$.
\eep
\proof 
With $e$ denoting the vector of all $1$s, we have
\beq
\sum_i (1 - \bm^2 (z) S)^{-1}_{ij}= e^T (1 - \mb(z)^2 S)^{-1} \delta_j = \delta_j^T (1 - S \bm(z)^2 )^{-1}e.
\eeq
On the other hand, differentiating the quadratic vector equation gives
\beq
\frac{\bm'_j (z) }{ \bm^2_j (z) } = 1 + [S \bm'(z) ]_j.
\eeq
Hence, if $v$ is the vector $v_j = \bm'_j (z) / \bm^2_j (z)$ then we have $(1- S \bm^2) v = e$.  Therefore,
\beq
((1- S \bm^2 )^{-1} e )_j = ( e^T ( 1 - \bm^2 S )^{-1} )_j =\sum_i (1 - \bm^2 S)^{-1}_{ij} = \frac{ \bm'_j (z) }{ \bm^2_j (z) }.
\eeq
We now use this identity to simplify various expressions appearing in $\hat{V}(f)$.
We have,
\beq
- \sum_{i, j} (1- \bm^2 (z) S)^{-1}_{ij} [S \bm' (w) \bm^2(z) ]_{jj}  =- \tr ( \bm' (z) S \bm' (w) ),
\eeq
where we used cyclicity of the trace. Similarly, 
 by symmetry in $a$ and $j$,
\begin{align}
& \frac{1}{N^2} \sum_{i, j, a} \sfo_{ja} (1- \bm^2 (z) S)^{-1}_{ij}  \mb_a (z) \mb^2_j (z) \partial_w ( \mb_a (w) \mb_j (w) ) \nonumber \\
=&  \frac{1}{2 N^2} \sum_{ja} \sfo_{aj} \partial_z \partial_w (\bm_a (z) \bm_j (z) \bm_a (w) \bm_j (w) ).
\end{align}
Using
\beq\label{e:Ssimplify}
(1 - S \bm (z) \bm (w) )^{-1} S \bm(z) \bm (w)
= (1 - S \bm (z) \bm (w) )^{-1} - 1,
\eeq
we also have
\begin{align}
& \sum_{i, j} (1 - \bm^2(z) S)^{-1}_{ij} \partial_w [ \mb (z) (1 - S \bm (z) \bm (w) )^{-1} S \bm(z) \bm (w) ]_{jj} \nonumber \\
= & \sum_{i, j} (1 - \bm^2(z) S)^{-1}_{ij} \partial_w [ \mb (z) (1 - S \bm (z) \bm (w) )^{-1}  ]_{jj} \nonumber\\
= & \tr( \bm'(z) \bm(z)^{-1} \partial_w(1-S \bm(z) \bm(w) )^{-1} ) \nonumber \\
= & \tr( \bm'(z) \bm(z)^{-1} (1 - S \bm(z) \bm(w))^{-1} S \bm(z) \bm'(w) (1-S \bm(z) \bm (w) )^{-1} ) =: g(z, w).
\end{align}
Note that this can be rewritten  as
\begin{align}
&\tr( \bm'(z) \bm(z)^{-1} (1 - S \bm(z) \bm(w))^{-1} S \bm(z) \bm'(w) (1-S \bm(z) \bm (w) )^{-1} )\nonumber \\
&= - \tr( \bm' (z) \bm(z)^{-1} (1- S \bm (z) \bm (w) )^{-1} \bm'(w) \bm(w)^{-1} )\nonumber \\
&+  \tr ( \bm' (z) \bm(z)^{-1} (1- S \bm (z) \bm (w) )^{-1} \bm'(w) \bm(w)^{-1}( 1- S \bm (z) \bm (w) )^{-1} ),
\end{align}
so it is a symmetric function of $z$ and $w$ by cyclic property of the trace.  
This completes the proof. \qed

The following estimate for the function $g(z, w)$ is straightforward.
\bel \label{lem:gzw-bd}
Let $C'>0$.  For $|z|, |w| \leq C'$,
\beq
|g(z, w) | \leq C \| \mb'(z) \|_\infty \| \mb'(w) \|_\infty \| (1 - S \mb(z) \mb(w) )^{-1} \|^2_{\ell^\infty \to \ell^\infty}
\eeq
\eel
\proof This follows from
\beq\label{tracebound}
| \tr (AB)| \leq N \sup_{i,j} |B_{ij} | \times \| A\|_{\ell^\infty \to \ell^\infty}
\eeq
with $B = S$ and $A$ the rest of the expression in the trace (recall \eqref{eqn:m-size}). \qed

We first collect a few straightforward bounds for various terms in $\hat{V} (f)$ for general test functions. We record the following lemma for use in Section~\ref{sec:rest}. In particular, for macroscopic test functions, this gives an estimate for the variance of constant order. 
\bel \label{lem:var-smooth} 
Let $f$ be a regular test function with data $(\|f''\|_1^{-1}, C', c')$.  Then,
\begin{align} \label{eqn:revision-4}
&\left| \int_{ \Oma \times \Oma} ( \bar{ \del}_w \tilf (w) ) ( \bar{\del}_z \tilf (z) )  g(z, w)  \right| + \left| \int_{ \Oma \times \Oma} ( \bar{ \del}_w \tilf (w) ) ( \bar{\del}_z \tilf (z) )  \tr ( \bm'(z) S \bm' (w) )  \right| \nonumber \\
+ &  \left| \int_{ \Oma \times \Oma} ( \bar{ \del}_w \tilf (w) ) ( \bar{\del}_z \tilf (z) )    \frac{1}{2 N^2} \sum_{ja} \sfo_{aj} \partial_z \partial_w (\bm_a (z) \bm_j (z) \bm_a (w) \bm_j (w) )  \right|  \leq C ( \|f''\|_1  +1 )^2.
\end{align}
In particular,
\beq
\mathrm{Var} ( \tr f (W) ) \leq C ( \|f''\|_1 + 1)^2.
\eeq
\eel
 \proof Note that $\| \mb \|_\infty$ and $\| \mb'\|_\infty$ are bounded on the set of $z$ and $w$ such that $\bar{ \del}_z \tilf (z)$ and $\bar{ \del}_w \tilf (w)$ are non-zero  by Proposition~\ref{prop:basic-m}. As we have used before, the set of such $z$ and $w$ lies in $\mathcal{D}_{c'}$ by Definition~\ref{def:reg-f}. The estimates for second and third terms on the LHS of \eqref{eqn:revision-4} then follow from the assumptions on $S$ and $s^{(4)}_{ij}$ in Definition~\ref{d:wignertype}, and \eqref{ftilde}.  From Proposition \ref{prop:stab-3} and Lemma \ref{lem:gzw-bd} we see that, for $z$ and $w$ such that  $\bar{ \del}_z \tilf (z)$ and $\bar{ \del}_w \tilf (w)$ are non-zero,
\beq \label{eqn:g-est-1}
|g(z, w) | \leq \frac{C}{ ( | \Im[z]| + | \Im[w] | )^2}.
\eeq
Expanding out the terms in the product $\bar{ \del}_z \tilf (z) \bar{\del}_w \tilf (w)$ using \eqref{ftilde}, and writing $z = x_1 + \i y_1$ and $w = x_2 + \i y_2$, we see that except for the $f''(x_1) \chi'(y_1) y_1 f''(x_2) \chi'(y_2) y_2$ term, the others are non-zero only when $g$ is bounded (i.e., either $y_1$ or $y_2$ is order $1$).  But for this term, $y_1 y_2 g(x_1 + \i y_1, x_2 + \i y_2)$ is bounded by \eqref{eqn:g-est-1}, so the final estimate follows. \qed

For some of the terms in $\hat{V} (f)$, we have a better estimate if $f$ is non-zero only in the bulk. We will see below (using these estimates) that these terms are subleading for the half-regular bump functions of Definition \ref{def:half-f}.
\bel \label{lem:var-sublead}
Let $f$ be a regular test function with data $(\|f''\|_1^{-1}, c', C')$.  Suppose that $f(x) \neq 0$ only for $x \in \Ikap$ for some $\kappa > 0$.  Then,
\begin{align}
& \left| \int_{ \Oma \times \Oma} ( \bar{ \del}_w \tilf (w) ) ( \bar{\del}_z \tilf (z) )  \tr ( \bm'(z) S \bm' (w) )  \right| \nonumber \\
+ &  \left| \int_{ \Oma \times \Oma} ( \bar{ \del}_w \tilf (w) ) ( \bar{\del}_z \tilf (z) )    \frac{1}{2 N^2} \sum_{ja} \sfo_{aj} \partial_z \partial_w (\bm_a (z) \bm_j (z) \bm_a (w) \bm_j (w) )  \right|  \leq C 
\end{align}
\eel
\proof The functions $\mb(z)$, $\mb'(z)$, $\mb(w)$, $\mb'(w)$ are bounded where $( \bar{ \del}_w \tilf (w) ) ( \bar{\del}_z \tilf (z) )  \neq 0$ by Proposition~\ref{prop:basic-m}, because the set of such $z$ and $w$ lies in $\mathcal{D}_{c'}$ by Definition~\ref{def:reg-f}.  For holomorphic $H(z, w)$, we have by  integration by parts in $x_1$ that
\begin{align}
& \left| \int_{\Oma \times \Oma } f''(x_1) y_1 \chi (y_1) (f(x_2)+ \i y_2 f'(x_2) ) \chi' (y_2) H(z, w) \d x_1 \d y_1 \d x_2 \d y_2 \right| \nonumber \\
= &\left| \int_{\Oma \times \Oma } f'(x_1) y_1 \chi (y_1) (f(x_2)+ \i y_2 f'(x_2) ) \chi' (y_2) \del_z H(z, w) \d x_1 \d y_1 \d x_2 \d y_2 \right|,
\end{align}
where we used the notation $z = x_1 + \i y_1$ and $w = x_2 + \i y_2$.  
If $H(z, w)$ is bounded in the upper half-plane, then $| \del_z H(z, w) | \leq C|y_1|^{-1}$ by the Cauchy integral formula (see Lemma \ref{lem:cauchy-int}).  Hence,
\beq
\left| \int_{\Oma \times \Oma } f'(x_1) y_1 \chi (y_1) (f(x_2)+ \i y_2 f'(x_2) ) \chi' (y_2) \del_z H(z, w) \d x_1 \d y_1 \d x_2 \d y_2 \right| \leq C.
\eeq
A similar calculation holds for the term with $f''(x_1) y_1 f''(x_2) y_2$, integrating by parts in both $x_1$ and $x_2$ and applying the estimate $|\partial_z \partial_w H(z, w) |\leq C |y_1 y_2|^{-1}$. \qed

We now turn to a more detailed analysis of the kernel $g(z, w)$.  First, we collect some facts about the perturbation theory of the operators arising in the definition of the function $g(z, w)$.
\bep \label{prop:F-pert}
For any $\kappa >0$ and $C' >0$ consider the domain $D_1 \subseteq \cc^2$ defined by
\beq
D_1 := \left( \Ikap + \i [-C', C'] \right)^2.
\eeq
Consider the symmetric matrix
\beq
F(z, w) = | \mb(z) \mb(w) |^{1/2} S |\mb(z) \mb(w)|^{1/2}.
\eeq
All of the estimates below hold uniformly for $(z, w) \in D_1$. 
The eigenvalue of $F(z, w)$ that is largest in magnitude is positive, and we denote it by
\beq
\lambda_1 (z, w) := \|F(z, w) \|_{\ell^2 \to \ell^2}.
\eeq
The associated $\ell^2$-normalized eigenvector, which we denote by $v(z, w)$, has positive entries.  We have
\beq
c \leq \sqrt{N} v_i (z, w) \leq C, \qquad c \leq \lambda_1 (z, w) \leq 1 -c ( \Im[z] + \Im[w] ).
\eeq
The operator has a uniform spectral gap,
\beq
\inf_{i \neq 1} \lambda_1 (z, w) - | \lambda_i (F(z, w)) | \geq c_g,
\eeq
for $c_g >0$. 
Decompose $F$ as
\beq
F(z, w) = \lambda_1 (z, w) v(z, w) v^T (z, w) + A (z, w).
\eeq
For $(z, w_1) \in D_1$ and $(z, w_2) \in D_1$, we have
\beq \label{eqn:pert-ub-c1}
\|v(z, w_1) - v(z, w_2) \|_2 + | \lambda_1 (z, w_1) - \lambda_1 (z, w_2) | \leq C|w_1 - w_2|,
\eeq
as well as
\beq \label{eqn:FA-ub}
\|F(z, w_1) - F(z, w_2) \|_{\ell^2 \to \ell^2 } + \| A(z, w_1) - A(z, w_2 ) \|_{\ell^2 \to \ell^2 } \leq C|w_1 -w_2|.
\eeq
The last two estimates also hold with the right side replaced by $\min\{ |w_1 - w_2| , |w_1 - \bar{w}_2 |\}$.

Furthermore, 
\begin{align} \label{eqn:eig-pert}
\lambda_1 (x + \i \eta, y - \i \eta ) &= \lambda_1 ( x + \i \eta, x - \i \eta)\left(1 + (y-x) \Re[ v^T(x+\i \eta, x - \i \eta) \mb'(z)/ \mb(z) v(x+ \i \eta, x- \i \eta) ] \right) \nonumber \\
&+ \O (|x-y|^2),
\end{align}
where $z = x + \i \eta$ above. 
\eep
\proof  The statements about the eigenvector entries, size of and positivity of $\lambda_1 (z, w)$ and spectral gap follow from Proposition \ref{prop:F-est}.  We may  assume that $z$ and $w$ are in the same half-space due to the symmetry of $F$.  For $w_1$ and $w_2$ as in the statement, consider
\beq
w_t := w_1 + t \frac{w_2-w_1}{ |w_2 - w_1|}
\eeq
Since $\mb(z)$ obeys $|| \mb'(z) ||_{\infty} \leq C$ as well as $c \leq | \mb_i (z) | \leq C$ in our domain by Proposition~\ref{prop:basic-m}, we have
\beq
| \dot{F}_{ij} (z, w_t )| \leq \frac{C}{N},
\eeq
where $\dot{F}(z, w_t) := \partial_t (F(z, w_t ) )$.  
Denote by $\Pp$ the orthogonal projection onto $v(z, w_t)^\perp$.  
We have the standard formula (see, e.g, Theorem 2.6 of Section VIII.2.3 of \cite{kato}),
\beq
\frac{ \d}{ \d t} v(z, w_t) = \frac{1}{ \lambda_1(z, w_t) - \Pp F \Pp } \Pp \dot{F} (z, w_t) v (z, w_t) ,
\eeq
and so by the spectral gap condition,
\beq \label{eqn:eig-pert-a2}
\| \partial_t v(z, w_t) \|_2 \leq C \| \dot{F}(z, w_t) \|_{\ell^2 \to \ell^2} \leq C.
\eeq
Similarly, 
\beq \label{eqn:eig-pert-a1}
\frac{ \d }{\d t} \lambda_1 (z, w_t) = v(z, w_t)^T \dot{F} (z, w_t ) v(z, w_t) = \O (1).
\eeq
The estimate for $A$ follows from the fact that
\beq
\|A(z, w_1) - A(z, w_2)\|_{\ell^2 \to \ell^2 } \leq \| F(z, w_1) - F(z, w_2) \|_{\ell^2 \to \ell^2} + \| [ \lambda_1 v v^T ](z, w_1) - [ \lambda_1 v v^T ] (z, w_2) \|_{\ell^2 \to \ell^2}.
\eeq
It remains to prove \eqref{eqn:eig-pert}.  From \eqref{eqn:eig-pert-a1}, \eqref{eqn:eig-pert-a2} the fact that $\| \mb'' (z) \|_\infty \leq C$ (which follows from Proposition \ref{prop:basic-m}) we have that the second derivative of $t \to \lambda_1 (z, w_t)$ is bounded by a constant.  To prove \eqref{eqn:eig-pert} it therefore remains to evaluate the derivative at the point $z=w$. 
We calculate, 
\begin{align}
&\frac{ \d}{ \d y} | \bm(x+ \i \eta) \bm(y- \i \eta) |^{1/2} S | \bm(x+\i \eta) \bm(y- \i \eta ) |^{1/2} \bigg\vert_{y =x} \nonumber \\
=&\frac{ \d}{ \d y} | \bm(x+ \i \eta) \bm(y+ \i \eta) |^{1/2} S | \bm(x+\i \eta) \bm(y+ \i \eta ) |^{1/2} \bigg\vert_{y =x} \nonumber \\
= & \frac{1}{2} \Re \left[ \frac{\mb'(x+\i \eta)}{ \mb (x+\i \eta ) } F(x+\i \eta, x+\i \eta) + F(x+\i \eta, x+\i \eta) \frac{\mb'(x+\i \eta)}{ \mb (x+\i \eta ) }  \right],
\end{align}
and the first equality in \eqref{eqn:eig-pert-a1}. Note the last line here follows from
\begin{align}
|f(t+\i \eta)|^{1/2}\partial_t |f(t+\i \eta) |^{1/2} &= \frac{|f(t+\i \eta)|^{1/2}}{2} \frac{\Re[f]\del_t \Re[f]+\Im[f] \del_t \Im[f]}{|f(t+\i \eta)|^{3/2}} \nonumber \\
&= \frac{ |f(t+\i \eta)|}{2} \frac{ \Re[f'(t+\i \eta) \bar{f} (t+\i \eta)] }{|f(t+\i \eta)|^2} \nonumber\\
&= \frac{ | f(t+\i \eta) |}{2} \Re\left[ \frac{ f'(t+\i\eta)}{ f(t+\i\eta)}\right],
\end{align}
which holds for any holomorphic $f$.
\qed

With the above proposition, we now turn to calculating the leading-order terms of the  functions $g(z, w)$ and $h(z, w)$ when $z$ and $w$ are close. We recall these functions were defined in Proposition~\ref{p:definegh}.

\bep \label{prop:F-bds}
Let $\kappa>0$ and suppose that $\Re[z], \Re[w] \in \Ikap$, and that $(\Im z)(\Im w) > 0$.  Then
\beq
|g(z, w) | \leq C.
\eeq
Now let $z = x + \i \eta$ and $w = y - \i \eta$ for $\eta>0$.  Suppose that $x, y \in \Ikap$.  Then there is an $r>0$ such that for $|x-y| < r$ and $0 < \eta < r$, we have
\begin{align}
h(z, w) &=   \tr ( \bm' (z) \bm(z)^{-1} ) + P(z, w),
\end{align}
where
\begin{align} \label{eqn:P-upper}
|P(z, w) | \leq \frac{C}{ \eta + |x-y|},
\end{align}
and the boundary values satisfy
\begin{align}\label{eqn:P-bv}
P(x + \i 0, y- \i 0 ) = \frac{1}{ y-x} + \O(1).
\end{align}
Furthermore, there is an $r_1>0$ such that if $\eta < r/2$ and $0 < \eta_1 < r_1$, then if $r>|x-y|>r/2$, then
\begin{align} \label{eqn:F-vert-bd}
|P(x+\i \eta, y - \i \eta- \i \eta_1)| \leq C.
\end{align}
\eep
\proof The first estimate follows from Lemmas \ref{lem:gzw-bd} and \ref{lem:stab-bulk}.  For the function $h(z, w)$ we recall here its definition,
\beq
 h (z, w) = \mathrm{tr}\left( \bm'(z) \bm(z)^{-1} (1 - S \bm(z) \bm(w) )^{-1} \right).
\eeq
 The remaining estimates will be consequences of the perturbation theory of the operator
\beq
F(x, y) = | \bm(z) \bm(w) |^{1/2} S | \bm(z) \bm(w) |^{1/2},
\eeq
which was summarized in the previous proposition.  Recall
 our notation $z = x+ \i \eta$ and $w = y - \i \eta$.   Fix $D_1$ as in Proposition \ref{prop:F-pert} with, say, $C'=1$. 
Define $U$ the diagonal unitary operator,
\beq
U = \frac{ \bm(z) \bm(w)}{ |\bm(z) \bm(w) |}.
\eeq
By Taylor expansion,
\beq \label{eqn:U-l2}
\| U - 1 \|_{ \ell^2 \to \ell^2 } \leq C|x-y|.
\eeq
For $h(z, w)$, we have
\beq
h(z, w) = \tr ( \bm'(z) ( \bm(z))^{-1} U^* (U^* - F)^{-1} ).
\eeq
Write $F = A + \lambda_1 v v^T$ with $\lambda_1$ and $v$ as in Proposition \ref{prop:F-pert}.  Note that,
\beq \label{eqn:A-gap}
\|A\|_{\ell^2 \to \ell^2} \leq 1 -c_g.
\eeq
From \eqref{eqn:A-gap}, we see that for any unitary $V$,
\beq \label{eqn:VA-ub}
\|(V-A)^{-1} \|_{\ell^2 \to \ell^2} \leq C .
\eeq
  Note also that $|A_{ij}| \leq C N^{-1}$ for all $i, j$ and  $z, w \in D_1$. By the Sherman--Morrison formula,
\begin{align} \label{eqn:hzw-sm}
\frac{1}{U^* - F} = \frac{1}{U^*-A} + \frac{ \lambda_1 (U^* - A)^{-1} v v^T (U^* - A)^{-1}}{1 - \lambda_1 v^T(U^*-A)^{-1} v }.
\end{align}
The contribution of the first term on the RHS of \eqref{eqn:hzw-sm} to $h(z, w)$ can be written as,
\begin{align}
\tr( \bm'(z) ( \bm(z) )^{-1} U^* (U^* - A)^{-1} ) &= \tr ( \bm' (z) \bm(z)^{-1} ) \nonumber \\
&+ \tr ( \bm'(z) ( \bm(z) )^{-1}  A U )\nonumber \\ &- \tr ( \bm'(z) ( \bm(z) )^{-1} A (U^* -A)^{-1} A U ).
\end{align}
The second and third lines are $\O (1)$, from the fact that $|A_{ij}| \leq CN^{-1}$, \eqref{eqn:VA-ub}, and Proposition \ref{prop:basic-m}. 
We now calculate the contribution of the second term of \eqref{eqn:hzw-sm} to $h(z, w)$.  By the cyclic property of trace, we have
\begin{align}
&\tr \left( \mb' (z)( \bm (z) )^{-1} U^* \frac{ \lambda_1 (U^* - A)^{-1} v v^T (U^* - A)^{-1}}{1 - \lambda_1 v^T(U^*-A)^{-1} v } \right) \nonumber \\
&=  \lambda_1 \frac{v^T (U^* - A)^{-1}  \mb'(z) ( \bm(z) )^{-1} U^* (U^* - A)^{-1} v }{1 - \lambda_1 v^T (U^*-A)^{-1} v}. \label{eqn:h-exp-1}
\end{align}

From \eqref{eqn:VA-ub}, \eqref{eqn:U-l2}, and \eqref{eqn:FA-ub} we have,
\begin{align} \label{eqn:UA1A-l2}
\| (U^*(z, w) - A(z, w))^{-1} - (1-A(z, z))^{-1} \|_{\ell^2 \to \ell^2} \leq C|x-y|.
\end{align}
Denote $\mff(z) = v(z, z)$ and $\mu(z) = \lambda_1 (z, z)$.  
The numerator of \eqref{eqn:h-exp-1} is
\begin{align}
&\left[\lambda_1 v^T (U^* - A)^{-1}  \mb'(z) ( \bm(z) )^{-1} U^* (U^* - A)^{-1} v \right] (z, w) \nonumber \\
=& \mu \mff^T (1 - A(z, z))^{-1} \mb'(z) ( \mb(z) )^{-1} (1-A(z, z))^{-1} \mff + \O (|x-y|) \nonumber \\
= & \mu \mff^T\mb'(z) ( \mb(z) )^{-1}  \mff + \O (|x-y|).
\end{align}
The notation with the square brackets in the first line is to indicate that $\lambda_1$, $U$, $A$, and $v$ are evaluated at $(z, w)$.  The second line is a consequence of \eqref{eqn:U-l2}, \eqref{eqn:pert-ub-c1}, \eqref{eqn:UA1A-l2}, and Proposition~\ref{prop:basic-m}.
The third line uses $(1-A(z,z))^{-1} \mff(z) = \mff(z)$. 

We now turn to the denominator of \eqref{eqn:h-exp-1}.  We begin with
\beq
\frac{1}{U^* - A} = \frac{1}{1-A} + \frac{1}{1-A} (1-U^*) \frac{1}{1-A} + \frac{1}{ U^* - A} (1- U^*)\frac{1}{1-A} (1-U^*) \frac{1}{1-A} .
\eeq
Then by \eqref{eqn:U-l2} and the fact that $(1-A)^{-1} v = v$,
\begin{align}\label{therefore2}
v^T (U^*-A)^{-1} v &=1 + v^T(1- U^*)v + \O (|x-y |^2).
\end{align}
For holomorphic $f$,
\begin{align}
\frac{ \bar{f} (t + \i \eta)}{ | f(t + \i \eta ) |}\partial_t \frac{ f(t + \i \eta) }{ | f ( t + \i \eta )| }=& \frac{ \bar{f} (t + \i \eta)}{ | f(t + \i \eta ) |} \left( \frac{f'(t+\i \eta)}{ |f ( t + \i \eta ) | } - \frac{f ( t + \i \eta ) }{ | f ( t + \i \eta )|^3} \Re[ f' ( t + \i \eta ) \bar{f} (t + \i \eta ) ]\right) \nonumber\\
= &\frac{ f'(t + \i \eta ) }{ f(t + \i \eta ) } - \Re\left[ \frac{ f'(t+\i \eta)}{ f ( t + \i \eta ) } \right] = \i \Im \left[ \frac{ f'(t+\i \eta)}{ f ( t + \i \eta ) } \right].
\end{align}
Therefore, 
\beq \label{eqn:U-der-b1}
\frac{ \d }{ \d y} \frac{ \mb(x- \i \eta ) \mb (y + \i \eta) }{ | \mb(x- \i \eta ) \mb (y + \i \eta) |} \bigg\vert_{y =x}= \i \Im \left[ \frac{ \mb'(x+ \i \eta)}{ \mb (x + \i \eta) } \right],
\eeq
from which it follows that,
\beq \label{eqn:U-der-b2}
\| 1 - U^*(z, w) - \i(x-y) \Im[\mb'(z) (\mb(z))^{-1}] \|_{\ell^2 \to \ell^2} \leq C|x-y|^2.
\eeq
We then have
\begin{align}\label{therefore1}
v^T(z, w) (1-U^*(z, w) ) v(z, w) &= \mff^T (1-U^*(z, w) ) \mff + \O (|x-y|^2) \nonumber\\
&= \i \Im\left[  (x-y )\mff^T \bm'(z) \bm(z)^{-1} \mff  \right] + \O ( |x-y|^2 ),
\end{align}
where in the first line we used \eqref{eqn:U-l2} and \eqref{eqn:pert-ub-c1}, and the second line follows from \eqref{eqn:U-der-b2}.  
From \eqref{eqn:eig-pert}, we have
\beq \label{eqn:pert-d1}
\lambda_1(z, w) = \mu (z)+ \mu (z) (y-x) \Re\left[ \mff^T \mb'(z) / \mb(z) \mff\right] + \O (|x-y|^2).
\eeq
For the denominator of \eqref{eqn:h-exp-1}, we  have therefore derived, using \eqref{therefore2}, \eqref{therefore1}, and \eqref{eqn:pert-d1}, that
\begin{align}
1 - \lambda_1 v^T (U^*-A)^{-1} v &= (1- \mu) + (x-y) \mu \overline{ \mff^T \mb'(z) / \mb(z) \mff } + \O (|x-y|^2 ).
\end{align}
We now turn to calculating the quadratic form
\beq
 \mff^T \mb'(z) / \mb(z) \mff,
\eeq    
which will be seen to have an explicit form on the real axis.  
 Taking the imaginary part of the quadratic vector equation and writing $\Ep = E+\i 0$, we see that
\beq \label{eqn:qve-cor-1}
\frac{ \Im[ \mb_i (\Ep ) ]}{ | \mb_i (\Ep )|^2}= (S \Im [ \mb (\Ep )] )_i,
\eeq
where now, as the notation suggests, we think of the right side as the matrix $S$ acting on the vector $\Im[ \mb (\Ep) ]$.  
Hence, the components of the vector $\mff_i(\Ep)$ are proportional to,
\beq\label{fi}
\mff_i ( \Ep ) \mbox{ } \propto \mbox{ } \frac{ \Im[ \mb_i (\Ep ) ]}{ | \mb_i (\Ep ) |}.
\eeq
Differentiating the quadratic vector equation, multiplying by $\Im [ \mb_i (\Ep ) ]$ and taking a sum, we derive
\begin{align}
\sum_i \frac{ \Im [ \mb_i ( \Ep ) ] \mb'_i (\Ep ) }{ \mb_i^2 (\Ep ) } &= \sum_i \Im [ \mb_i (\Ep ) ] + \Im[\mb (\Ep )]^T S \mb' (\Ep) \nonumber \\
&=   \sum_i \Im [ \mb_i (\Ep ) ] + \sum_i \frac{ \Im [\mb_i (\Ep ) ] \mb_i' (\Ep ) }{ | \mb_i (\Ep ) |^2} .
\end{align}
In the second line we used the symmetry of $S$ and \eqref{eqn:qve-cor-1}.  Since
\beq
\frac{1}{ \mb_i^2 (\Ep ) } - \frac{1}{ | \mb_i (\Ep ) |^2 } = - 2 \i \frac{ \Im [ \mb_i (\Ep ) ] }{ \mb_i ( \Ep ) | \mb_i (\Ep )|^2},
\eeq
we see that
\beq
\sum_i  \frac{ \Im [ \mb_i ( \Ep ) ]^2 \mb_i' ( \Ep ) }{ \mb_i ( \Ep ) | \mb_i ( \Ep ) |^2 } = \frac{\i}{2} \sum_i \Im [ \mb_i ( \Ep ) ].
\eeq
Using \eqref{fi} and the previous equation, we have
\beq
\mff^T \mb'( \Ep ) \mb(\Ep)^{-1} \mff = \frac{ \i}{2} \frac{ \sum_i \Im [ \mb_i ( \Ep ) ] }{ \sum_i \Im[ \mb_i ( \Ep ) ]^2 | \mb_i (\Ep)|^{-2} }, 
\eeq
and so,
\beq
\Re\left[ \mff^T \mb'( \Ep ) \mb(\Ep)^{-1} \mff \right] = 0, \qquad \Im \left[ \mff^T \mb'( \Ep ) \mb(\Ep)^{-1} \mff \right] \asymp \frac{1}{ \rho (E) } \asymp 1.
\eeq
By \eqref{eqn:pert-ub-c1} we have
\beq
\| \mff(E+ \i 0) - \mff (E+\i \eta ) \|_2 \leq C \eta ,
\eeq 
and so we see that there is an $r>0$ such that
\beq
\Im\left[ \mff(x+\i \eta)^T \mb'(x+\i \eta) / \mb(x+\i \eta) \mff(x+\i \eta)\right] \geq c
\eeq
for $x \in \Ikap$ and $0 < \eta < r$.  By \eqref{eqn:F-norm-bulk} we have $\mu \leq 1- c \eta$, so
\beq
| 1 - \lambda_1 v^T (U^*-A)^{-1} v  | \geq c( \eta + |x-y| )
\eeq
for $|x-y| <r$ and $0 < \eta < r $, for some $r>0$.  The statement \eqref{eqn:P-bv} about boundary values follows from the above results, as we have derived the leading-order expansions in $(x-y)$ of the numerator and denominator of \eqref{eqn:h-exp-1}. 

 For \eqref{eqn:F-vert-bd}, we note that if $|x-y| > r/2$, the denominator of \eqref{eqn:h-exp-1} is bounded below by
\beq
| 1 - \lambda_1 v^T (U^*-A)^{-1} v  | \geq cr.
\eeq
On the other hand,
\beq
\left| [\lambda_1 v^T (U^*-A)^{-1} v] (x+ \i \eta, y - \i \eta ) - [\lambda_1 v^T (U^*-A)^{-1} v ] (x + \i \eta , y - \i ( \eta + \eta_1 )) \right| \leq C \eta_1
\eeq
by Proposition \ref{prop:F-pert}.  Hence, taking $r_1$ small enough yields \eqref{eqn:F-vert-bd}. 
\qed

We now use the previous result to take boundary values and derive the following.  Recall the definition of $\Oma$ in \eqref{eqn:oma-def}. 
\bel \label{lem:var-lead}  Let $f$ be a regular test function with data $(\|f''\|_1^{-1}, c', C')$,  Suppose that $f$ is supported in $\Ikap$ for some $\kappa >0$.  
Fix $\mathfrak a >0$ and assume that
\beq
\log(N) N^{\mfa-1} \|f''\|_1  \leq 1
\eeq
for large enough $N$ (depending only on $\mathfrak a$).
 There is an $r>0$ (depending only on the $\kappa$ in $\Ikap$) such that if the support of $f$ is contained in an interval of length $r$, with midpoint $E_0$, then
\begin{align}
 \frac{2}{ \pi^2}\int_{\Oma^2} ( \bar{\del}_z \tilf (z) ) ( \bar{\del}_w \tilf (w) )g(z, w) = \frac{1}{ \pi^2} \int_{[E_0-r, E_0+r]^2} \frac{ f(y) - f(x) }{y-x} f'(y) \d x \d y + \O(1).
\end{align}
\eel

\proof Fix $0 < \eta < N^{\mfa-1}$.  We will take $\eta \dto 0$ at the end of the proof to obtain boundary values of the integrals above.  Let the support of $f$ be contained in an interval $[E_0 -r/2, E_0+r/2]$ with $r$ small enough so that the estimates of Proposition \ref{prop:F-bds} hold for $10r$. Fix the domains
\beq
D^\pm_\eta :=\{ x + \i y : x \in [E_0-r, E_0 +r], \eta < \pm y < 4L_* \}, \qquad D_\eta = D^+_\eta \cup D^-_\eta,
\eeq
where $L_*$ is as in Definition \ref{def:dc}.  Recall the estimate $|g(z, w) | \leq C / ( |\Im[z]| + | \Im[w] | )^2$ from \eqref{eqn:g-est-1}.  From this, we have by direct computation that
\beq
\int_{ \eta < y_1 < N^{\mfa-1}}\int (f''(x_1) y_1) ( f(x_2) + \i f'(x_2) y_2 ) \chi'(y_2) g(x_1 + \i y_1, x_2 + \i y_2) \d x_1 \d x_2 \d y_1 \d y_2 = \O (N^{\mfa-1} \|f''\|_1 ) ,
\eeq
 where the $x_i$ integration is over $\Ikap^2$, as this is where $f$ is non-zero, and the $y_2$ integration is only over where $\chi'(y_2) \neq 0$.

Similarly, we have
\begin{align}
\int_{\eta < |y_1|, |y_2| < N^{\mfa-1}}|f''(x_1) y_1 f''(x_2) y_2 | |g(x_1 + \i y_1, x_2 + \i y_2) | \d x_1 \d x_2 \d y_1 \d y_2 \leq C ( N^{\mfa-1} \|f''\|_1 )^2.
\end{align}
Applying Lemma \ref{lem:H} with $s=2$ in the integration in $z =x_1 + \i y_1$, we have
\begin{align}
\left|\int_{z \in \Oma, \eta < |y_2| < N^{\mfa-1}} f''(x_1) f''(x_2) y_1 y_2 g(x_1 + \i y_1, x_2 + \i y_2) \d x_1 \d x_2 \d y_1 \d y_2 \right|\leq C \log(N) (N^{\mfa-1} \|f''\|_1 )^2.
\end{align}
Hence, we obtain the estimate
\begin{align}
\frac{2}{ \pi^2}\int_{\Oma^2} ( \bar{\del}_z \tilf (z) ) ( \bar{\del}_w \tilf (w) )g(z, w) \d x_1 \d x_2 \d y_1 \d y_2 =& \frac{2}{ \pi^2}\int_{D_\eta^2 } ( \bar{\del}_z \tilf (z) ) ( \bar{\del}_w \tilf (w) )g(z, w) \d x_1 \d x_2 \d y_1 \d y_2 \notag\\
+ &  \O (  N^{\mfa-1} \|f''\|_1  ).
\end{align}
Let $P(z, w)$ be as in Proposition \ref{prop:F-bds}.  From the formula
\beq
P(z, w) = \tr \left( \mb' (z)(1- S \mb(z) \mb(w))^{-1} S \mb(w) \right),
\eeq
we see from Proposition \ref{prop:stab-3} and \eqref{tracebound} that
\beq \label{eqn:F-bd-2}
|P(z, w) | \leq \frac{C}{ | \Im[z]| + | \Im[w] | }.
\eeq
We now have
\begin{align}
 & \frac{2}{ \pi^2}\int_{D^+_\eta \times D^-_\eta } ( \bar{\del}_z \tilf (z) ) ( \bar{\del}_w \tilf (w) )g(z, w)  \d x_1 \d x_2 \d y_1 \d y_2 \notag\\
 =&  \frac{2}{ \pi^2}\int_{D^+_\eta \times D^-_\eta } ( \bar{\del}_z \tilf (z) ) ( \bar{\del}_w \tilf (w) )\del_w P(z, w)  \d x_1 \d x_2 \d y_1 \d y_2 \nonumber \\
=&-  \frac{2}{ \pi^2}\int_{D^+_\eta \times D^-_\eta } ( \bar{\del}_z \tilf (z) ) ( \bar{\del}_w \tilf' (w) ) P(z, w) \d x_1 \d x_2 \d y_1 \d y_2 \nonumber \\
= & -  \frac{2}{ \pi^2}\int_{D^+_\eta \times D^-_\eta} ( \bar{ \del}_z  \bar{\del}_w ) [ \tilf(z) \tilf' (w)  P(z, w)] \d x_1 \d x_2 \d y_1 \d y_2 \nonumber \\
=&- \frac{2}{ \pi^2}\int_{D^+_\eta \times D^-_\eta } ( \bar{ \del}_z  \bar{\del}_w ) [(\tilf(z) - \tilf(w) ) \tilf' (w) ) P(z, w)] \d x_1 \d x_2 \d y_1 \d y_2 .
\end{align}
In the second line we integrated by parts (by considering $\del_w P(z, w) = \del_{x_2} P (z, w)$ and integrating by parts in the real variable $x_2$).  We also used  that $P(z, w)$ is a holomorphic function of $z$ in the domain $\D_\eta$, so $\bar{\del}_z P(z, w) = 0$, and the same statement in the $w$ variable.  We now apply Green's theorem, which in complex notation reads, for any $F \in C^1(\mathbb{C})$, 
\beq \label{eqn:green-thm}
\int_{\Omega} \bar{\del}_z F(z) \d x \d y = - \frac{\i}{2} \int_{\del \Omega} F(z) \d z ,
\eeq
the RHS being a line integral over $\del \Omega$ with the usual counter-clockwise orientation. 
The main contribution of the integral over the boundary of $(z, w) \in D^+_\eta \times D^-_\eta$ comes from the lines $(x+\i \eta) \times (y-\i \eta)$.  However, there is an additional contribution over the remaining boundary of $D^+_\eta$ due to the $- \tilf(w) \tilf'(w) P(z, w)$ term being non-zero there.  To fix notation, let
\beq
\del D^\pm_\eta = \Gamma_1^\pm \cup \Gamma_2^\pm, \qquad \Gamma_1^\pm = \{ x \pm \i \eta : x \in [E_0 -r, E_0 + r ] \},
\eeq
and $\Gamma_2^\pm$ is defined so that the union is disjoint.  For $(z, w) \in \Gamma_2^+ \times \Gamma_1^-$, we see that $|P(z, w)| \leq C$ if $\tilf(w) \neq 0$, by \eqref{eqn:F-vert-bd} and \eqref{eqn:F-bd-2}. Hence,
\begin{align}
\left| \int_{ \Gamma_2^+ \times \Gamma_1^- } \tilf(w) \tilf'(w) P(z, w) \d z \d w \right| &\leq C \int_{E_0-r}^{E_0+r} \left| (f(x) + \i \eta f'(x) )( f'(x) + \i \eta f''(x) ) \right| \d x \notag\\
&\leq C \int_{E_0-r}^{E_0+r} |f'(x) | + \eta(1+ \| f'' \|_\infty )(1+| f' \|_\infty) \d x  \notag\\
&\leq C + \O ( \eta),
\end{align}
where we do not track the $N$-dependence of the $\eta$ term as it goes to $0$ at the end of the proof. Note we used that $\| f \|_\infty \leq C$ and $\| f' \|_1 \leq C$ which both hold by assumption.  

For $(z, w) \in \Gamma_1^+ \times \Gamma_1^-$ we have
\beq
|\tilf(z) - \tilf(w) | \leq ||f'||_\infty( |\Re[z] - \Re[w] | + \eta),
\eeq
and so from \eqref{eqn:P-upper} 
we see that 
\beq
\left| ( \tilf(z) - \tilf (w) ) \tilf'(w) P(z, w) \right|
\eeq
is bounded uniformly in $\eta$ for $(z, w) \in \Gamma_1^+ \times \Gamma_1^-$.  Hence, we can take $\eta \dto 0$ and obtain
\begin{align}
&\lim_{\eta \dto 0 } - \frac{2}{ \pi^2}\int_{D^+_\eta \times D^-_\eta } ( \bar{ \del}_z  \bar{\del}_w ) ( (\tilf(z) - \tilf(w) ) \tilf' (w) ) P(z, w) \d x_1 \d x_2 \d y_1 \d y_2  \nonumber \\
&= \frac{1}{2 \pi^2} \int_{[E_0-r, E_0+r]^2} \frac{ f(y) - f(x)}{ y-x} f'(y) \d x \d y + \O(1).
\end{align}
Since $g(z, w)$ is uniformly bounded for $z, w$ in the same half-plane, the integrals over $D^+_\eta \times D^+_\eta$ and $D^-_\eta \times D^-_\eta$ contribute only constant order.  By symmetry the contribution over $D^+_\eta \times D^-_\eta$ is the same as $D^-_\eta \times D^+_\eta$.  This completes the proof. 
\qed

The following lemma provides a calculation of the quadratic form we found in the previous proposition for smoothed step functions. The proof is a simple application of the fundamental theorem of calculus and so is deferred to Appendix \ref{sec:h12-calc}.

\bel \label{lem:h12-calc}
Let $t \in (0, 10^{-2})$ and $M \geq 1$.  Let $1/10 > r >0$.  
Let $\phi$ be a function obeying the following: we have
$
\phi'(x) \neq 0 
$
only if $|x| \leq t M$, and
\beq\label{e:phiderivbound}
 | \phi'(x)| \leq \frac{Kt}{t^2 + x^2},
\eeq
for some $K >0$.  
Assume $tM \leq r/2$, $\phi(-1) = 0$, and $\phi(1) = 1$.  Then 
\begin{align}
\int_{[ -r, r]^2} \left( \frac{ \phi (x) - \phi (y) }{x-y} \right)^2 \d x \d y &= 2 | \log (t) | + (K+1)^2 \O ( | \log | \log t | |).
\end{align}
Above, the big $\O$ notation hides dependence in the error on $r$.
\eel

With all of the above preparations, we can finally calculate the quantity $\hat{V} (f)$ for half-regular bump functions, as in Definition \ref{def:half-f} to leading order.
\bel \label{lem:half-reg-var-final}
Let $\kappa>0$.  
There is an $r>0$, depending on $\kappa$ such that the following holds.  
Let $f$ be a half-regular bump function with data $(t, M, E_0, E_1, C', c')$.  Denote $t = N^{-\omega}$ with $0 < \omega < 1/5$.  Assume $0 < \mfa < 1 - \omega$.  Assume $c' <r/10$ and $|E_0-E_1| \leq r/2$ and $t M \leq c'/100$. Assume $E_0, E_1 \in \Ikap$.  Then,
\begin{align}
\hat{V} (f) &= \frac{1}{ \pi^2} | \log(t) | + \O( | \log  |\log(t)||).
\end{align}
\eel
\proof From Lemmas \ref{lem:var-sublead} and  \ref{lem:var-lead}, it follows that
\begin{align}
\hat{V} (f) =  \frac{1}{ \pi^2} \int_{[E_0-r, E_0+r]^2} \frac{ f(y) - f(x) }{y-x} f'(y) \d x \d y + \O(1).
\end{align}
Using integration by parts and  the fact that the support of $f$ is restricted to $[E_0-3r/4, E_0+3r/4]$, we have
\begin{align}
 &\int_{[E_0-r, E_0+r]^2} \frac{ f(y) - f(x) }{y-x} f'(y) \d x \d y =  \int_{[E_0-r, E_0+r]^2} \frac{ f(y) - f(x) }{y-x} \frac{\d}{ \d y} (f(y) - f(x) ) \d x \d y \nonumber \\
=& - \int_{[E_0-r, E_0+r]^2} \frac{ f(y) - f(x) }{y-x} f'(y) \d x \d y + \int_{[E_0-r, E_0+r]^2} \frac{ (f(y) - f(x) )^2}{ (x-y)^2} \d x \d y + \O(1).
\end{align}
The claim then follows from Lemma \ref{lem:h12-calc}, after noting that
\beq
 \int_{[E_0-r, E_0+r]^2} \frac{ (f(y) - f(x) )^2}{ (x-y)^2} \d x \d y  =  \int_{[E_0-c'/10, E_0+c'/10]^2} \frac{ (f(y) - f(x) )^2}{ (x-y)^2} \d x \d y  + \O(1)
\eeq
and that the restriction of $f$ to $[E_0-c'/10, E_0+c'/10]$,  satisfies the hypotheses of Lemma \ref{lem:h12-calc}.  \qed

\subsection{Characteristic function of LSS coming from homogenization}

The following result compiles many of our calculations above into a single result about the characteristic function of linear spectral statistics corresponding to half-regular bump functions (as in Definition \ref{def:half-f}).  It is used to calculate the distribution of the linear spectral statistic coming from the application of the homogenization theorem.

\bet  \label{thm:clt-homog}
Let $\kappa>0$.  There is an $r>0$ depending on $\kappa$ such that the following holds.  Let $f$ be a half-regular bump function with data $(t, M, E_0, E_1, C', c')$, with $E_0, E_1 \in \Ikap$.    Assume $t = N^{-\omega}$ with $0 < \omega < 1/5$.  Assume that $|E_0 -E_1| < r/2$, and that $tM \leq c'/100$ and that $c'<r/10$.  There is a $b>0$, and a $V$ that satisfies
\beq
V = \frac{1}{ \pi^2} |\log(t)| + \O ( \log \log(N) ),
\eeq
so that for $| \lambda| \leq N^{b}$ we have
\beq
\ee\left[ \exp \left( \i \lambda( \tr f(W) - \ee[ \tr f (W) ] ) \right) \right] = \exp\left[ - \lambda^2 V/2 \right] + \O (N^{-b} ),
\eeq
and so for $| \lambda| \leq \log(N)^{1/4}$ we have
\beq
\ee \left[ \exp \left( \i \lambda \frac{ \tr f (W) - \ee[ \tr f (W) ] }{ \sqrt{ \log(N) }} \right) \right] = \exp\left[ - \lambda^2 \frac{\omega}{2 \pi^2} \right] + \O ( (\log(N))^{-1/4} ).
\eeq
\eet
\proof The first estimate follows from Proposition \ref{prop:char-func} with an appropriate choice of $\mfa$, with $V = \hat{V} (f)$ calculated in Lemma \ref{lem:half-reg-var-final}. The second estimate follows from rescaling the first.  \qed

\subsection{Calculation of expectation}

Consider again $W$ a matrix of general Wigner-type satisfying Assumptions \ref{it:ass-1} and \ref{it:ass-2}.  In this section we fix  a regular test function $f$ as in Definition \ref{def:reg-f}.  We will calculate the leading-order corrections to the expectation of the linear spectral statistic corresponding to $f$.  This is similar to the second half of Section 4.3 of \cite{meso}, and easier than what we have already seen in Section \ref{sec:char-calc} so the proof is deferred to Appendix \ref{sec:lss-expect}.

\bel \label{lem:lss-expect}
Let $f$ be a regular test function with data $(\|f''\|_1^{-1}, C', c')$.  Suppose that $\|f''\|_1 \leq N^{1-\delta}$, for some $\delta >0$.  We have
\begin{align}
\ee[ \tr f(W) ] - N\int f(E) \rho (E) \d E &= \frac{1}{\pi} \int_{\Oma} ( \bar{\partial}_z \tilf (z) )  \nonumber \\
&\times \bigg\{ - \tr (S \mb \mb' ) + \tr ( (1 - S \mb^2 )^{-1} S \mb'  \mb ) + \frac{1}{N^2} \sum_{j, a} \sfo_{aj} \mb_j' \mb_j \mb_a^2 \bigg\} \nonumber\\
&+ N^{\eps} \O ( N^{\mfa-1} \|f''\|_1 +N^{-1/2} (1 + \|f''\|_1 )^{1/2} ).  \label{eqn:expect-expand}
\end{align}
Moreover,
\beq
\left| \ee[ \tr f (W) ] - N\int f(E) \rho (E) \d E \right| \leq C.
\eeq
\eel

\section{DBM of mesoscopic spectral statistics} \label{sec:meso}

In this section we consider the evolution of mesoscopic linear statistics under Dyson Brownian motion (DBM).   For some fixed initial distribution of eigenvalues $\{ \lambda_i (0) \}_{i=1}^N$, Dyson Brownian motion is the following system of stochastic differential equations:
\beq
\d \lambda_i (t) = \sqrt{\frac{2}{N}} \d B_i + \frac{1}{N} \sum_{j \neq i } \frac{1}{ \lambda_i (t) - \lambda_j (t) } \d t .
\eeq
In our applications we will always assume that the initial data are the eigenvalues of a matrix of general Wigner-type that satisfies Assumptions \ref{it:ass-1} and \ref{it:ass-2}.
 We denote the spectral measure associated with $W$ through its variance matrix $S$ by $\rho(E)$, and denote the free-convolution of $\rho$ with the semicircle distribution at time $t$ by $\rho_t$. From Lemma \ref{lem:free-conv-aa1}, $\rho_t$ in fact comes from a variance matrix $S_t$ that satisfies Assumptions \ref{it:ass-1} and \ref{it:ass-2} as well, and moreover that the edges of the supports of all of the $\rho_t$ are within $\O(t)$ of each other.  

 The $N$-quantiles of $\rho_t$ are denoted by $\gamma_i (t)$.  In this section we fix a $\mfc >0$ and interval $I_\mfc$,
\beq
I_\mfc := [ \alpha + \mfc, \beta - \mfc],
\eeq
as in \eqref{eqn:Ikap-def}. 
Let $f$ be a function satisfying
\beq
\|f\|_1 + \|f'\|_1 \leq C, \qquad \|f''\|_1 \leq C \eta_*^{-1}
\eeq
for some $\eta_* = N^{\delta_*} / N$.  Assume that $f$ is $0$ outside of $I_\mfc$.  We will consider the free convolution for times $0 \leq t \leq t_0 = N^{-\tau_0}$. The main idea of this section is that if $t_0 \gg \eta_*$, then a linear spectral statistic of the eigenvalues living on the scale $\eta_*$ and evolved to time $t = t_0$  in fact has the same distribution as the independent sum of  a Gaussian (with a universal variance) and a linear spectral statistic of the initial data on the larger scale $t_0$, up to negligible errors.

We have the following. To state we require  auxilliary exponents $\mfb, \delta >0$ whose roles are apparent in the proof. Choose $\delta < \delta_* / 100$ and $\mfb$ such that $\delta_* - \mfb > \delta/2$, and $\mfb > \delta/2$, and $\mfb < \delta_*/10$. 
\bet \label{thm:meso-dbm} Let $f$, $\delta>0$, and $\mfb >0$ be as above.  
With overwhelming probability,
\begin{align}
\frac{1}{N} \sum_i f ( \lambda_i (t_0) ) - \int f(x) \rho_{t_0} (x) \d x &= \frac{1}{N} \sum_i g ( \lambda_i (0) ) - \int g (x) \rho_0(x) \d x +\frac{1}{N}X \\
&+ \O \left( N^{-\mfb+\delta/2-1} + \frac{N^{\delta+\mfb} }{N^2 \eta_*} + \frac{N^{\delta} t_0}{N}+ \frac{ N^{\delta- 2 \mfb} \eta_*}{N t_0} \right),
\end{align}
where
\beq
g(E) := \frac{1}{\pi} \int f(x) \Im \frac{1}{ E -x- t_0 m_{t_0} (x + \i 0)} \d x ,
\eeq
and $X$ is a centered, real-valued Gaussian random variable.  If the function $f$ is a half-regular bump function with data $(\eta_*, M, E_0, E_1, C', c)$ the variance of $X$ is 
\beq \label{eqn:var-X}
\Var(X) = \frac{1}{ \pi^2} |\log (t_0 / \eta^*) | + \O ( \log \log (N) ).
\eeq
\eet
The proof is given in Section \ref{sec:dbm-proofs}.

Below we develop some properties of the function $g$ that appears in Theorem \ref{thm:meso-dbm}.  In particular, if $f$ is a half-regular bump function on the scale $\eta_*$, then up to negligible errors, the function $g$ is also a half-regular bump function, but on the larger scale $t_0$.

\bel   \label{lem:meso-functs}
Let $f$ be a half-regular bump function with data $(\eta_*, M, E_0, E_1, C', c')$, as defined in Definition~\ref{def:half-f}.  Let the function $g$  be as above.  Fix $M' = N^{q}$ with $0 < q < \tau_0$.  Then, on the event that $\tau_\delta >t_0$,
\begin{align} 
\frac{1}{N} \sum_i g ( \lambda_i (0) ) - \int g(x) \rho_0 (x) \d x = \frac{1}{N} \sum_i\hat{h} ( \lambda_i (0) ) - \int \hat{h} (x)  \rho_0 (x) \d x + N^{-1+\delta} \log(N) \O ( 1/M' +t_0) ,
\end{align}
where the function $\hat{h}$ is as follows.  It is smooth and compactly supported,  and  denoting $\hatE_0 := E_0 + t_0 \Re[ \mto (E_0)]$, we have
\beq
| \hat{h}'(x) | \leq \frac{C t_0}{ (x-\hatE_0)^2 + t_0^2}, \qquad |\hat{h}''(x) | \leq \frac{1}{ t_0} \frac{C t_0}{ (x-\hatE_0)^2 + t_0^2}.
\eeq
We also have $\hat{h}'(x) = 0$ for $|x-\hatE_0 | > t_0 M'$, unless $|x-E_1| < 3c'/4$, in which case $|\hat{h}'(x)|  + |\hat{h}''(x) | \leq C$.  Moreover, $\hat{h}(E_1 - 3c'/4) =1$.
\eel
 The proof is given in Section \ref{sec:meso-functs-proof}.

\section{Single eigenvalue fluctuations for Gaussian divisible ensembles} \label{sec:rest}

In this section we prove our main result Theorem \ref{thm:main}.  Before doing so, we provide in each of the following subsections proofs of the remaining results in Section \ref{sec:technical} that are used in the proof of Theorem \ref{thm:main}.

\subsection{Proof of Theorem \ref{thm:final-clt-homog}} \label{sec:final-clt-homog}

Let $p(x)$ and $E_0$ be as in the theorem statement.  As in the theorem statement, assume $|\lambda| \leq \log(N)^{1/4}$.  Fix a small constant $c_1 >0$ satisfying $c_1 < r/1000$, where $r$ is as in the statement of Theorem \ref{thm:clt-homog}.   Let $\chi(x)$ be a smooth bump function satisfying $\chi(x) = 0$ for $|x-E_0 | > c_1$ and $\chi(x) =1 $ for $|x-E_0| < 9 c_1/10$.  

We write
\beq
p(x) = \chi(x) p(x) + (1- \chi(x) ) p(x) =: p_1(x) + p_2(x).
\eeq
The function $p_2(x)$ is regular with data $(c, c, C)$ for some constants $c, C$, so by Lemmas \ref{lem:var-smooth} and \ref{lem:lss-expect},
\beq
\ee\left[ \left( \tr p_2 (W_{t_0} ) - N \int p_2 (x) \hat{\rho}_{t_0} (x) \right)^2 \right] \leq C,
\eeq
where $\hat{\rho}_{t_0}$ denotes the spectral measure associated with the variance matrix of $W_{t_0}$; this is $S+ t_0 e e^T +t_0/N$, where $e$ is the vector whose entries are all $1 / \sqrt{N}$. By Lemma \ref{lem:diag-S} we have,
\beq
N \left|  \int p_2 (x) \hat{\rho}_{t_0} (x)  \d x - \int p_2(x) \rho_{t_0} (x) \d x \right| \leq  C t_0.
\eeq
Therefore
\begin{align}
& \ee \left[ \exp\left( \i  ( \log(N) )^{-1/2} \lambda \left( \tr p ( W_{t_0}) - N \int p(x) \rho_{t_0} (x) \d x \right) \right) \right] \nonumber \\
= &\ee \left[ \exp\left( \i  ( \log(N) )^{-1/2} \lambda \left( \tr p_1 ( W_{t_0}) - N \int p_1(x) \rho_{t_0} (x) \d x \right) \right) \right] + \O ( | \lambda| / \sqrt{ \log(N) } ).
\end{align}
The function $p_1(x)$ is half-regular with data $(t_1, N^{\delta_M}, E_0, E_0+3c_1/4, c_1/10, C)$ for some $C>0$.  
Conditionally on $W$, the distribution of the eigenvalues of $W + \sqrt{t_0} G$ has the same distribution as the process $\{ \lambda_i (t_0) \}_i$ of Theorem \ref{thm:meso-dbm}.  Therefore, with $X$ denoting the centered Gaussian random variable of the statement of Theorem \ref{thm:meso-dbm}, we have,
\begin{align}
&\ee\left[ \exp\left( \i  ( \log(N) )^{-1/2} \lambda \left( \tr p_1 ( W_{t_0}) - N \int p_1(x) \rho_{t_0} (x) \d x \right) \right) \right] \nonumber \\
= & \ee\left[ \exp \left( \i ( \log(N ))^{-1/2} \lambda \left( \tr g (W) - N \int g (x) \rho (x) \d x \right) \right) \right] \ee\left[\exp[ \i \log(N)^{-1/2} \lambda X] \right] + \O (|\lambda| )N^{-c} )
\end{align}
for some $c >0$.  Here $g$ is as in the statement of Theorem \ref{thm:meso-dbm}.  The variance of the Gaussian is given to leading order in \eqref{eqn:var-X}, so
\beq\label{combine1}
\ee\left[ \exp[ \i \log(N)^{-1/2} \lambda X ] \right] = \exp\left[ - \frac{ \lambda^2}{2 }\frac{|\log (t_1 /t_0) |}{ \pi^2 \log(N)} \right] + \O \left( ( \log (N) )^{-1/4} \right).
\eeq
Let $M' =N^{q}$, with $q < \tau_0 /100$, and $h$ as in Lemma \ref{lem:meso-functs}.  Then,
\begin{align}
 &\ee \left[ \exp \left( \i ( \log(N ))^{-1/2} \lambda \left( \tr g (W) - N \int g (x) \rho (x) \d x \right) \right) \right] \nonumber \\
= & \ee \left[ \exp \left( \i ( \log(N ))^{-1/2} \lambda \left( \tr h (W) - N \int h (x) \rho (x) \d x \right) \right) \right] + \O (N^{-q/2} ).
\end{align}
From the estimates of Lemma \ref{lem:meso-functs}, it follows that $h$ is a half-regular bump function with data $(t_0, M', \hatE_0, E_1, c_1/2, C')$ for some $C'>0$.  By our choice of $c_1$, it follows that $h$ satisfies the assumptions of Theorem \ref{thm:clt-homog}, and therefore 
by Theorem \ref{thm:clt-homog} and Lemma \ref{lem:lss-expect}, we have
\begin{align}\label{combine2}
\ee \left[ \exp \left( \i ( \log(N ))^{-1/2} \lambda \left( \tr h (W) - N \int h (x) \rho (x) \d x \right) \right) \right] =& \exp\left[ -\frac{ \lambda^2}{2} \frac{ |\log (t_0 ) | }{\pi^2 \log(N)} \right] \nonumber \\
+& \O( ( \log(N) )^{-1/4} ).
\end{align}
Combining \eqref{combine1} and \eqref{combine2} completes the proof of Theorem \ref{thm:final-clt-homog}. \qed

\subsection{Proof of Theorem \ref{thm:gde-univ}}  \label{sec:gde-univ}

From Theorem \ref{thm:homog} and Fourier inversion, we have
\begin{align} \label{eqn:gde-b1}
 \ee \left[ F \left( \frac{ N \rho_{t_0+t_1} ( \gamma_{i_0, t_0+t_1} )}{ \sqrt{ \log(N) }} ( \lambda_{i_0} (W_{t_0+t_1} ) - \gamma_{i_0, t_0+t_1}  )\right) \right] =  \int \hat{F} ( \lambda) \ee[ \e^{ \i \lambda X_y } ] \ee[ \e^{ \i  \lambda Z_1 } ] + \O (N^{-c} ).
\end{align}
Here
\beq
\sqrt{ \log(N) } X_y :=\rhosc (0) N y_{N/2} (t_1) + \sum_j  f ( \rhosc (0) y_j (0)) - \int f (\rhosc(0) s) \rhosc (s) \d s ,
\eeq
where $\{ y_j \}_j$ is the process from the  statement of Theorem \ref{thm:homog}, and
\beq
\sqrt{ \log(N) } Z_1 := \sum_j f ( \rho_{t_0} ( \gamma_{i_0, t_0 } ) ( \lambda_j ( W_{t_0} ) - \gamma_{i_0, t_0 } ) - \int f ( \rho_{t_0} ( \gamma_{i_0, t_0 } ) (s - \gamma_{i_0, t_0 } ) ) \d \rho_{t_0} (s) \d s.
\eeq
 In \eqref{eqn:gde-b1}, we also used the fact that, as a consequence of Proposition \ref{prop:stab-1},
\beq
 \rho_{t_0+t_1} (\gamma_{i_0, t_0+t_1} ) = \rho_{t_0} (\gamma_{i_0, t_0+t_1} ) + \O (\sqrt{t_1} ).
\eeq
Applying the above argument, but to the process $\{ z_i (t) \}_{i=1}^N$ of the statement of Theorem \ref{thm:homog} instead, gives,
\begin{align}
\ee\left[ F \left( \frac{ N \rhosc(0) }{ \sqrt{ \log(N) } }  \lambda_{N/2} (\sqrt{1+a^2 t_1 } G ) \right)\right] =  \int \hat{F} ( \lambda) \ee[ \e^{ \i \lambda X_y } ] \ee[ \e^{ \i  \lambda Z_2 } ] + \O (N^{-c} )
\end{align}
where $a$ is as in the theorem statement, $X_y$ is the same as above, and
\beq
\sqrt{ \log(N) }Z_2 = \sum_j  f ( \rhosc (0) \lambda_j (G)) - \int f (\rhosc(0) s) \rhosc (s) \d s .
\eeq
Note that by rigidity,
\beq\label{wehave2}
\ee\left[ F \left( \frac{ N \rhosc(0) }{ \sqrt{ \log(N) } }  \lambda_{N/2} (\sqrt{1+a^2 t_1 } G ) \right)\right] = \ee\left[ F \left( \frac{ N \rhosc(0) }{ \sqrt{ \log(N) } }  \lambda_{N/2} ( G ) \right)\right] + \O( N^{\eps} t_1 )
\eeq
for any $\eps >0$.  
By Theorem \ref{thm:final-clt-homog}, we have for $|\lambda| \leq \log(N)^{1/4}$ that
\begin{align}\label{wehave3}
\left| \ee[ \e^{ \i \lambda Z_1 } ] - \ee[ \e^{ \i \lambda Z_2 } ] \right| \leq  C\log(N)^{-1/4}.
\end{align}
Since $\hat{F}$ is Schwartz, we can restrict the integration in $\lambda$ in the above formulas to, e.g., $|\lambda| \leq (\log(N))^{1/10}$ at an error of $\O ( ( \log(N) )^{-C} )$ for any $C>0$.  Combining \eqref{eqn:gde-b1} and \eqref{wehave2} and using \eqref{wehave3} completes the proof of Theorem \ref{thm:gde-univ}.  \qed

\section{DBM proofs} \label{sec:dbm-proofs}

In this section we will prove the various results of Section \ref{sec:meso}.  We first record a few preliminary results and observations.

It is proven in Appendix \ref{a:free-2} that $\rho_t$ in fact comes from a variance matrix $S_t$ that satisfies Assumptions \ref{it:ass-1} and \ref{it:ass-2} as well, and moreover that the edges of the supports of all of the $\rho_t$ are within $\O(t)$ of each other.

The Stieltjes transform $m_t(z)$ of the free convolution $\rho_t (E)$ satisfies the complex Burgers' equation  \cite{pastur,rogershi1993}, 
\beq \label{eqn:burg}
\del_t m_t = m_t \del_z m_t.
\eeq
It has an implicit solution
\beq
m_t (z) = m_0 (z + t m_t (z) ),
\eeq
which is uniformly bounded.

The complex Burgers' equation \eqref{eqn:burg} has characteristics $z_t$ that solve
\beq
\del_t z_t = - m_t (z_t).
\eeq
That is,
\beq
\del_t (m_t (z_t)) = 0.
\eeq
The characteristic $z_t$ with final condition $z$ has the explicit formula
\beq \label{eqn:char-final}
z_t = z + (t_0 - t) m_{t_0} (z),
\eeq
and from this we see that any characteristic that ends with real part in $I_\mfc$ lies within, say, $I_{\mfc/2}$ (recall the definition \eqref{eqn:Ikap-def}) for all times $0 \leq t \leq t_0$.  The following lemma is clear from the above discussion.

\bel
Consider the map $z \to z_t(z) := z + (t_0 - t) m_{t_0} (z)$.  For $z$ with $\Re[z] \in I_\mfc$, we have
\beq
\del_{z} z_t (z) = 1 + \O (t_0 ).
\eeq
Also,
\beq
| \del^k z_t (z) | \leq C_k (\delta_{k1} + t_0 ), \qquad | \del_z^k m_t (z) | \leq C_k
\eeq
for any $k$ and $\Re[z] \in I_{\mfc/2}$.   Further, for $\Re[z] \in I_{\mfc/2}$ and $| \Im[z]| \leq 10$, we have
\beq
c \leq | \Im[ m_t (z) ] \leq C ,
\eeq
as well as 
\beq  \label{eqn:meso-dbm-t1}
\Im [z_t (z) ] \asymp \Im[z] + (t_0 - t) .
\eeq
\eel

We denote 
\beq
m_{N, t} (z) := \frac{1}{N} \sum_{i=1}^N \frac{1}{ \lambda_i (t) - z } .
\eeq

We will also repeatedly use the following, which is a simple consequence of the Cauchy integral formula.
\bel
On the event $\tau_\delta >t_0$, we have for every $k\geq 0$ that
\beq \label{eqn:mN-der-bd}
| m_{N, t}^{(k)} (z) - m^{(k)}_t (z) | \leq C_k \frac{N^{\delta/100}}{N |\Im[z]|^{k+1} }
\eeq
for $| \Im[z ] | \geq  N^{\delta/50-1}$.
\eel

We require a stopping time, which we now define.  Fix $\delta >0$ satisfying (this is the same $\delta >0$ that appears in the statement of Theorem \ref{thm:meso-dbm}),
\beq
0 < \delta < \delta_* /100,
\eeq
and let
\begin{align}
\tau_{\delta} := \inf_t \bigg\{  & t: \exists i : | \lambda_i (t) - \gamma_i (t) | > N^{\delta/100} \frac{1}{ N^{2/3} \min\{ i^{1/3}, (N+1 - i)^{1/3} \} } \\
\mbox{ or } &\exists z : \Im[z] \geq N^{\delta/100-1} : |m_{N, t}(z)  - m_t (z) | > N^{\delta/100} (N \eta)^{-1} \bigg\}.
\end{align}
From Appendix \ref{a:free-2}, the following lemma is clear.
\bel
With overwhelming probability, $\tau_{\delta} > t_0$. 
\eel

By the Helffer--Sj{\"o}strand formula we have on the event $\tau_\delta > t_0$ that (see (4.12) of \cite{meso}), 
\begin{align} \label{eqn:dbm-hs}
\sum_i f ( \lambda_i (t_0 ) ) - N \int f (E) \rho_{t_0} (E) \d E &= \frac{1}{\pi} \int_{|y| > N^{-\mfb} \eta_*} (\bar{\del}_z \tilf (z) ) N(m_{N, t } - m_t ) \d z + \O (N^{-\mfb+\delta/10}),
\end{align}
where we choose $\mfb$ such that $\delta_* - \mfb > \delta/2$, and $\mfb > \delta/2$, and $\mfb < \delta_*/10$. This is the same $\mfb >0$ as in the statement of Theorem \ref{thm:meso-dbm}.  Here,
\beq
\tilf(z) =( f(x) + \i y f'(x) ) \chi(y), \qquad \bar{\del}_z \tilf (z) = \frac{1}{2} \left( \i y \chi(y) f''(x) + \i (f(x) + \i f'(x) y ) \chi'(y) \right)
\eeq
and $\chi(y)$ is an even cut-off function such that $\chi(y) =1$ for $|y| \leq 1/2$ and $\chi(y) = 0$ for $|y|>1$.

\subsection{Proof of Theorem \ref{thm:meso-dbm}}

First, by the It{\^{o}} lemma,
\beq\label{eqn: DBM-mN}
\d m_{N, t} (z) = m_{N, t} (z) \del_z m_{N, t} (z) + \frac{1}{N^2}  \sum_{i} \frac{1}{ ( \lambda_i - z)^3}  -\sqrt{2} \frac{1}{ N^{3/2}} \sum_i \frac{ \d B_i}{ ( \lambda_i - z)^2}.
\eeq
Putting in the characteristic $z_t$  and using $\del_t z_t = - m_t (z_t)$, we get
\begin{align}\label{eqn: DBM-adjusted}
\d m_{N, t} ( z_t ) =& (m_{N, t} (z_t ) - m_t (z_t ) ) \del_z ( m_{N, t} (z_t) - m_t (z_t ) ) + (m_{N, t} (z_t ) - m_t (z_t) ) \del_z (m_t (z_t )) \nonumber \\
&+  \frac{1}{N^2}  \sum_{i} \frac{1}{ ( \lambda_i - z_t)^3}  - \sqrt{2 } \frac{1}{ N^{3/2}} \sum_i \frac{ \d B_i}{ ( \lambda_i - z_t)^2}.
\end{align}
We will now use $z_t(z)$ to denote the characteristic with final condition $z_{t_0} (z) = z$ as in \eqref{eqn:char-final}. When the context is clear we omit the argument $z$ and write $z_t = z_t (z)$.

Using \eqref{eqn: DBM-adjusted} in \eqref{eqn:dbm-hs} (recall $\del_t m_t (z_t) = 0$), we obtain
\begin{align}
& \int_{|y| > N^{-\mfb} \eta_*} \del_{\bar{z}}f(z) \left( m_{N, t_0} (z) - m_{t_0} (z) \right) \d x \d y 
 -  \int_{|y| > N^{-\mfb} \eta_*} \del_{\bar{z}}f(z) \left( m_{N, 0} (z_0) - m_{0} (z_0) \right) \d x \d y  \label{eqn:err6} \\
 = & \int_{0}^{ t_0} \int_{|y| > N^{-\mfb } \eta_*} ( \del_{\bar{z}} f ) (m_{N, t} (z_t) - m_t (z_t ) )  ( m_{N, t}' (z_t ) - m_t' (z_t ) ) \,\d x \d y \d t \label{eqn:err1} \\
 + & \int_{0}^{ t_0} \int_{|y| > N^{-\mfb } \eta_*} ( \del_{\bar{z}} f ) (m_{N, t} (z_t ) - m_t (z_t ) )  m_t' (z_t)  \label{eqn:err2}\, \d x \d y \d t \\
 + & \frac{1}{N^2}    \sum_{i} \int_{0}^{ t_0} \int_{|y| > N^{-\mfb } \eta_*} ( \del_{\bar{z}} f )  \frac{1}{ ( \lambda_i - z_t)^3}  \, \d x \d y \d t \label{eqn:err3} \\
 - &   \sqrt{ 2} \frac{1}{ N^{3/2}} \sum_i \int_{|y| > N^{-\mfb } \eta_*}\int_{0}^{ t_0} ( \del_{\bar{z}} f ) \frac{ \d B_i}{ ( \gamma_i(t) - z_t)^2} \, \d x \d y \label{eqn:err4} \\
 +&   \sqrt{ 2} \frac{1}{ N^{3/2}} \sum_i \int_{|y| > N^{-\mfb } \eta_*}\int_{0}^{ t_0} ( \del_{\bar{z}}f ) \left( \frac{ 1} { ( \gamma_i(t) - z_t)^2} - \frac{1}{ ( \lambda_i (t) - z_t )^2} \right) \d B_i\, \d x \d y \label{eqn:err5}. 
\end{align}
The second term of \eqref{eqn:err6} will be seen to contribute to the function $g$ in the theorem statement.  The term \eqref{eqn:err4} is the Gaussian variable $N^{-1} \pi X$ whose variance is calculated in Lemmas \ref{lem:gaussian-var-calc-1} and \ref{lem:gaussian-var-calc-2} below.  Since $\chi(y)$ is even, $X$ is real.  The remaining terms will be shown to be error terms.   We start with the martingale term \eqref{eqn:err5}, which we rewrite as
\begin{align}
 &\frac{1}{ N^{3/2}} \sum_i  \int_{0}^{ t_0} \int_{|y| > N^{-\mfb } \eta_*} ( \del_{z}f )\left( \frac{ 1} { ( \gamma_i(t) - z_t)^2} - \frac{1}{ ( \lambda_i (t) - z_t )^2} \right) \d B_i \d x \d y \\
 = & \sum_{i=1}^N \frac{1}{ N^{3/2}}  \int_{0}^{ t_0} \int_{|y| > N^{-\mfb } \eta_*} ( \i \chi' (y) (f (x) + \i y f' (x) )) \left( \frac{ 1} { ( \gamma_i(t) - z_t)^2} - \frac{1}{ ( \lambda_i (t) - z_t )^2} \right) \, \d x \d y \d B_i  \label{eqn: quad-1} \\
 + & \sum_{i=1}^N \frac{1}{ N^{3/2}} \int_{0}^{  t_0} \int_{|y| > N^{-\mfb } \eta_*} \i y \chi (y) f'' (x)   \left( \frac{ 1} { ( \gamma_i(t) - z_t)^2} - \frac{1}{ ( \lambda_i (t) - z_t )^2} \right) \, \d x \d y \d B_i. \label{eqn:err5b}
\end{align}
We define martingales $M_1 $ and $M_2$ in the obvious way so that on the event $\tau_\delta > t_0$, the terms \eqref{eqn: quad-1} and \eqref{eqn:err5b} are $M_1 (t_0 \wedge \tau_\delta)$ and $M_2 ( t_0 \wedge \tau_\delta)$, respectively.  Due to the Burkholder--Davis--Gundy inequality, it suffices to bound the associated quadratic variation processes (we apply the inequality to $M_i ( t_0 \wedge \tau_\delta)$). We start with \eqref{eqn: quad-1}.  Up to a constant factor, the quadratic variation is bounded above by
\begin{align}\label{eqn: quad-int}
\frac{1}{N^3} \int_0^{t_0 \wedge \tau_\delta} \int & |\chi' (y_1) \chi' (y_2 ) | | f(x_1) + \i y f' (x_1) | | f(x_2) + \i y f' (x_2) | \nonumber \\
& \times \sum_{i} \left| \frac{1}{ ( \gamma_i (t) - z_t)^2 } - \frac{1}{ (\lambda_i (t) - z_t )^2 } \right| \left| \frac{1}{ ( \gamma_i (t) - w_t)^2 } - \frac{1}{ (\lambda_i (t) - w_t )^2 } \right| \,\d x_1 \d x_2 \d y_1 \d y_2 \d t,
\end{align}
where we defined $z = x_1 + \i y_1$ and $w=x_2 + \i y_2$ and $z_t$ and $w_t$ in the obvious manner.  Since $\chi' (y) \neq 0$ only for $|y| > 1/2$, we have
\begin{align}
\frac{1}{N^3} \sum_{i} \left| \frac{1}{ ( \gamma_i (t) - z_t)^2 } - \frac{1}{ (\lambda_i (t) - z_t )^2 } \right| \left| \frac{1}{ ( \gamma_i (t) - w_t)^2 } - \frac{1}{ (\lambda_i (t) - w_t )^2 } \right| \leq \frac{1}{N^3} \sum_i \frac{N^{\delta}}{N^{4/3} i^{2/3} } \leq C \frac{N^{\delta}}{N^4},
\end{align}
using the rigidity estimates that hold before $\tau_\delta$. 
For the term \eqref{eqn:err5b}, since $f(x)$ is compactly supported, we first integrate by parts in $x$ to obtain
\beq
 M_2 (t_0 \wedge \tau_\delta  ) = -\frac{1}{N^{3/2}} \sum_i \int_0^{t_0\wedge \tau_\delta} \int_{ |y| > N^{-\mfb} \eta_*} \i y \chi (y) f' (x) \left( \frac{1}{ ( \gamma_i (t) - z_t )^3 }- \frac{1}{ ( \lambda_i (t) - z_t )^3 } \right)2 \del_x z_t \,\mathrm{d}x\mathrm{d}y\mathrm{d}B_i.
\eeq
Since $| \del_x z_t | \leq C$, we see that the quadratic variation is bounded up to a constant factor by
\begin{align}
& \frac{1}{N^3} \int_0^{t_0} \int_{N^{-\mfb} \eta_*< |y_1|, |y_2| < 1} | f'(x_1) f'(x_2 )y_1 y_2 | \nonumber \\
& \times \sum_{i=1}^N   \left| \frac{1}{ ( \gamma_i (t) - z_t)^3 } - \frac{1}{ (\lambda_i (t) - z_t )^3 } \right| \left| \frac{1}{ ( \gamma_i (t) - w_t)^3 } - \frac{1}{ (\lambda_i (t) - w_t )^3 } \right| \,\mathrm{d}t.
\end{align}
For $x_1, x_2$ in the support of $f'$, we have
\begin{align}
 \frac{1}{N^3} & \sum_i \left| \frac{1}{ ( \gamma_i (t) - z_t)^3 } - \frac{1}{ (\lambda_i (t) - z_t )^3 } \right| \left| \frac{1}{ ( \gamma_i (t) - w_t)^3 } - \frac{1}{ (\lambda_i (t) - w_t )^3 } \right| \nonumber \\
\leq & \frac{N^{\delta/2}}{N^5} \sum_i \frac{1}{ | \gamma_i(t) - z_t |^4 |\gamma_i (t) - w_t|^4} + \frac{N^{\delta/2}}{N^3} \sum_i \frac{1}{N^{4/3} i^{2/3}}.
\end{align}
Now, for $x \in I_{\mfc/2}$ and $|y| \geq N^{-1}$, we have
\beq \label{eqn:dbm-a1}
\frac{1}{N} \sum_{i} \frac{1}{ | \gamma_i (t) - (x+ \i y ) |^2} \leq \frac{C}{ |y|} | \Im[ m_t (x+ \i y) ] | \leq \frac{C}{ |y|},
\eeq
and so for $x_1, x_2$ in the support of $f'(x)$,
\begin{align}
&\int_{0}^{t_0} \int_{ N^{-\mfb} \eta_*< |y_1|, |y_2| < 1}  | f'(x_1) f'(x_2) y_1 y_2 |   \frac{N^{\delta/2}}{N^5} \sum_i \frac{1}{ | \gamma_i(t) - z_t |^4 |\gamma_i (t) - w_t|^4}  \d x_1 \d x_2 \d y_1 \d y_2 \d t \nonumber \\ 
& \leq C \int_{0}^{t_0} \int_{N^{-\mfb} \eta_*< |y_1|, |y_2| < 1}   |y_1 y_2 f'(x_1) f'(x_2) | \frac{N^{\delta}}{N^4} \frac{1}{ | \Im [z_t] |^4 | \Im[w_t]|^3}  \d x_1 \d x_2 \d y_1 \d y_2 \d t \nonumber  \\
&\leq C \int_{0}^{t_0} \int_{ N^{-\mfb} \eta_*< y_1, y_2 < 1} |f'(x_1) f'(x_2) | \frac{N^{\delta}}{N^4} \frac{1}{ | y_1 + (t_0-t) |^3 |y_2 + (t_0 - t) |^2}   \d x_1 \d x_2 \d y_1 \d y_2 \d t \nonumber\\ 
&\leq C \int_{0}^{t_0} \int_{ N^{-\mfb} \eta_*< y_1, y_2 < 1} \frac{N^{\delta}}{N^4} \frac{1}{ | y_1 + (t_0-t) |^3 |y_2 + (t_0 - t) |^2}   \d y_1 \d y_2 \d t \nonumber \\
& \leq \frac{N^{ \delta+ 2\mfb}}{ N^2 (N \eta_* )^2}.
\end{align}
Note that we used \eqref{eqn:meso-dbm-t1}, as well as a similar estimate for $\Im[w_t]$. 
From the above calculations, we see that the term \eqref{eqn:err5} is
\beq
\O \left( \frac{N^{\delta+ \mfb}  }{ N (N \eta_* ) } \right)
\eeq
with overwhelming probability on the event $\tau_\delta > t_0$.

We now turn to the remaining terms \eqref{eqn:err1}, \eqref{eqn:err2}, and \eqref{eqn:err3}.   Recall the form of the derivative $\bar{\partial}_z \tilf(z)$,
\beq
\bar{\partial}_z \tilf (z) = \i f''(x) y\chi (y) + ( \i f'(x) y + \i f(x) ) \chi'(y).
\eeq
We treat the contribution of the $\chi'(y)$ term in \eqref{eqn:err1}, \eqref{eqn:err2}, and \eqref{eqn:err3} first.  Note that for $y \geq 1/2$, we have from \eqref{eqn:mN-der-bd} that
\beq
(m_{N, t} (z_t) - m_t (z_t ) )  ( m_N' (z_t ) - m_t' (z_t ) ) = \O (N^{\delta-2} )
\eeq
on the event $\tau_\delta > t_0$, and so the $\chi'(y)$ component of \eqref{eqn:err1} contributes $\O (t_0 N^{\delta-2})$.  Since the term $m' (z_t)$ is bounded, we similarly see that the $\chi'(y)$ component of \eqref{eqn:err2} contributes $\O (t_0 N^{\delta-1})$.  A similar estimate clearly holds for the $\chi'(y)$ contribution of \eqref{eqn:err3}.

We now turn to the $f''(x) y \chi(y)$ contributions to  \eqref{eqn:err1}, \eqref{eqn:err2}, and \eqref{eqn:err3}.  They require an integration by parts in the $x$ variable due to the fact that $\|f''\|_1$ is large.  Beginning with \eqref{eqn:err2} we have,
\begin{align}
& \left| \int_{0}^{t_0} \int_{  N^{-\mfb} \eta_*\leq |y| \leq 1} f''(x) y \chi(y) (m_{N, t} (z_t) - m_t (z_t ) ) m' (z_t) \d t \d x \d y\right| \nonumber \\
\leq & C \int_0^{t_0} \int_{  N^{-\mfb} \eta_*\leq |y| \leq 1} |f' (x) y | | \del_z [( m_{N, t} (z_t) - m_t (z_t ) ) m_t' (z_t ) ] | \d x \d y \d t \nonumber \\
\leq & C\int_0^{t_0} \int_{  N^{-\mfb} \eta_*\leq y \leq 1}  | f'(x) y | \frac{N^{\delta/2}}{N ( y+ (t_0 - t ) )^2} \d x \d y \d t \nonumber \\
\leq & C \frac{ N^{\delta} t_0 }{N}. 
\end{align}
In passing from the first to second line we used integration by parts, that $f'(x)$ is compactly supported, and that $\partial_x \varphi (z) = \del_z \varphi (z)$ for analytic $\varphi$.  The second inequality uses \eqref{eqn:mN-der-bd} and  \eqref{eqn:meso-dbm-t1}.   The same argument using the fact that (again from \eqref{eqn:mN-der-bd})
\beq
\del_z [ ( m_{N, t} (z_t ) - m_t (z_t) )  ( m'_{N, t} (z_t ) - m_t' (z_t) )] = \O ( N^{\delta/2-2} ( \Im[z_t] )^{-4} )
\eeq
gives an estimate of $\O ( N^{\delta+\mfb} (N^2 \eta_* )^{-1} )$ for the $ y \chi (y) f''(x)$ contribution of \eqref{eqn:err1}.  For the $y \chi (y)f''(x)$ component of \eqref{eqn:err3} we first observe that
\beq
\frac{1}{N}\sum_i \frac{1}{(\lambda_i-z_t)^3}= \frac{1}{2} m''_{N, t} (z_t) .
\eeq
We write 
\beq
N^{-1} m''_{N, t} (z_t) =N^{-1}  m''_t (z_t )  + N^{-1}(m_{N, t}- m_t)''(z_t).
\eeq
Since $m_t (z)$ has bounded derivatives, we have using integration by parts in $x$, as above,
\begin{align}
\frac{1}{N} \left| \int_0^{t_0} \int_{ |y| >N^{-\mfb} \eta_* } f''(x) y \chi (y)  m_t''(z_t) \d t \d x \d y \right| \leq  C t_0  N^{-1}.
\end{align}
  The remaining part gives
\beq
\frac{1}{N} \left| \int_0^{t_0} \int_{ |y| >N^{-\mfb} \eta_* } f''(x) y \chi (y) (m_{N, t}'' (z_t) - m''(z_t )) \d t \d x \d y \right|  \leq C \frac{N^{\delta+\mfb}}{N^2 \eta_* } 
\eeq
after again integrating by parts and using $\partial_z (m_{N, t}'' (z_t) - m''_t (z_t ) ) = \O (N^{\delta} N^{-1} \Im[z_t]^{-4} )$, which holds by \eqref{eqn:mN-der-bd}.  The variance of the Gaussian is calculated in Lemmas \ref{lem:gaussian-var-calc-1} and \ref{lem:gaussian-var-calc-2} below.  Furthermore, it is clear that
\begin{align}
\int_{ |y|< N^{-\mfb} \eta_* } \bar{\del}_z \tilf (z) ( m_{N, 0} (z_0) - m_0 (z_0) ) \d x \d y = \O \left( N^{\delta- 2 \mfb} \eta_*(N t_0)^{-1} \right).
\end{align}
We have
\begin{align}
\frac{1}{ \pi} \int ( \bar{\del}_z \tilf (z) ) ( m_{N, 0} (z_0) - m_0 (z_0) ) = \frac{1}{N} \sum_i g ( \lambda_i (0) ) - \int g (x) \rho_0(x) \d x.
\end{align}
This completes the proof. \qed

\subsection{Calculation of variance of $X$}

The variance of the Gaussian random variable $X$ is
\begin{align}
\mathrm{Var}(X) &= \frac{2}{ \pi^2} \frac{1}{N} \sum_i \int_{0}^{t_0} \int_{ \eta_* N^{-\mfb} < |y_1|, |y_2| } ( \bar{\del}_z \bar{\del}_w \tilf (z) \tilf (w) ) \frac{1}{ ( \gamma_i (t) - z_t)^2 ( \gamma_i (t) - w_t)^2 } \d t \d x_1 \d x_2 \d y_1 \d y_2.
\end{align}
In the case of half-regular bump functions, this quantity is calculated to leading order in the two following lemmas.

\bel \label{lem:gaussian-var-calc-1}
If $f$ is a half-regular bump function with data $(\eta_*, M, E_0, E_1, C', c')$ then,
\begin{align}
& \frac{2}{\pi^2} \frac{1}{N} \sum_i \int_{0}^{t_0} \int_{ |y_1|, |y_2| > N^{-\mfb } \eta_* } ( \bar{\del}_z \tilf (z) ) ( \bar{ \del}_w \tilf (w) ) \frac{1}{( \gamma_i (t) - z_t )^2 ( \gamma_i (t) - w_t )^2} \d x_1 \d x_2 \d y_1 \d y_2  \d t \nonumber \\
= &\frac{1}{2} \int_{0}^{t_0}\int f'(x_1) f'(x_2)  ( \pto (x_1) + \pto (x_2 ) ) \frac{ s 4\pto (E_0 ) }{ (x_1 - x_2)^2 + (s2 \pi \pto (E_0 ) )^2 } \d s \d x_1 \d x_2 \nonumber  \\
+ &\log(N)^2 \O\left( t_0 + M \eta_* + \|f''\|_1 \eta_* N^{-\mfb} + (N \eta_* )^{-1} N^{\mfb} \right).
\end{align}
\eel
\proof We first note that, since $|y| > 1/2$ if $\chi' (y) \neq 0$,
\begin{align}
\frac{1}{N} \sum_i \int_0^{t_0} \int_{ |y_1|, |y_2| > N^{-\mfb } \eta_* } & ( \chi'(y_1) \chi'(y_2) ) ( f'(x_1) y + \i f(x_1 ) )( f'(x_2) y + \i f (x_2 ) ) \notag\\
& \times \frac{1}{( \gamma_i (t) - z_t )^2 ( \gamma_i (t) - w_t )^2} \d x_1 \d x_2 \d y_1 \d y_2 \d t = \O (t_0  ).
\end{align}
   Integrating by parts in $x_2$,
\begin{align}
&\left| \frac{1}{N} \sum_i \int_0^{t_0} \int_{|y_2| > N^{-\mfb} \eta_* } ( \chi'(y_1) \chi (y_2) ) ( f' (x_1) y_1+ \i f(x_1) )(f''(x_2) y_2 )   \frac{1}{( \gamma_i (t) - z_t )^2 ( \gamma_i (t) - w_t )^2}  \right| \nonumber  \\
= & \left| \frac{1}{N} \sum_i \int_0^{t_0} \int_{|y_2| > N^{-\mfb} \eta_* } ( \chi'(y_1) \chi (y_2) ) ( f' (x_1) y_1+ \i f(x_1) )(f'(x_2) y_2 )  \del_w \frac{1}{( \gamma_i (t) - z_t )^2 ( \gamma_i (t) - w_t )^2}   \right| \nonumber \\
\leq  & C\int_{0}^{t_0} \int_{ 1 > y_2 > N^{-\mfb} \eta_* } \frac{y_2}{ (y_2 + (t_0 - t) )^2} \leq C t_0 \log(N),
\end{align}
where we used \eqref{eqn:dbm-a1} in the first inequality as well as \eqref{eqn:meso-dbm-t1}.  
For small $c_1= c'/10$, we have that the steep $\eta_*$-scale part of $f$ fits inside $|E -E_0 | < c_1/2$ and $c_1 < |E_0 -E_1 |/10$.  
Integrating by parts gives
\begin{align}
& \left| \frac{1}{N} \sum_i \int_0^{t_0} \int_{|y_1|, |y_2| > N^{-\mfb} \eta_*, |x_1 -E_0| > c_1 } ( \chi'(y_1) \chi (y_2) ) ( y_1 f''(x_1) )(f''(x_2) y_2 )   \frac{1}{( \gamma_i (t) - z_t )^2 ( \gamma_i (t) - w_t )^2}  \right| \nonumber \\
= & \left| \frac{1}{N} \sum_i \int_0^{t_0} \int_{|y_1|, |y_2| > N^{-\mfb} \eta_*, |x_1 -E_0| > c_1 } ( \chi'(y_1) \chi (y_2) ) ( y_1 f''(x_1) )(f'(x_2) y_2 ) \partial_{w}   \frac{1}{( \gamma_i (t) - z_t )^2 ( \gamma_i (t) - w_t )^2}  \right| \nonumber \\
\leq & C \int_{0}^{t_0} \int_{1 > y_2, y_1  > N^{-\mfb} \eta_* } \frac{ |y_1 y_2|}{ (y_1 + (t_0 -t))^2 (y_2 + (t_0 - t))^2} \leq C t_0 \log(N)^2.
\end{align}
Above, we used \eqref{eqn:dbm-a1}, \eqref{eqn:meso-dbm-t1} and that the $L^1$ norm of $f''(x) \1_{ \{ |x - E_0 | > c_1 \}}$ is $\O (1)$ by assumption.  We also arrive at a similar estimate for the integral over $|x_1 - E_0| , |x_2 - E_0 | > c_1$ (no integration by parts is required).

Summarizing the above we have,
\begin{align} \label{eqn:gauss-var-1}
&\frac{1}{N} \sum_i \int_{0}^{t_0} \int_{ |y_1|, |y_2| > N^{-\mfb } \eta_* } ( \bar{\del}_z \tilf (z) ) ( \bar{ \del}_w \tilf (w) ) \frac{1}{( \gamma_i (t) - z_t )^2 ( \gamma_i (t) - w_t )^2} \d x_1 \d y_1 \d x_2 \d y_2 \d t  \nonumber \\
= & \frac{-1}{4N}\sum_i \int_{0}^{t_0} \int_{ |x_1 - E_0|, |x_2 - E_0 | < c_1 }  \int_{ |y_1|, |y_2| > N^{-\mfb } \eta_* } f''(x_1) f''(x_2) y_1 y_2 \chi(y_1) \chi(y_2)   \frac{ \d x_1 \d y_1 \d x_2 \d y_2 \d t }{( \gamma_i (t) - z_t )^2 ( \gamma_i (t) - w_t )^2} \nonumber \\
+ & \O \left(t_0 \log(N)^2 \right).
\end{align}
Note that by assumption $f$ is a half-regular bump function and $f'(E_0 \pm c_1 ) =0$, so the boundary terms vanish in various integration by parts that we perform below.  

We now rewrite
\beq
\frac{1}{ ( \gamma_i (t) - z_t )^2(\gamma_i (t) - w_t)^2 } = \frac{1}{ \del_z z_t \del_w w_t } \del_z \del_w \left( \frac{1}{ w_t - z_t } \left( \frac{1}{\gamma_i (t) -z_t } - \frac{1}{ \gamma_i (t) - w_t } \right) \right).
\eeq
If $z$ and $w$ are in opposite half-planes, we have
\begin{align}
\frac{1}{ w_t -z_t } \left( \frac{1}{N} \sum_i \frac{1}{\gamma_i (t) -z_t } - \frac{1}{ \gamma_i (t) - w_t } - (m_t (z_t) -m_t (w_t ) ) \right) = \O\left( \frac{1}{N \Im [z_t]^2 } + \frac{1}{ N \Im [w_t ]^2} \right).
\end{align}
If they are in the same half-plane, we write
\begin{align}
&\frac{1}{ w_t -z_t } \left( \frac{1}{N} \sum_i \frac{1}{\gamma_i (t) -z_t } - \frac{1}{ \gamma_i (t) - w_t } - (m_t (z_t) -m_t (w_t ) ) \right) \nonumber \\
&= \int_{0}^1  \frac{\d }{ \d u} \left( \frac{1}{N} \sum_i \frac{1}{\gamma_i (t) -z_t +u } - m_t (z_t + u)  \right)\bigg\vert_{u = s(z_t -w_t )} \d s \nonumber \\
&=  \O\left( \frac{1}{N \Im [z_t]^2 } + \frac{1}{ N \Im [w_t ]^2} \right).
\end{align}
Then, integrating by parts in $x_1$ and $x_2$, we have
\begin{align}
\bigg| &  \int_0^{t_0} \int_{ |x_1 - E_0|, |x_2 - E_0 | < c_1 } \int_{|y_1|, |y_2| > N^{-\mfb } \eta_* } y_1 y_2 \chi(y_1) f''(x_1) \chi (y_2) f''(x_2) \nonumber \\
&\times  \frac{1}{ \del_z z_t \del_w w_t } \del_z \del_w \frac{1}{ w_t -z_t } \left( \frac{1}{N} \sum_i \frac{1}{\gamma_i (t) -z_t } - \frac{1}{ \gamma_i (t) - w_t } - (m_t (z_t) -m_t (w_t ) ) \right) \d x_1 \d x_2 \d y_1 \d y_2 \d t \bigg| \nonumber \\
& \leq \frac{C }{N} \int_{0}^{t_0} \int_{|y_1|, |y_2| > N^{-\mfb } \eta_* }  \frac{ |f'(x_1) f'(x_2) y_1 y_2 | }{| \Im [z_t ] |^4 | \Im [w_t ] |^2 }  \d x_1 \d x_2 \d y_1 \d y_2 \d t \leq  \frac{ C \log(N)}{  (N \eta_* N^{-\mfb } )},
\end{align}
where we used again \eqref{eqn:meso-dbm-t1}. 
Now, $\partial_z z_t =  1 + \O (t_0)$, and
\begin{align}
\frac{ m_t (z_t ) - m_t (w_t ) }{ z_t - w_t } = \O\left( \frac{1}{ | \Im[z_t] | + | \Im [w_t ] | }\right),
\end{align}
so
\begin{align}
\bigg| &\int_0^{t_0} \int_{ |x_1 - E_0|, |x_2 - E_0 | < c_1 } \int_{|y_1|, |y_2| > N^{-\mfb } \eta_* } y_1 y_2 \chi(y_1) f''(x_1) \chi (y_2) f''(x_2) \nonumber \\
& \times \left( \del_z \del_w \frac{m_t (z_t) - m_t (w_t ) }{ w_t -z_t } \right)\left( \frac{1}{ \del_z z_t \del_w w_t }  -1 \right)  \d x_1 \d x_2 \d y_1 \d y_2 \d t \bigg| \nonumber \\
& \leq C \int_{0}^{t_0} \int_{|y_1|, |y_2| > N^{-\mfb } \eta_* }  \frac{ t_0 |y_1 y_2| }{| \Im[z_t] |^2 | \Im[w_t]|^2} \frac{|f'(x_1) f'(x_2)|}{ | \Im[z_t]| + | \Im [w_t] | }  \d x_1 \d x_2 \d y_1 \d y_2 \d t \leq C t_0  \log(N)^2.
\end{align}
When $z$ and $w$ are in the same half-planes, since
\beq
\frac{ m_t (z_t ) - m_t (w_t ) }{ z_t - w_t } = \O (1),
\eeq
we have
\begin{align}
 \bigg| \int_0^{t_0} & \int_{ |x_1 - E_0|, |x_2 - E_0 | < c_1 } \int_{y_1, y_2 > N^{-\mfb } \eta_* } y_1 y_2 \chi(y_1) f''(x_1) \chi (y_2) f''(x_2) \nonumber \\
& \times  \left( \del_z \del_w \frac{m_t (z_t) - m_t (w_t ) }{ w_t -z_t } \right)   \d x_1 \d x_2 \d y_1 \d y_2 \d t  \bigg|   \leq C t_0 \log(N)^2.
\end{align}
We have so far reduced the proof to the calculation of the double integral of
\beq
f''(x_1) y_1 \chi (y_1 ) f'' (x_2) y_2 \chi (y_2) \del_z \del_w \frac{ m_t (z_t) - m_t (w_t) }{z_t - w_t}
\eeq
over $|x_1 - E_0|, |x_2 - E_0 | < c_1$ and $|y_1|, |y_2| > N^{-\mfb} \eta_*$, and $z$ and $w$ lie in the same half-planes.  We will eventually calculate this quantity using Green's theorem. In order to do so, we must restore the other components (involving $f'$ and $f$) to $\tilf (z)$ and $\tilf (w)$, as well as restore the small $y$ region $|y_i|< N^{-\mfb} \eta_*$. 

We will require the following notation in order to make the calculation of the boundary integral easier.  We let now,
\beq
\hatz_t = z + (t_0 - t) m_{t_0} (E_0 \pm \i 0),
\eeq
with the $\pm$ corresponding to whether or not $z$ lies in the upper or lower half-plane. Note that the argument of $m_{t_0}$ is fixed and does not depend on $z$. We fix similar notation $\hatw_t$.  
Then, for $x_1$ and $x_2$ satisfying $|x_1 - E_0 |, |x_2 - E_0 | <c_1$ and $f' (x_1), f'(x_2) \neq 0$ (i.e., $|x_i - E_0 | < M \eta_*$), we have when $z$ and $w$ lie in opposite half planes that
\beq
\frac{ m_t (z_t ) - m_t (w_t ) }{ z_t - w_t }  - \frac{ m_t (z_t ) - m_t (w_t ) }{ \hatz_t - \hatw_t} = \O\left( \frac{(t_0  - t) (M \eta_* + |y_1| + |y_2|)}{  (| \Im [z_t]| + | \Im[w_t] | )^2} \right),
\eeq
since $|m_t'| \leq C$.
Then by direct calculation,
\begin{align}
& \left| \del_z^2 \del_w^2 \left(\frac{ m_t (z_t ) - m_t (w_t ) }{ z_t - w_t }  - \frac{ m_t (z_t ) - m_t (w_t ) }{ \hatz_t - \hatw_t} \right) \right| \nonumber \\
 \leq & C \frac{(t_0  - t) (M \eta_* + |y_1| + |y_2|)}{  (| \Im [z_t]| + | \Im[w_t] | )^2 | \Im[z_t] |^2 | \Im[w_t]|^2} + C \frac{t_0-t}{  (| \Im [z_t]| + | \Im[w_t] | )^5}.
\end{align}
Therefore,
\begin{align}
\bigg| &  \int_0^{t_0} \int_{ |x_1 - E_0|, |x_2 - E_0 | < c_1 } \int_{y_1, -y_2 > N^{-\mfb } \eta_* } y_1 y_2 \chi(y_1) f''(x_1) \chi (y_2) f''(x_2) \nonumber \\
& \times  \del_z \del_w \left(  \frac{m_t (z_t) - m_t (w_t ) }{ z_t -w_t } - \frac{ m_t (z_t ) - m_t (w_t ) }{ \hatz_t - \hatw_t}   \right)  \d x_1 \d x_2 \d y_1 \d y_2 \d t \bigg|\nonumber\\
& \leq C \int_{0}^{t_0}   \int_{1 > |y_1|, |y_2| > N^{-\mfb } \eta_* }  \frac{| f'(x_1) f'(x_2) y_1 y_2 | }{| \Im[z_t] \Im[w_t]|^2} \left( \frac{(t_0  - t) (M \eta_* + |y_1| + |y_2|)}{  (| \Im [z_t]| + | \Im[w_t]| )^2} + \frac{t_0-t}{  (| \Im [z_t]| + | \Im[w_t] | )}\right) \nonumber \\
&\leq C (M \eta_* \log(N)+ t_0 \log(N)^2).
\end{align}
Now, the same arguments leading to \eqref{eqn:gauss-var-1} give
\begin{align}
 &\frac{-1}{4} \int_0^{t_0} \int_{ |x_1 - E_0|, |x_2 - E_0 | < c_1 } \int_{y_1, -y_2 > N^{-\mfb } \eta_* } y_1 y_2 \chi(y_1) f''(x_1) \chi (y_2) f''(x_2)  \left( \del_z \del_w \frac{m_t (z_t) - m_t (w_t ) }{ \hatz_t -\hatw_t } \right) \nonumber \\
&= \int_0^{t_0} \int_{ y_1, -y_2 > N^{-\mfb } \eta_* }  ( \bar{\del}_z \tilf (z) \bar{\del}_w \tilf (w) ) \left( \del_z \del_w \frac{m_t (z_t) - m_t (w_t ) }{ \hatz_t -\hatw_t } \right) \nonumber  \\
&+ \O \left( t_0 \log(N)^2 \right).
\end{align}
We now wish to restore the integration region $|y_1|, |y_2| < N^{-\mfb} \eta_*$ in preparation for using Green's theorem.  We have
\begin{align}
 &\left| \int_0^{t_0}  \int_{y_1 > N^{-\mfb } \eta_* > |y_2| } y_1 y_2 \chi(y_1) f''(x_1) \chi (y_2) f''(x_2)  \left( \del_z \del_w \frac{m_t (z_t) - m_t (w_t ) }{ \hatz_t -\hatw_t } \right) \d x_1 \d x_2 \d y_1 \d y_2 \d t\right|  \nonumber \\
= &\left|  \int_0^{t_0}  \int_{y_1 > N^{-\mfb } \eta_* > |y_2| } y_1 y_2 \chi(y_1) f'(x_1) \chi (y_2) f''(x_2) \partial_z  \left( \del_z \del_w \frac{m_t (z_t) - m_t (w_t ) }{ \hatz_t -\hatw_t } \right) \d x_1 \d x_2 \d y_1 \d y_2 \d t\right|  \nonumber \\
\leq & C \|f''\|_1 \eta_* N^{-\mfb} \log(N).
\end{align}
A similar estimate holds for the contribution with $\chi'(y_1) (f(x_1) + \i y_1 f'(x_1)) f''(x_2)$ with $y_1 > N^{-\mfb } \eta_* > |y_2|$.  The contribution of the region $|y_1|, |y_2| < N^{-\mfb } \eta_*$ is $\O ( \|f''\|_1^2 \eta_*^2 N^{-2 \mfb} \log(N)  )$.  We integrate by parts twice to obtain
\begin{align}
& \int_0^{t_0} \int_{y_1 , -y_2 >0}  \bar{\del}_z \bar{\del}_w \tilf (z) \tilf (w)  \left( \del_z \del_w \frac{m_t (z_t) - m_t (w_t ) }{ \hatz_t -\hatw_t } \right) \d x_1 \d x_2 \d y_1 \d y_2 \d t \nonumber \\
= &\int_0^{t_0} \int_{y_1 , -y_2 >0}  \bar{\del}_z \bar{\del}_w \tilf' (z) \tilf' (w)  \left(  \frac{m_t (z_t) - m_t (w_t ) }{ \hatz_t -\hatw_t } \right) \d x_1 \d x_2 \d y_1 \d y_2 \d t.
\end{align}
We now examine the boundary value of the kernel.  Recall $m_t (z_t) = m_{t_0} (z)$.  We denote the boundary values by
\beq
m_{t_0} (x \pm \i 0)  = \pi H \pto (x)  \pm \i \pi \pto (x) .
\eeq
Then
\begin{align}
=&\frac{ \mto (x_1 + \i 0 ) - \mto (x_2 - \i 0 ) }{ x_1 + s m_{t_0} (E_0 + \i 0) - (x_2 + s m_{t_0} (E_0 - \i 0) ) } +  \frac{ \mto (x_1- \i 0 ) - \mto (x_2 + \i 0 ) }{ x_1 + s m_{t_0} (E_0 - \i 0) - (x_2 + s m_{t_0} (E_0 + \i 0) ) } \nonumber \\
=& \pi \frac{ H \pto (x_1) + \i \pto (x_1) - H \pto (x_2) + \i \pto (x_2) }{x_1 - x_2 + s \i 2 \pi \pto (E_0 ) } + \pi  \frac{ H \pto (x_1) - \i \pto (x_1) - H \pto (x_2) - \i  \pto (x_2) }{x_1 - x_2 - s \i  2 \pi \pto (E_0 ) } \nonumber \\
=& 2 \pi \frac{ (H \pto (x_1) - H \pto (x_2 ))(x_1 -x_2) } {(x_1-x_2)^2 + ( s \pi 2 \pto (E_0 ) )^2} + \pi^2( \pto (x_1) + \pto (x_2 ) ) \frac{ s 4 \pto (E_0 ) }{ (x_1 - x_2)^2 + (s \pi 2 \pto (E_0 ) )^2 }. \label{eqn:ker-a1}
\end{align}
Since the first term on the last line of \eqref{eqn:ker-a1} is a bounded function, we obtain from Green's theorem \eqref{eqn:green-thm} that
\begin{align}
& \int_0^{t_0} \int_{y_1 , -y_2 >0}  \bar{\del}_z \bar{\del}_w \tilf' (z) \tilf' (w)  \left(  \frac{m_t (z_t) - m_t (w_t ) }{ \hatz_t -\hatw_t } \right) \nonumber \\
&= \frac{\pi^2}{4} \int_{0}^{t_0}\int f'(x_1) f'(x_2)  ( \pto (x_1) + \pto (x_2 ) ) \frac{ s 4 \pto (E_0 ) }{ (x_1 - x_2)^2 + (s2 \pi \pto (E_0 ) )^2 } + \O (t_0).
\end{align}
This yields the claim.  \qed

We now further examine the integral arising in Lemma \ref{lem:gaussian-var-calc-1}.

\bel \label{lem:gaussian-var-calc-2} 
We have, 
\begin{align}
& \frac{1}{2} \int_{0}^{t_0}\int f'(x_1) f'(x_2)  ( \pto (x_1) + \pto (x_2 ) ) \frac{ s  4 \pto (E_0 ) }{ (x_1 - x_2)^2 + (s \pi 2 \pto (E_0 ) )^2 } \d x_1 \d x_2 \d s \nonumber \\
= &\frac{1}{ \pi^2} | \log(t_0/\eta_*) | + \O ( \log \log(N) ).
\end{align}
\eel 
\proof We can explicitly do the $t$ integration,
\begin{align}
 &\int_{0}^{t_0}\int f'(x_1) f'(x_2)  ( \pto (x_1)  ) \frac{ s 4 \pto (E_0 ) }{ (x_1 - x_2)^2 + (s \pi  2 \pto (E_0 ) )^2 }  \d x_1 \d x_2 \d s  \nonumber  \\
=& \frac{1}{ 2 \pi^2 \pto (E_0) } \int f'(x_1) f'(x_2) \pto (x_1) \log \left(1 + \frac{ (t_0 2 \pi \pto (E_0) )^2}{ (x_1 -x_2)^2 } \right) \d x_1 \d x_2.
\end{align}
Fix a small $c_1 >0$ such that $c_1 < c' /10$, where $c'$ is from the data of $f$.  We have that $f'(x) \neq 0$ only if $|x-E_0| < c_1$ or $|x-E_1| < c'/2$.  If $|x_1 - E_0 | <c_1$ and $|x_2 - E_1 | < c'/2$ then
\beq
 \log \left(1 + \frac{ (t_0 2 \pi \pto (E_0) )^2}{ (x_1 -x_2)^2 } \right) = \O (t_0^2),
\eeq
so
\begin{align}
\int_{ |x_1 - E_0| < c_1, |x_2 - E_1 | < c'/2} f'(x_1) f'(x_2) \pto (x_1) \log \left(1 + \frac{ (t_0 2 \pi \pto (E_0) )^2}{ (x_1 -x_2)^2 } \right) \d x_1 \d x_2 = \O (t_0^2).
\end{align}
If $|x_1 -E_1| < c'/2$ and $|x_2 - E_1 | < c'/2$ then we use that $f'$ is bounded, so
\begin{align}
&\left| \int_{ |x_1 - E_1| < c'/2, |x_2 - E_1 | < c'/2} f'(x_1) f'(x_2) \pto (x_1) \log \left(1 + \frac{ (t_0 2 \pi \pto (E_0) )^2}{ (x_1 -x_2)^2 } \right) \d x_1 \d x_2 \right| \nonumber\\
\leq & C  \int_{ |x_1 - E_1| < c'/2, |x_2 - E_1 | < c'/2}  \log \left(1 + \frac{ (t_0 2 \pi \pto (E_0) )^2}{ (x_1 -x_2)^2 } \right) \d x_1 \d x_2 \nonumber\\
\leq & C \int_{ |x_1 -x_2 | < t_0 ,  |x_2 - E_1 | < c'/2}( |\log (t_0) | + | \log (x_1-x_2) | ) \d x_1 \d x_2 + C \leq C (t_0 \log(N)+1).
\end{align} 
Hence,
\begin{align}
&\int f'(x_1) f'(x_2) \pto (x_1) \log \left(1 + \frac{ (t_0 2 \pi \pto (E_0) )^2}{ (x_1 -x_2)^2 } \right) \d x_1 \d x_2 \nonumber \\
= & \int_{ |x_1 - E_0|, |x_2 -E_0| < c_1} f'(x_1) f'(x_2) \pto (x_1) \log \left(1 + \frac{ (t_0 2 \pi \pto (E_0) )^2}{ (x_1 -x_2)^2 } \right) \d x_1 \d x_2 + \O ( t_0 \log(N) ).
\end{align}
Now when $|x_1 - E_0 | <c_1$, $f'(x) \neq 0$ only if $|x_1 -E_0| \leq M \eta_*$, and for such $x$, $|\pto(x_1) - \pto (E_0)| \leq C M \eta_*$. Then
\begin{align}
 & \left| \int_{ |x_1 - E_0|, |x_2 -E_0| < c_1} f'(x_1) f'(x_2)( \pto (x_1)-\pto (E_0)) \log \left(1 + \frac{ (t_0 2 \pi \pto (E_0) )^2}{ (x_1 -x_2)^2 } \right) \d x_1 \d x_2 \right| \nonumber \\
 \leq&  C M \eta_* \log(N) \int_{ |x_1 -x_2 | > N^{-100} } |f'(x_1) f'(x_2) | \d x_1 \d x_2 \nonumber\\
+ & C M \eta_*\|f'\|^2_{\infty} \int_{ |x_1 -x_2 | < N^{-100}, |x_2 - E_0| < c_1 } ( | \log(t_0) | + | \log (x_1 - x_2) | ) \d x_1 \d x_1 \nonumber\\
\leq & C M \eta_* \log(N),
\end{align}
where we used, for example, $\|f'\|_\infty \leq \|f''\|_1 \leq N$ because $f'$ is of compact support.  
 Hence,
\begin{align}
&\frac{1}{  2 \pi^2 \pto (E_0) } \int f'(x_1) f'(x_2) \pto (x_1) \log \left(1 + \frac{ (t_0 2 \pto (E_0) )^2}{ (x_1 -x_2)^2 } \right) \d x_1 \d x_2 \nonumber \\
= & \frac{1}{2 \pi^2} \int_{ |x_1 - E_0 |, |x_2 - E_0| <c_1} f'(x_1) f'(x_2) \log \left(1 + \frac{ (t_0 2 \pto (E_0) )^2}{ (x_1 -x_2)^2 } \right) \d x_1 \d x_2 + \O( t_0 \log(N) ).
\end{align}
Now, 
\begin{align}
 \log \left(1 + \frac{ (2t_0 \pto (E_0) )^2}{ (x_1 -x_2)^2 } \right) -  \log \left( \frac{ (2 t_0 \pto (E_0) )^2}{ (x_1 -x_2)^2 } \right) =  \log \left(1 + \frac{ (x_1 -x_2)^2 }{ (2 t_0 \pto (E_0) )^2} \right).
\end{align}
If both $f'(x_1), f'(x_2) \neq 0$ and $|x_1-E_0|, |x_2 - E_0 | < c_1$, then the right side is $\O ( (M \eta_* /t_0 )^2 )$, and
\begin{align}
&\frac{1}{2 \pi^2} \int_{ |x_1 - E_0 |, |x_2 - E_0| <c_1} f'(x_1) f'(x_2) \log \left(1 + \frac{ (2 t_0 \pto (E_0) )^2}{ (x_1 -x_2)^2 } \right) \d x_1 \d x_2 \nonumber \\
= & \frac{1}{2 \pi^2} \int_{ |x_1 - E_0 |, |x_2 - E_0| <c_1} f'(x_1) f'(x_2) \log \left( \frac{ (2 t_0 \pto (E_0) )^2}{ (x_1 -x_2)^2 } \right) \d x_1 \d x_2  + \O ( (M \eta_*  /t_0 )^2 ) \nonumber  \\
= &\frac{1}{ \pi^2} \log (t_0 ) - \frac{1}{ \pi^2} \int_{ |x_1 - E_0 |, |x_2 - E_0| <c_1} f'(x_1) f'(x_2) \log |x_1 - x_2| \d x_1 \d x_2 +  \O (1)\end{align}
We may assume that $\eta_* M < c_1/2$.  Denote $R = [E_0 - c_1, E_0+c_1]^2$.   Then
\begin{align}
\int_R f'(x_1) f'(x_2) \log |x_1 - x_2| \d x_1 \d x_2  &= \int_{R} f'(x_1) \frac{\d}{ \d x_2} (f (x_2 ) - f(x_1 ) ) \log |x_1 - x_2|   \d x_1 \d x_2  \nonumber \\
& = -\int_{ R} f'(x_1) \frac{ f(x_2) - f(x_1) }{ x_2 - x_1}  \d x_1 \d x_2  + \O (1) ,
\end{align}
where we used that the boundary term in the integration by parts is $\O(1)$ because $f'(x_1) \neq 0$ only for $|x_1 -E_0| < c_1/2$.  Integrating by parts again,
\begin{align}
\int_{ R} f'(x_1) \frac{ f(x_2) - f(x_1) }{ x_2 - x_1}  \d x_1 \d x_2   =& \int_{ R} \frac{ \d }{ \d x_1} (f(x_1) - f(x_2 ) ) \frac{ f(x_2 ) - f(x_1) }{ (x_2 - x_1 ) }  \d x_1 \d x_2   \nonumber \\
= \int_{ R}  & \frac{  ( f(x_2 ) - f( x_1 ) )^2}{ (x_2 - x_1 )^2}  \d x_1 \d x_2   - \int_{R} f'(x_1) \frac{ f(x_2) - f(x_1) }{ x_2 - x_1}   \d x_1 \d x_2   + \O (1).
\end{align}
We then conclude the claim from Lemma \ref{lem:h12-calc}. \qed

\subsection{Proof of Lemma \ref{lem:meso-functs}} \label{sec:meso-functs-proof}

\proof Recall
\beq
g(E) := \frac{1}{ \pi} \int f(x) \Im \frac{1}{ E -(x+ t_0 \mto (x) )} \d x ,
\eeq
where the argument $x$ in $m_{t_0}$ is understood as $m_{t_0} ( x + \i 0)$. 
For any sufficiently small constant $c_2 >0$, with $c_2 < c'/100$, let $\chi (E)$ be a smooth bump function that is $1$ for $E \in [E_0 - c_2/4, E_1+c'/2+c_2/4]$ and  $0$ for $E \notin [E_0-c_2/2, E_1+c'/2+c_2/2]$.  For $E \notin [E_0-c_2/4, E_1+c'/2+c_2/4]$ we easily see that 
$|g'(E)| + |g(E)| \leq Ct_0$, due to the fact that $|E - (x+t_0 \mto (x) ) | \geq c$ for such $E$ and $x$ such that $f(x) \neq 0$.   Hence, 
\beq
\frac{1}{N} \sum_i g ( \lambda_i (0) ) - \int g(x) \rho_0 (x) \d x = \frac{1}{N} \sum_i \chi (\lambda_i (0) )g ( \lambda_i (0) ) - \int \chi (x) g(x) \rho_0 (x) \d x + \O( t_0 N^{\delta-1} ).
\eeq
Let $\hatE_0 := E_0 + \Re[ t_0 \mto (E_0) ]$.  Denote $q(x) = g(x) \chi (x)$ and define $h(x)$ by
\beq
h(E) = q(\hatE_0) + \int_{\hatE_0}^{E} q'(x) [ ( \chi_1 ( (x-\hatE_0) / (M' t_0 ) )) + \chi_2 (x ) ] \d x,
\eeq
where $\chi_1$ is a smooth bump function such that $\chi_1(x) =1$ if $|x| \leq 1/2$ and $\chi_1(x) =0$ for $|x| \geq 1$, and $\chi_2$ is a smooth bump function such that $\chi_2 =1$ for $|x-E_1| < 5c'/8$ and $\chi_2 =0$ for $|x-E_1| > 3c'/4$.   Note that $q'(x) =0$ for $x > E_1 + 5c'/8$.  Note that these choices of bump functions reflect the constraints on where $\hat{h}'(x) \neq 0$ in the statement of the lemma.  The function $h(E)$ will be seen to be essentially $\hat{h}(E)$ after accounting for some rescalings. 

We first obtain some derivative bounds for $q(E)$, first with the goal of showing that the error in replacing $q(E)$ by $h(E)$ is small, and then eventually with the goal of proving the claimed derivative bounds for $\hat{h}(E)$.

 Let $E_2$ be the left-most point where $\chi_2 \neq 1$.  We consider derivative bounds of $q(x)$ for $E_3 < E <E_2$, where $E_3$ is the first point where where $\chi (E) = 1$.  Note that $|q'(x) | \leq C t_0$ for $E <E_3$.  For $E_3 < E <E_2$ we have
\begin{align}
\pi q'(E) &= \int f(x) \Im \left[  \frac{\d}{ \d E} \frac{1}{ E - (x + t_0 \mto (x) ) } \right] \d x \nonumber \\
&= -\int f(x) \Im\left[ \frac{1}{ 1 + t_0 \mto' (x) } \frac{\d}{ \d x} \frac{1}{E - (x+t_0 \mto (x) ) } \right] \d x \nonumber \\
&= \int f'(x) \Im \left[ \frac{1}{ 1 + t_0 \mto' (x) } \frac{1}{ E - (x + t_0 \mto (x) ) } \right] \d x \nonumber \\
&+ \int f(x) \Im\left[ \frac{ t_0 \mto''(x) }{ (1+t_0 \mto'(x) )^2 (E - (x + t_0 \mto (x) ) )} \right] \d x. \label{eqn:q-der-bd-1}
\end{align}
The last  line is easily seen to be $\O (t_0 \log (N ) )$.  We write the third line as
\begin{align}
& \int f'(x) \Im \left[ \frac{1}{ 1 + t_0 \mto' (x) } \frac{1}{ E - (x + t_0 \mto (x) ) } \right]  \d x \nonumber \\
= & \int f'(x) \Im \left[ \frac{1}{ 1 + t_0 \mto' (x) }\right] \Re\left[ \frac{1}{ E - (x + t_0 \mto (x) ) } \right] \d x\nonumber \\
+ & \int f'(x) \Re \left[ \frac{1}{ 1 + t_0 \mto' (x) }\right] \Im\left[ \frac{1}{ E - (x + t_0 \mto (x) ) } \right] \d x. \label{eqn:q-der-bd-2}
\end{align}
Note for both terms, regardless of the value of $E$, we always have the estimate $\O ( 1 / t_0)$ since $\Im [\mto (x) ] \geq c$ for all $x$ where $f'(x) \neq 0$, and $\|f'\|_1 \leq C$.  Now if $E \in [E_3, E_2]$ satisfies $|E - E_0 - t_0 m_{t_0} (E_0) | \geq t_0$ we claim that for all $x $ such that $f'(x) \neq 0$, we have $|E - x - t_0 m_{t_0} (x) | \geq c |E - \hatE_0|$ for some $c>0$. If $|x - E_0| \leq \eta_* M \leq \frac{1}{2} t_0$ for large enough $N$ then this follows immediately from the reverse triangle inequality. If $|x - E_1 | \leq c'/2$ then this follows since $t_0 = o (1)$ and $E_2 < E_1 - 5 c'/8$. Therefore, if $|E- \hatE_0 |\geq t_0$, we have
\begin{align}
&\bigg|  \int f'(x) \Im \left[ \frac{1}{ 1 + t_0 \mto' (x) }\right] \Re\left[ \frac{1}{ E - (x + t_0 \mto (x) ) } \right]  \d x\nonumber  \\
+ & \int f'(x) \Re \left[ \frac{1}{ 1 + t_0 \mto' (x) }\right] \Im\left[ \frac{1}{ E - (x + t_0 \mto (x) ) } \right] \bigg| \d x \nonumber \\ 
\leq & \int |f'(x) |\left( \frac{ t_0}{ |E -\hatE_0|} + \frac{t_0}{ |E-\hatE_0|^2} \right) \d x\leq \frac{C t_0}{ |E- \hatE_0|^2 + t_0^2}. \label{eqn:dbm-a2}
\end{align}
Additionally, we see that the final estimate \eqref{eqn:dbm-a2} holds for any $E \in [E_3, E_2]$ by using the previously mentioned estimate $\O(1/t_0)$ for $|E- \hatE_0| \leq t_0$. Therefore, 
\beq \label{eqn:meso-dbm-t2}
| h'(E) - q'(E)|  \leq \1_{ \{ |E-\hatE_0 | > M' t_0/2 \} } \frac{t_0}{ t_0^2 + |E- \hatE_0 |^2} + t_0 \log(N),
\eeq
and by rigidity,
\beq
\frac{1}{N} \sum_i (h( \lambda_i (0) ) - q ( \lambda_i (0) ) )- \int ( h ( x) - q(x) ) \rho (x) \d x = N^{-1} \O ( N^{\delta} (t_0 + 1/M' ) ).
\eeq
Now, $h'(E)$ has compact support, which lies in the bulk of $\rho$.  The function $q(E)$ is also of compact support and $q ( \hatE_0 ) = h ( \hatE_0 )$. By integrating the derivative estimate \eqref{eqn:meso-dbm-t2} we conclude that $|h(-\infty)| + | h (+\infty)| = \O (t_0 \log(N) + 1 / M')$.  

Let $\varphi$ be a smooth step function such that $\varphi'(x) \neq 0$ only for $|x-E_1| < c_2$.  Then the function $[h(-\infty) - h (+\infty) ] \varphi$ satisfies
\beq
\frac{1}{N} \sum_i [h(-\infty) - h (+\infty) ] \varphi ( \lambda_i (0) ) - \int [h(-\infty) - h (+\infty) ] \varphi (x) \rho (x) \d x = \O (N^{-1+\delta} ( t_0 \log(N)+1/M') ).
\eeq
We set $\tilde{h}(x) = h(x) -h(-\infty) +  [h(-\infty) - h (+\infty) ] \varphi(x)$.  The function $\tilde{h}$ is almost the function $\hat{h}$ we seek; the latter will be obtained from the former by rescaling by a constant close to $1$.  We will do this later; first we will prove the claimed derivative estimates for $\hat{h}$ by proving them first for $\tilde{h}$. 

Clearly $\tilde{h}(x)$ and its derivatives satisfy the support conditions stated in the lemma.  We only have to check that its derivatives obey the required estimates.  For this, we need only to estimate the derivatives of $h(x)$.  From \eqref{eqn:dbm-a2} we see that
\beq
|h'(E)| \leq |q'(E) | \leq  \frac{C t_0}{ |E- \hatE_0|^2 + t_0^2}
\eeq
for $|E - \hatE_0| \leq M' t_0$.   For $|E-E_1| \leq 3c'/4$, we have using the representation \eqref{eqn:q-der-bd-1} and \eqref{eqn:q-der-bd-2},
\begin{align}
|q'(E)| & \leq C \int ( |f'(x) | + |f(x)|) \d x \nonumber \\
& + \int_{ |x-E_1| < c'/2} \bigg\{ \left| \Im \left[ \frac{1}{ 1 + t_0 \mto' (x) }\right] \Re\left[ \frac{1}{ E - (x + t_0 \mto (x) ) } \right] \right| \nonumber \\
& + \left| \Re \left[ \frac{1}{ 1 + t_0 \mto' (x) }\right] \Im\left[ \frac{1}{ E - (x + t_0 \mto (x) ) } \right]  \right| \bigg\} \d x \nonumber \\
&\leq C+ C \int_{|x-E_1| <c'/2}  \frac{t_0}{ (E- x - t_0 \Re[ \mto (x) ] )^2 + t_0^2 } \d x. \label{eqn:dbm-b-1}
\end{align}
The term in the first line accounts for the contribution of the region $|x-E_0| \leq M \eta_*$ where $|E-x- t_0 \mto (x) | \geq c$ as well as the last line of \eqref{eqn:q-der-bd-1}.  Now, there is a $C_1 >0$ so that if $x$ is such that $|E- x| > C_1 t_0$, then $|E- x - t_0 \Re[ \mto (x) ] | \geq c  |E-x|$.  Hence, dividing the region of integration into $|E-x| > C_1 t_0$ and $|E-x| \leq C_1 t_0$, we see that the integral on the last line of \eqref{eqn:dbm-b-1} is bounded above by a constant $C$, which finishes the required estimates of $h'(E)$.  

We now turn to estimates of the second derivative.  First, assume $|E-\hatE_0| < M' t_0$.  Then, by definition,
\beq
|h''(E)| \leq  C \frac{|q'(E)|}{M' t_0} + |q''(E)| \leq \frac{C}{ |E-\hatE_0|^2 + t_0^2} + |q''(E) | ,
\eeq
where we used our previous established estimates for $q'(E)$. 
For $|E- \hatE_0| \leq M't_0$ we consider the following formula for $q''(E)$, which follows from differentiating \eqref{eqn:q-der-bd-1} and \eqref{eqn:q-der-bd-2},
\begin{align}
\pi q''(E) =& -\int f(x)  \Im\left[ \frac{ t_0 \mto''(x) }{ (1+t_0 \mto'(x) )^2 (E - (x + t_0 \mto (x) ) )^2} \right] \d x \nonumber \\
-& \int f'(x) \Im \left[ \frac{1}{ 1 + t_0 \mto' (x) }\right] \Re\left[ \frac{1}{ (E - (x + t_0 \mto (x) ))^2 } \right]  \d x\nonumber \\
- & \int f'(x) \Re \left[ \frac{1}{ 1 + t_0 \mto' (x) }\right] \Im\left[ \frac{1}{ (E - (x + t_0 \mto (x) ))^2 } \right] \d x.
\end{align}
The first line is bounded, as the quantity in the brackets is bounded by $C t_0 |E- (x+t_0 \mto (x))|^{-2}$ which we have already seen has a bounded integral.  For the second and third lines, we always have $| (E - (x + t_0 \mto (x) )| \geq c t_0$ and so $|q''(E) | \leq C / t_0^2$.  We have seen above that if $|E- \hatE_0 | \geq  t_0$, then $|E- x - t_0 \mto (x)| \geq c |E- \hatE_0|$ for $x$ such that $f'(x) \neq 0$.  Since $\|f'\|_1 \leq C$, we therefore have proven that
\beq
|q''(E)| \leq \frac{C}{ |E- \hatE_0|^2 + t_0^2 }\quad \text{for} \quad |E-\hatE_0| \leq t_0 M'.
\eeq
We now need to estimate $q''(E)$ for $|E-E_1| \leq 3c'/4$.  
We have 
\begin{align}
&\pi q''(E) \nonumber\\
&= \int f(x) \Im \left[ \frac{\d^2}{ ( \d E)^2} \frac{1}{ E- (x+ t_0 \mto (x) )} \right] \d x \nonumber \\
&= \int_{|x-E_1| < 7c'/8} f(x) \Im \left[ \frac{\d^2}{ ( \d E)^2} \frac{1}{ E- (x+ t_0 \mto (x) )} \right] \d x +\O(1)  \nonumber \\
&= \int_{|x-E_1| < 7c'/8} f(x) \Im \left[ \frac{1}{ 1 + t_0 \mto'(x) } \frac{\d}{\d x} \frac{1}{ 1+ t_0 \mto'(x) } \frac{ \d }{\d x} \frac{1}{ E- (x+ t_0 \mto (x) )} \right] \d x + \O(1) \nonumber  \\
&= \int_{|x-E_1| < 7c'/8}  \Im\left[ \frac{1}{ (E- (x + t_0 \mto (x) ))} \right]  \nonumber \\
& \times  \Re\left[ f'' (x) \left( \frac{1}{ 1 + t_0 m'(x) } \right)^2 - \frac{ 3 f'(x) t_0 \mto''(x)}{ (1 + t_0 \mto'(x) )^3} + f \frac{t_0^2 (\mto''(x))^2-t_0 \mto'''(x) (1+ t_0 \mto (x) )}{ (1+t_0 \mto'(x) )^4} \right] \d x \label{eqn:dbm-a3} \\
&+ \int_{|x-E_1| < 7c'/8}  \Re\left[ \frac{1}{ (E- (x + t_0 \mto (x) ))} \right] \nonumber \\
& \times  \Im\left[ f'' (x) \left( \frac{1}{ 1 + t_0 m'(x) } \right)^2 - \frac{ 3 f'(x) t_0 \mto''(x)}{ (1 + t_0 \mto'(x) )^3} + f \frac{t_0^2 (\mto''(x))^2-t_0 \mto'''(x) (1+ t_0 \mto (x) )}{ (1+t_0 \mto'(x) )^4} \right] \d x + \O(1). \label{eqn:dbm-a4}
\end{align}
In the second equality we used that $| E- (x+ t_0 \mto (x) ) | \geq c$ for $|E-E_1 | < 3c'/4$ and $|x-E_1 | > 7c'/8$.  The final equality follows from integration by parts, in which we absorbed some boundary terms into the $\O(1)$ term.  The real part on the line \eqref{eqn:dbm-a3} is bounded and so the first integral is bounded, again by the integrability of $\Im[ (E-(x+t_0 \mto (x) )^{-1} ]$ that we have already used.  The imaginary part on the line \eqref{eqn:dbm-a4} is $\O(t_0)$ so the second integral is bounded by
\beq
\int_{ |x-E_1| < 7c'/8} \frac{ t_0}{ |E- (x+t_0 \mto (x) ) |^2} \d x \leq C
\eeq
which follows from our earlier reasoning.  This completes our estimates of the derivatives of $h(x)$.  Now let $E_4 = E_0+c'/10$.  Consider
\begin{align}
g(E_4) &= \frac{1}{ \pi} \int f(x) \Im \frac{1}{ E_4 - (x+t_0 \mto (x) ) } \d x \nonumber \\
&= \frac{1}{ \pi} \int_{|x-E_4| < c_2 } \Im \frac{1}{ E_4 - (x+t_0 \mto (x) ) } \d x  + \O (t_0).
\end{align}
Now,
\begin{align}
& \left| \int_{|x-E_4| < c_2 }  \frac{1}{ E_4 - (x+t_0 \mto (x) ) } \d x - \int_{|E-E_4| < c_2 }  \frac{1}{ E_4 - (x+t_0 \mto (E_4) ) } \right| \nonumber \\
& \leq \int_{ |x-E_4| < c_2} \frac{ t_0 |E_4-x|}{ |E_4 - (x+t_0 \mto (x) )| |E_4 - (x+t_0 \mto (E_4) | } \d x
\end{align}
If $|x-E_4 | > C t_0$ for some large $C>0$, then the denominator is bounded below  by $c|x-E_4|^2$ and so this part of the integral contributes $\O (t_0 \log(N))$.  The remaining region of integration $|x-E_4| < C t_0$ contributes $\O (t_0)$ because the denominator is bounded below by $c t_0^2$.  Hence,
\beq
q(E_4) = g(E_4) = \frac{1}{ \pi}  \int_{|E-E_4| < c_2 } \Im\left[ \frac{1}{ E_4 - (x+ t_0 \mto (E_4) ) } \right] + \O (t_0 \log(N) ) = 1 + \O (t_0 \log(N) ),
\eeq
so 
\beq
\tilde{h} (E_4) = 1 + \O (t_0 \log(N) + 1/M').
\eeq
Finally, set $\hat{h} (E) = \tilde{h} (E) / \tilde{h} (E_4)$.  Then
\beq
\frac{1}{N} \sum_i \tilde{h} ( \lambda_i (0) ) - \hat{h} ( \lambda_i (0) )  - \int ( \tilde{h} ( x) - \hat{h} (x) ) \rho (x) \d x = N^{-1} \O ( N^{\delta} (t_0 + 1 / M' ) ).
\eeq
This completes the proof. \qed

\section{Homogenization application: proof of Theorem \ref{thm:homog}} \label{sec:homo}

Let $W$ be a matrix of general Wigner-type.  Let $G$ be a GOE matrix independent of $W$.  We recall our notation that $\rho_t$ is the free convolution at time $t$ of the spectral measure $\rho(E)$ associated with $W$ through $S$, with $N$-quantiles denoted by $\gamma_{ i,t}$.  Fix an index $i_0$ as in the theorem statement.  Define the constants $a$ and $b$ by
\beq
a := \frac{ \rho_{t_0} ( \gamma_{i_0, t_0 } ) }{\rhosc(0) }, \qquad b :=   \gamma_{i_0, t_0}
\eeq
We use the convention of \cite{fixed} that the quantiles of the semicircle distribution are defined so that $\gamsc_{N/2} = 0$.
Define the process $\hatx_i (t)$ as follows.  For $t=0$, the initial data are the following rescaled eigenvalues of $W+\sqrt{t_0}G$:
\beq
\hatx_i (0) = a \left(  \lambda_i ( W + \sqrt{t_0} G ) -b \right),
\eeq
and for $t >0$, $\hatx_i (t)$ is DBM,
\beq
\d \hatx_i (t) = \sqrt{ \frac{2}{ N }} \d B_i (t) + \frac{1}{N} \sum_{j \neq i } \frac{1}{ \hatx_i (t) - \hatx_j (t) } \d t.
\eeq
Here, the $B_i (t)$ denote a family of independent Brownian motions.  
Consider two auxiliary processes, $\haty_i(t)$ and $\hatz_i (t)$, such that at $t=0$ they are both distributed as independent GOE matrices (independent from $G$ and $W$ as well as each other) and for $t >0$, they satisfy
\beq
\d \haty_i (t) = \sqrt{ \frac{2}{ N }} \d B_{i+i_0-N/2} (t) + \frac{1}{N} \sum_{j \neq i } \frac{1}{ \haty_i (t) - \haty_j (t) } \d t
\eeq
and
\beq
\d \hatz_i (t) =  \sqrt{ \frac{2}{ N }} \d B_{i+i_0-N/2}  (t) + \frac{1}{N} \sum_{j \neq i } \frac{1}{ \hatz_i (t) - \hatz_j (t) } \d t.
\eeq
Here, we extended the family $B_i (t)$ to a larger family of independent Brownian motions, $\{ B_i \}_{i=-N}^{2N}$.  
From Theorem 3.1 of \cite{fixed} we have that there is a function $\zeta$ and a constant $\gamma$ satisfying
\begin{align}\label{76}
&(\hatx_{i_0} (t_1)  - \gamma- \haty_{N/2} (t_1) ) \nonumber \\
= & \frac{1}{N} \sum_{|j| \leq N^{ \tau_1 + (\tau_0/2-\tau_1)/3 }} \zeta \left( j N^{-1}, t_1 \right) [ \hatx_{i_0+j} (0) - \haty_{N/2+j} (0) ] + \O (N^{-\tau_1/100-1} ),
\end{align}
with overwhelming probability.  The function $\zeta$ is smooth, and obeys the following estimates due to Proposition 3.2 of \cite{fixed}:
\beq \label{eqn:zeta-bd} 
\int \zeta(x, t) = 1, \qquad 0 \leq \zeta(x, t) \leq \frac{C t}{ x^2 + t^2}, \qquad | \zeta^{(k)} (x, t) | \leq \frac{C_k}{ t^{k}} \frac{t}{x^2 + t^2}.
\eeq
  The $\gamma$ is defined as
\beq
\gamma := a \left( \gamma_{i_0, t_0+t_1 a^{-2} } - \gamma_{i_0, t_0} \right). 
\eeq
We now undo the scaling applied to the process $\hatx_i (t)$ and moreover rewrite the sum involving $\zeta$ in the form of a linear spectral statistic.  
Consider
\beq
\tilx_i (t) := \frac{1}{a} \hatx_i ( a^2 t ) +b.
\eeq
Note that the process $\tilx_i (t)$ has, for each fixed time $t$, the same distribution as the eigenvalues of $W + \sqrt{t_0 + a^2 t} G$. 
By \eqref{76},
\begin{align}\label{710}
\tilx_i (t_1) =& \gamma_{i_0, t_0+t_1} 
+ \frac{1}{a} \haty_{N/2} ( a^2 t_1 ) \nonumber \\
+&\frac{1}{a} \frac{1}{N} \sum_{|j| \leq N^{ \tau_1 + (\tau_0/2-\tau_1)/3 }} \zeta ( j N^{-1}, a^2 t_1 ) (a \tilx_{i_0+j} (0) - ab ) \nonumber \\
-&\frac{1}{a} \frac{1}{N}  \sum_{|j| \leq N^{ \tau_1 + (\tau_0/2-\tau_1)/3 }} \zeta ( j N^{-1},  a^2 t_1 ) \haty_{N/2+j } + \O(N^{-1-\tau_1/100}).
\end{align}
By our assumptions on the spectral density of $S$ (see Appendix~\ref{a:free-2}), we have for $|j| \leq N/ \log(N)$ that
\beq \label{eqn:gamfc-j}
\gamma_{i_0+j, t_0} = \gamma_{i_0, t_0} + \frac{j}{N \rho_{ t_0} ( \gamma_{i_0, t_0 } )} + \O \left( j^2 N^{-2} \right),
\eeq
and similarly
\beq \label{eqn:gamsc-j}
\gamsc_{N/2+j} = \gamsc_{N/2}+ \frac{j}{N \rhosc (0) } + \O \left( j^2 N^{-2} \right).
\eeq
Therefore
\begin{align}
 &\frac{1}{N} \sum_{|j| \leq N^{ \tau_1 + (\tau_0/2-\tau_1)/3 }} \zeta ( j N^{-1}, a^2 t_1 ) ( a \gamma_{ i_0, t_0 } - \gamsc_{N/2 } ) \nonumber \\
= & \frac{1}{N} \sum_{|j| \leq N^{ \tau_1 + (\tau_0/2-\tau_1)/3 }} \zeta ( j N^{-1}, a^2 t_1 ) ( a \gamma_{i_0+j, t_0} - \gamsc_{N/2+j } ) + \O ( N^{-3/2} ). \label{eqn:homo-a1}
\end{align}
Fix $0 < \omega < 1/100$.  We have, with overwhelming probability, the following computation for the two sums on the last two lines of \eqref{710}:
\begin{align}
&  \sum_{|j| \leq N^{ \tau_1 + (\tau_0/2-\tau_1)/3 }} N^{-1} \zeta (jN^{-1}, a^2 t_1 ) (a \tilx_{i_0+j} (0) - ab ) - \sum_{|j| \leq N^{ \tau_1 + (\tau_0/2-\tau_1)/3 }} N^{-1} \zeta ( j N^{-1}, a^2 t_1 ) \haty_{N/2+j }(0) \nonumber  \\
= &\sum_{|j| \leq N^{ \tau_1 + (\tau_0/2-\tau_1)/3 }} N^{-1}\zeta (j N^{-1}, a^2 t_1 ) (a \tilx_{i_0+j} (0) -a \gamma_{ i_0+j, t_0} ) \nonumber \\
-& \sum_{|j| \leq N^{ \tau_1 + (\tau_0/2-\tau_1)/3 }} N^{-1} \zeta ( j N^{-1}, a^2 t_1 ) (\haty_{N/2+j } - \gamsc_{N/2+j } ) + \O (N^{-3/2} ) \nonumber \\
&= \sum_{|j| \leq N^{\tau_1+ \omega } } N^{-1}\zeta (j N^{-1}, a^2 t_1 ) (a \tilx_{i_0+j} (0) -a \gamma_{i_0+j, t_0} ) \nonumber  \\
-&\sum_{|j| \leq N^{\tau_1+\omega } } N^{-1} \zeta ( j N^{-1} , a^2 t_1 ) (\haty_{N/2+j } - \gamsc_{N/2+j } ) + \O (N^{-1+\eps-\omega} ).
\end{align}
Here we applied the rigidity estimates \eqref{eqn:dbm-quant} and the bound \eqref{eqn:zeta-bd} for $\zeta$ in the last line.  
Let $\chi (x) \geq 0$ be a smooth bump function,
\beq
\chi (x) = \begin{cases} 1, & |x| \leq 1\\
0,& |x| > 2 \end{cases}
\eeq
and define $f(x)$ by
\beq
f(x) = \int_{-\infty}^x \chi ( x / (t_1 N^{2 \omega} ) ) \zeta (x, a^2 t_1 ) \d x.
\eeq
Then we have, with overwhelming probability,
\begin{align}
&\frac{1}{N}\sum_{|j| \leq N^{\tau_1+ \omega } } \zeta (j/N, a^2 t_1 ) ( \tilx_{i_0+j} (0) - \gamma_{i_0+j, t_0} ) \nonumber \\
=& \frac{1}{N} \sum_{j} f' ( j/N ) ( \tilx_{i_0+j}(0) - \gamma_{i_0+j, t_0} ) + \O ( N^{-1-\omega+\eps } ) \nonumber \\
=& \frac{1}{N}  \sum_j \int_{\gamma_{ i_0+j, t_0}}^{ \tilx_{i_0+j} (0) } f' ( \rho_{ t_0} ( \gamma_{i_0, t_0} ) (s - \gamma_{i_0, t_0} )) \d s + \O ( N^{-1-\omega+\eps } +N^{-1-\tau_1+\eps} ),
\end{align}
where the last line again uses rigidity as well as the estimates \eqref{eqn:zeta-bd} for $|\zeta'|$. 
Now,
\begin{align}\label{collect1}
&\sum_j \int_{\gamma_{i_0+j,t_0}}^{ \tilx_{i_0+j} (0) } f' ( \rho_{ t_0} ( \gamma_{i_0, t_0} ) (s - \gamma_{i_0, t_0} )) \d s  \nonumber  \\
= & \frac{1}{  \rho_{ t_0} ( \gamma_{i_0, t_0} ) } \sum_{j} f (  \rho_{ t_0} ( \gamma_{i_0, t_0} ) ( \tilx_{i_0+j} (0) - \gamma_{i_0, t_0} ) ) \nonumber \\
- & \frac{1}{  \rho_{ t_0} ( \gamma_{i_0, t_0} ) } \sum_{j} f (  \rho_{t_0} ( \gamma_{i_0, t_0} ) ( \gamma_{i_0+j,t_0} - \gamma_{i_0, t_0} ) ).
\end{align}
A similar argument yields
\begin{align}\label{e:collect2}
\frac{1}{N} \sum_{|j| \leq N^{\tau_1+ \omega } } \zeta (j/N, a^2 t_1 )  ( \haty_{N/2+j}(0) - \gamsc_{N/2+j} ) &= \frac{1}{N} \frac{1}{ \rhosc (0) } \sum_j f ( \rhosc(0) ( \haty_{N/2+j}(0) ))- f ( \rhosc ( 0)(\gamsc_{N/2+j } ))\nonumber  \\
&+\O  ( N^{-1-\omega+\eps } +N^{-1-\tau_1+\eps} ),
\end{align}
with overwhelming probability.  
Similarly to how \eqref{eqn:homo-a1} follows from \eqref{eqn:gamfc-j} and \eqref{eqn:gamsc-j}, we have
\begin{align}\label{collect3}
&\frac{1}{N} \sum_j f ( \rhosc (0) \gamsc_{N/2+ j } ) - \frac{1}{N} \sum_{j} f (  \rho_{ t_0} ( \gamma_{i_0, t_0} ) ( \gamma_{i_0+j} - \gamma_{i_0, t_0} ) ) \nonumber \\
= & \int f ( \rhosc(0) x ) \rhosc (x) \d x -\int f (  \rho_{ t_0} ( \gamma_{i_0, t_0} ) \cdot (s -\gamma_{i_0, t_0}) ) \rho_{ t_0 } (s) \d s + \O (N^{-3/2} ).
\end{align}
Collecting the above, we see that
\begin{align}
&  \rho_{t_0} ( \gamma_{i_0, t_0} ) [\tilx_{i_0} (t_1) - \gamma_{i_0, t_0+t_1 } ]- \rhosc (0) \haty_{N/2} (a^2 t_1) \nonumber \\
= & \frac{1}{N}  \left( \sum_j f (  \rho_{ t_0 } (\gamma_{i_0, t_0 } ) ( \tilx_j (0) - \gamma_{ i_0, t_0 } ) ) - \int ( f (\rho_{t_0 } (\gamma_{i_0, t_0 } ) ( s - \gamma_{ i_0, t_0 } ) )  \rho_{ t_0} (s)\d s \right) \nonumber \\
- & \frac{1}{N} \left( \sum_j f ( \rhosc (0) \haty_j(0) ) - \int f(\rhosc(0) s) \rhosc ( s) \d s\right) + \O (N^{-1-\tau_1/100} + N^{-1-\omega} ).
\end{align}
Now, note that $f ( N^{2 \omega} (t_1 ) ) = 1 + \O (N^{-\omega} )$.  Dividing $f$ by $f( N^{2 \omega} (t_1) )$ contributes an additional error of $\O(N^{-1-\omega+\eps})$.  We take $x_i (t) = \tilx_i (t)$ and $y_i (t)= \haty_i (a^2t)$, and apply a similar argument with $z_i (t)$ in place of $x_i (t)$ to complete the proof. \qed 

\section{Removal of Gaussian component: proof of Theorem \ref{thm:main}} \label{sec:proof-of-main}

Let $W$ be a matrix of general Wigner-type satisfying Assumptions \ref{it:ass-1} and \ref{it:ass-2}.  Denote its variance matrix by $S$.  Fix $t_0 = N^{-1/1000}$.  By a standard construction (see \cite[Lemma 16.2]{erdos2017dynamical}), for each $i \leq j$, we can find random variables $x_{ij}$ such that,
\beq
\ee[ x_{ij}^k ] = \ee[ W_{ij}^k], \qquad k=1, 2, 3
\eeq
and
\beq
\left| \ee [x_{ij}^4 ] - \ee[ W_{ij}^4] \right| \leq \frac{C t_0}{N^2},
\eeq
and such that we have the distributional equality
\beq
x_{ij} \stackrel{d}{=} z_{ij} + (1+ \delta_{ij} )\sqrt{t_0/N} g_{ij},
\eeq
where the $g_{ij}$ are standard Gaussian random variables independent of $z_{ij}$.  
Moreover, the moments of $N^{1/2} x_{ij}$ are bounded. 
Denote by $\hat{S}$ the matrix of variances of the $z_{ij}$.  Then,
\beq
\max_{i, j} |S_{ij} - \hat{S}_{ij} | \leq C \frac{t_0}{N}
\eeq
and by Proposition \ref{prop:stab-1},  $\hat{S}$ satisfies Assumptions \ref{it:ass-1} and \ref{it:ass-2}.  Now consider the free convolution  $\hatp_t$ 
of the spectral measure $\hat{\rho}$ associated with $\hat{S}$ with the semicircle distribution at time $t$, with quantiles $\hat{\gamma}_{i, t}$.  

By Theorem \ref{thm:gde-univ}, we have
\beq
\left| \ee[ F ( N \hatp_{t_0} (\hat{\gamma}_{i_0, t_0} ) ( \lambda_{i_0} (X) - \hat{\gamma}_{i_0, t_0} ) / \sqrt{ \log(N) } ) - \ee[ F ( N \rhosc (0) \lambda_{N/2} G / \sqrt{ \log(N) } )] \right| \leq C ( \log(N) )^{-1/10}.
\eeq
Now, note that by the calculations in Appendix \ref{a:free-2}, $\hatp_{t_0}$ is the spectral measure associated with the matrix
\beq
\hat{S}_{t_0} = \hat{S} + t_0 e e^T /N,
\eeq
where $e$ is the constant vector of all ones.  By construction,
\beq
\hat{S}_{t_0} = S - \1 t_0 /N,
\eeq
and by Lemma \ref{lem:diag-S},
\beq
\left| \ee[ F ( N \hatp_{t_0} (\hat{\gamma}_{i_0, t_0} ) ( \lambda_{i_0} (X) - \hat{\gamma}_{i_0, t_0} ) / \sqrt{ \log(N) } ) - \ee[ F ( N \rho (\gamma_{i_0} ) ( \lambda_{i_0} (X) - \gamma_{i_0}) / \sqrt{ \log(N) } ) \right| \leq N^{-c}
\eeq
some $c>0$. 

In order to compare the single eigenvalue fluctuations of $X$ to $W$, we apply the Lindeberg strategy in which we replace the matrix entries of $X$ by $W$ one at a time and estimate the error.  This was carried out in the case of single eigenvalue fluctuations for Wigner matrices in Corollary 18 of \cite{tao2011random}, and the proof technique carries over to the class of matrices considered here.  

For completeness, we give an alternative (but of course highly related) approach based on a construction \cite{comparison} that has the advantage of giving somewhat explicit error estimates.

 For any $\delta >0$, Lemma 3.2 of \cite{comparison} constructs functions $ \tillam_i$ (depending on $\delta >0$) on the space of symmetric matrices such that with overwhelming probability,
\beq
| \tillam_i (H) - \lambda_i | \leq \frac{N^{\eps}}{N^{1+\delta}} ,  \qquad |\del_{ab}^k \tillam_i (H) | \leq C_k \frac{ N^{\eps+(k-1) \delta}}{N}.
\eeq
In the paper \cite{comparison}, $H$ was a generalized Wigner matrix, but the construction carries over to matrices that obey optimal rigidity and eigenvector delocalization with respect to a fixed spectral measure whose Stieltjes transform obeys similar bounds to the semicircle law (in this case, it is only required that $|| \bm||_\infty$ is bounded); see Section 4 of \cite{comparison}. 

We see that
\beq
\left| \ee[ F ( N \rho (\gamma_{i_0} ) ( \lambda_{i_0} (X) - \gamma_{i_0} ) / \sqrt{ \log(N) } ) - \ee[ F ( N \rho (\gamma_{i_0} ) ( \tillam_{i_0} (X) - \gamma_{i_0} ) / \sqrt{ \log(N) } )] \right| \leq N^{\eps-\delta}
\eeq
for any $\eps >0$.  A similar estimate holds for $W$.  We have derivative bounds that hold with overwhelming probability:
\beq
 \left| \del^k_{ab} F ( N \rho (\gamma_{i_0} ) ( \tillam_{i_0} (X) - \gamma_{i_0} ) / \sqrt{ \log(N) } \right| \leq N^{\eps+(k-1) \delta}.
\eeq
In replacing the entries of $X$ by $W$ one-by-one, we encounter $\O(N^2)$ replacements. 
  Suppose that we are at an intermediate stage in the replacement strategy and wish to replace the $(a, b)$th entry of $H_1$ by that of $H_2$, and these two matrices have the same entries elsewhere.  The error is estimated by Taylor expanding the function $H \to F ( N \rho (\gamma_{i_0} ) ( \tillam_{i_0} (H) - \gamma_{i_0} ) / \sqrt{ \log(N) }$ around $H$ where $H$ is $H_1$ with $(a, b)$th entry set to $0$, which is the same matrix as $H_2$ with $(a, b)$th entry set to $0$.  This expansion is done to fifth order.  The first three terms in the Taylor expansion cancel exactly.  The fourth order error is
\beq
\O (t_0 N^{-2+\eps+3 \delta } )
\eeq
and the fifth order term is $\O(N^{-2-1/2+\eps+4 \delta })$.  Since there are $N^2$ entries, this is acceptable.   

We find
\begin{align}
&\bigg| \ee[ F ( N \rho (\gamma_{i_0} ) ( \lambda_{i_0} (X) - \gamma_{i_0} ) / \sqrt{ \log(N) } ) ] \nonumber \\
- &\ee[ F ( N \rho (\gamma_{i_0} ) ( \lambda_{i_0} (W) - \gamma_{i_0} ) / \sqrt{ \log(N) } )] \bigg| \leq C N^{\eps} ( N^{-\delta} + t_0 N^{3\delta}),
\end{align}
which yields the claim after taking $\delta >0$  sufficiently small. \qed

\section{Calculation of expectation of single eigenvalue} \label{sec:expect-calc}

This section is dedicated to calculating the first order correction to the density of states of the expectation of a linear statistic of a matrix of general Wigner-type.  In this section Assumptions \ref{it:ass-1} and \ref{it:ass-2} are in force.  In this section we will use the notational convention that boundary values of $\mb(z)$ and related quantities on the real axis will be denoted by $\mb (E ) := \mb (E + \i 0)$, etc.

The following is a consequence of (c) of Corollary A.1 of \cite{qve}.
\bel \label{lem:im-m}
There exists $c>0$ such that the following holds.  Let $z = E + \i \eta$, for $|z| \leq c$ and $\eta >0$.  If $E >0$, we have for every $i$,
\beq
c \frac{\eta}{|z|^{1/2} } \leq \Im[ \mb_i ( \beta + z) ] \leq \frac{1}{c} \frac{ \eta}{ |z|^{1/2}}.
\eeq
If $ E<0$, then for every $i$,
\beq
c |z|^{1/2} \leq \Im [ \mb_i (\beta + z ) ] \leq \frac{|z|^{1/2}}{c}.
\eeq
Similar estimates hold near $\alpha$.
\eel
We have the following for the real part of $\mb$ near the spectral edges.
\bel \label{lem:Rem-edge}
There is $ c>0$ such that the following holds for $|z-\beta| < c$.  For every $i$,
\beq
- \Re[ \mb_i (z) ] \geq c
\eeq
\eel
\proof From \eqref{eqn:rho-i} we see that $\Re[ \mb_i ( \beta) ] \leq - c$ for some $c>0$.  The lemma then follows from the general estimate $\| \mb(z) - \mb(w) \|_\infty \leq C |z-w|^{1/3}$, a consequence of (7.4) of \cite{qve}. \qed

In this section we use the notation,
\beq
F(z) := |\mb (z) | S | \mb (z) |, \qquad U(z) := \frac{ \mb(z)^2}{ | \mb(z)|^2}.
\eeq
We denote the largest eigenvalue of $F$ by $\mu (z)$ and its eigenvector by $v (z)$.  We also introduce the operator $A(z)$ by
\beq
F(z) = A(z) + \mu (z) v(z) v(z)^T.
\eeq
From Proposition \ref{prop:F-pert} we have that uniformly over $z$ in any compact set,
\beq\label{Aspectralgap}
 \|A(z)\|_{\ell^2 \to \ell^2} \leq 1-c
\eeq
for some $c>0$.  Since $|v| \leq CN^{-1/2}$ we see that $|A_{ij}| \leq CN^{-1}$ for all $i, j$ and uniformly over $z$ in any compact set.   In a similar manner to Proposition \ref{prop:stab-3} we then see that,
\beq \label{eqn:A-inf-bd}
\| (U^* - A)^{-1} \|_{\ell^\infty \to \ell^\infty} + \| (1-A)^{-1} \|_{\ell^\infty \to \ell^\infty} \leq C.
\eeq

We collect in the following proposition several estimates on these quantities that will be used in the rest of the section.
\bep \label{prop:expect-prelim}
There is a constant $C>0$ such that the following estimates hold.  We have
\beq \label{eqn:U-bd}
\|U^* -1\|_{\ell^2 \to \ell^2} +\|U^* -1\|_{\ell^\infty \to \ell^\infty}  \leq C | \Im [ m (z) ] |,
\eeq
and
\beq \label{eqn:U-bd-t1}
\| (U^* - A)^{-1} v - v \|_{2} + \sqrt{N} \| (U^* - A)^{-1} v - v \|_{\infty} \leq C | \Im [ m (z) ] |.
\eeq
For some $ c>0$ we have that the following holds for $|z-\beta| < c$,
\beq \label{eqn:U-denom-1}
|1 - \mu v^T (U^*-A)^{-1} v | \geq c|z - \beta|^{1/2},
\eeq
and
\beq \label{eqn:U-denom-t1}
| 1 - \mu + v^T (U^* -1)v | \geq c|z- \beta|^{1/2},
\eeq
and
\beq \label{eqn:imU-a1}
|\Im[v^T (U^* - 1) v ] | \asymp | \Im [ m(z) ] |,
\eeq
and
\beq \label{eqn:ReUv}
| \Re[ v^T (U^*-1) v] | \leq C | \Im [ m(z) ] |^2,
\eeq
and 
\beq \label{eqn:est-aa1}
\left| \left( 1 - \mu + v^T (U^* -1)v  \right) - \left( 1 - \mu v^T (U^*-A)^{-1} v \right) \right| \leq C \Im[ m(z) ]^2.
\eeq
Similar estimates hold near $\alpha$.
\eep
\proof The first estimate follows from the fact that $U^*-1$ is diagonal with $i$th element equal to
\beq \label{eqn:Uii}
U^*_{ii} - 1 = 2 \frac{ - \i \Im [ \mb_i(z) ]\Re[ \mb_i (z) ] - \Im[\mb_i (z) ]^2}{ | \mb_i (z) |^2}.
\eeq This expression also yields \eqref{eqn:ReUv}.  The estimate \eqref{eqn:imU-a1} follows from Lemma \ref{lem:Rem-edge} and \eqref{eqn:Uii}.  
The second estimate of the proposition follows from the identity
\beq
(U^*-A)^{-1} v - v = (U^*-A)^{-1} v - (1-A)^{-1} v = (U^*-A)^{-1}(1-U^*)(1-A)^{-1} v.
\eeq
We have the expansion
\begin{align} \label{eqn:vUAv}
v^T \frac{1}{ U^*-A} v - 1 &= v^T \frac{1}{ U^* - A} v - v^T \frac{1}{ 1 - A} v \nonumber\\
&= v^T \frac{1}{ 1 - A} (1-U^*) \frac{1}{ 1 - A} v + v^T \frac{1}{ U^*-A} (1-U^*) \frac{1}{ 1 - A} (1- U^*) \frac{1}{ 1 - A} v \nonumber\\
&= v^T (1- U^*) v + \O ( \|1-U^*\|_{\ell^2 \to \ell^2}^2 ).
\end{align}
Hence,
\begin{align}
1 - \mu v^T (U^* -A)^{-1} v &= 1- \mu + \mu v^T (U^* - 1) v + \O( \Im [ m(z) ]^2 ).
\end{align}
For $1-\mu$ we have the identity (5.20) of \cite{qve},
\beq
1- \mu = |\eta| \frac{ \langle | \mb(z) | v(z) \rangle }{ \langle | \Im[\mb(z)]| | \mb(z)|^{-1} v(z) \rangle } \asymp \frac{| \eta | }{ | \Im[ m(z) ] |}.
\eeq
The estimates \eqref{eqn:U-denom-1} and \eqref{eqn:U-denom-t1} now follow from the above expansions and the behavior of $\Im [m (z) ]$ given by Lemma \ref{lem:im-m}. 
\qed

We first establish some estimates on the derivative $\mb'(z)$.

\bel \label{lem:mder-edge} 
There is a $ c>0$ and $C>0$ such that for $|z - \beta | \leq c$ we have,
\beq \label{eqn:m-der-edge-bd}
\| \mb'(z) \|_{\infty} \leq \frac{C}{ \sqrt{ |E- \beta| + | \eta | } }.
\eeq
A similar estimate holds near the left edge $\alpha$. 
For $ \alpha < E < \beta$ we have,
\beq
\| \Re[ \mb' (E) ] \|_\infty \leq C.
\eeq
\eel
\proof We have
\beq
\mb' = \frac{1}{ 1 - \mb^2 S }\mb^2 = | \mb| \frac{1}{ U^* - F} | \mb|^{-1} U^* \mb^2= |\mb| \frac{1}{ U^* - F} | \mb|.
\eeq
where $U$ and $F$ are as above. As above, let
\beq
F = A + \mu v v^T.
\eeq
By the Sherman--Morrison formula,
\beq
\frac{1}{ U^* - F} = \frac{1}{U^* - A } + \mu \frac{(U^*-A)^{-1} v v^T(U^*-A)^{-1}}{1 - \mu v (U^*-A)^{-1} v}
\eeq
By \eqref{eqn:A-inf-bd},
\beq
 \| | \mb | (U^* - A)^{-1} | \mb| \|_\infty \leq C. 
\eeq
The estimate \eqref{eqn:m-der-edge-bd} now follows from \eqref{eqn:U-denom-1}.  

We turn to the second estimate of the lemma.  
 From the above as well as Proposition \ref{prop:expect-prelim}, we have
\begin{align} \label{eqn:m-der-exp}
\mb' &= |\mb| \frac{1}{ U^* -A} |\mb| + \mu | \mb| \frac{ (U^*-A)^{-1} v v^T (U^*-A)^{-1} | \mb|}{1 - \mu v^T (U^*-A)^{-1} v } \nonumber \\
&= \mu |\mb| v \frac{ v^T |\mb| }{ 1- \mu + \mu v^T (U^*-1) v } + \O(1).
\end{align}
On the real axis $\mu (E) =1 $ for $\alpha < E < \beta$ and so taking the real part of the first term above we conclude
\beq
| \Re[ \mb'(E) ] | \leq C,
\eeq
from \eqref{eqn:imU-a1} and  \eqref{eqn:ReUv}.
This completes the proof. \qed

We need some elementary estimates from perturbation theory.

\bep \label{prop:F-edge-pert}
There are $c>0$ and $C>0$ such that the following estimates hold for $|z - \beta | \leq c$.  We have
\beq
\|v(z) - v(\beta)\|_2 + N^{1/2}\|v (z) - v ( \beta) \|_\infty \leq C | \beta - z|^{1/2},
\eeq
and
\beq \label{eqn:mu-bd-t1}
| \mu(z) - \mu (\beta) | = | \mu(z) - 1 | \leq C |\beta-z|^{1/2}.
\eeq
\eep
\proof We have from \eqref{eqn:m-der-edge-bd} that
\beq \label{eqn:m-perturb-edge}
\| \mb(z) - \mb ( \beta ) \|_\infty \leq C | z - \beta |^{1/2},
\eeq
and so
\beq
\| F(z) - F ( \beta) \|_{\ell^2 \to \ell^2} + \| F(z) - F ( \beta) \|_{\ell^\infty \to \ell^\infty}\leq C | z- \beta |^{1/2}.
\eeq
In particular,
\beq\label{F12}
| \mu(z) - \mu ( \beta) | \leq \| F(z) - F(\beta) \|_{\ell^2 \to \ell^2} \leq C | z- \beta |^{1/2},
\eeq
where the first inequality follows from the elementary  $| \lambda_i (A) - \lambda_i (B) | \leq \| A - B \|_{\ell^2 \to \ell^2} $.
Now, 
\beq
(F(\beta) - \mu(\beta) ) (v(\beta) - v(z) ) = ( F(z) - F( \beta ) )v ( z) - ( \mu ( z ) - \mu (\beta) ) v(z).
\eeq
Multiplying on both sides by $\Pp (\beta)$, the projection onto the subspace perpendicular to $v(\beta)$, we have
\beq\label{multboth}
\Pp(\beta) ( v(z) - v ( \beta) ) = \frac{1}{ \mu(\beta) - \Pp(\beta) F (\beta) \Pp ( \beta ) } \left(  ( F(z) - F( \beta ) )v ( z) - ( \mu ( z ) - \mu (\beta) ) v(z) \right).
\eeq
Note
\beq\label{Aid}
\frac{1}{ \mu(\beta) - \Pp(\beta) F (\beta) \Pp ( \beta ) } = \frac{1}{ 1 - A(\beta)},
\eeq
and recall the spectral gap inequality \eqref{Aspectralgap}.
Using these facts and \eqref{F12} in \eqref{multboth},
we have
\beq
\| \Pp ( \beta) v(z) \|_{2} = \| \Pp ( \beta) (v (z) - v ( \beta ) )\|_2 \leq C |z- \beta|^{1/2}.
\eeq
 Therefore, since $1 = \| v(z) \|_2^2 =  \langle v ( \beta ), v ( z) \rangle^2 + \| \Pp ( \beta) v(z) \|_{2}^2$ we have
\beq
| \langle v ( \beta ), v ( z) \rangle^2 -1 | \leq C|z-\beta|
\eeq
which implies $| \langle v ( \beta ), v ( z) \rangle -1 | \leq C |z - \beta| $ where the sign of the inner product is determined to be positive since $v(z)$ has positive entries for all $z$. 
Hence, we see that
\beq
\| v(z) - v( \beta ) \|_2 =  \sqrt{ 2 -2 \langle v ( \beta ) , v ( z) \rangle }  \leq C | z - \beta |^{1/2}.
\eeq 

By \eqref{Aid} and \eqref{eqn:A-inf-bd},
\beq
\|\Pp ( \beta ) v(z) \|_\infty \leq C ( \|F(z) - F(\beta)\|_{\ell^\infty \to \ell^\infty}\|v(z)\|_\infty + | \mu(z) - \mu ( \beta) | \|v(\beta)\|_\infty ) \leq CN^{-1/2} |z-\beta|^{1/2}.
\eeq
Now, $v(z) = p(z) v(\beta) + \Pp ( \beta) v(z)$ for some coefficient $p(z)$ satisfying $p(z) = 1 + \O (|z-\beta| )$.  Hence
\beq
\|v(z) - v(\beta) \|_\infty \leq |1-p(z) | \|v(\beta)\|_\infty + \| \Pp ( \beta) v(z)\|_\infty,
\eeq
which yields the last claim of the proposition.
\qed

We now obtain some estimates on contributions to the expectation corrections.

\bel
Let $|z - \beta | = r$.  There is a $c>0$ such that for $0 < r < c$, we have
\beq \label{eqn:bdry-1-S-a1}
\tr\left(  ( 1-S \mb^2 )^{-1} S \mb' \mb \right) =  \frac{ 1}{4 (\beta- z)} + \O ( r^{-1/2} \log ( \eta ) ).
\eeq
Moreover, for $\alpha < E < \beta$, we have
\beq \label{eqn:expect-t1}
\left| \Im\left[ \tr\left(  ( 1-S \mb^2 )^{-1} S \mb' \mb \right)  \right] \right| \leq \frac{C}{\sqrt{ |E- \alpha | |E- \beta | }}.
\eeq
\eel
\proof Using the cyclic property of trace,  we have the formula
\begin{align} \label{eqn:expand-bb}
\tr\left(  ( 1-S \mb^2 )^{-1} S \mb' \mb \right) 
&= \tr \left( (U^* - F)^{-1} |\mb| S \mb' | \mb| \mb^{-1} \right) \nonumber\\
&= \tr\left( (U^* - A)^{-1} |\mb| S \mb' | \mb| \mb^{-1} \right) \nonumber\\
&+ \mu \frac{v^T (U^*-A)^{-1}  |\mb| S \mb' | \mb| \mb^{-1} (U^*-A)^{-1} v}{1 - \mu v^T(U^*-A)^{-1}v}.
\end{align}
In the last line we used the Sherman--Morrison formula.
Recall the general estimate
\beq \label{eqn:gen-est-t1}
| \tr(X Y) | \leq C N \left( \sup_{i, j} |X_{ij} | \right) ||Y||_{\ell^\infty \to \ell^\infty}.
\eeq
From this, Propositions \ref{prop:basic-m} and \ref{prop:stab-3}, we obtain the estimate \eqref{eqn:expect-t1} for $E$ away from $\alpha$ or $\beta$. 

Using \eqref{eqn:gen-est-t1} 
we have, for $z$ near $\beta$ (using the estimate with $Y= (U^*  - A)^{-1}$), 
\beq
\left| \tr\left( (U^* - A)^{-1} |\mb| S \mb' | \mb| \mb^{-1} \right) \right| \leq \frac{C}{ |z-\beta|^{1/2}},
\eeq
using Lemma \ref{lem:mder-edge}. 
We now focus on the second estimate for $z$ near $\beta$; the case of $z$ near $\alpha$ is similar.  On the real axis $ \alpha <E < \beta$, where $\mu =1$,  the  expression  \eqref{eqn:expand-bb}simplifies to
\beq
\tr\left(  ( 1-S \mb^2 )^{-1} S \mb' \mb \right)  = \frac{ v^T |\mb|S \mb' |\mb| \mb^{-1} v}{1 - v^T (U^* - A)^{-1} v } + \O ( ( |E-\alpha||E- \beta |)^{-1/2} ) ,
\eeq
 where we also used \eqref{eqn:U-bd-t1}.

  We have
\begin{align}
\left| \Im\left[ \frac{ v^T |\mb|S \mb' |\mb| \mb^{-1} v}{1 - v^T (U^* - A)^{-1} v } \right] \right| &\leq C \left| \Im [ v^T |\mb|S \mb' |\mb| \mb^{-1} v] \Re[ (1 - v^T (U^* - A)^{-1} v)^{-1} ] \right| \nonumber\\
&+ C \left| \Re [ v^T |\mb|S \mb' |\mb| \mb^{-1} v] \Im[ (1 - v^T (U^* - A)^{-1} v)^{-1} ] \right| .
\end{align}
From \eqref{eqn:est-aa1}, \eqref{eqn:ReUv}, and \eqref{eqn:U-denom-1} we have
\beq
\left| \Re \left[ (1 - v^T (U^* - A) v )^{-1} \right] \right| \leq C,
\eeq
and so for $E$ near $\beta$,
\beq
\left| \Im [ v^T |\mb|S \mb' |\mb| \mb^{-1} v] \Re[ (1 - v^T (U^* - A)^{-1} v)^{-1} ] \right|  \leq \frac{C}{\sqrt{  |E- \beta | }}.
\eeq
By \eqref{eqn:U-denom-1},
\beq
\left| \Im \left[ (1 - v^T (U^* - A) v )^{-1} \right] \right| \leq C\frac{C}{\sqrt{  |E- \beta | }}.
\eeq
From Lemma \ref{lem:mder-edge} we have for all $i$ that
\beq
\left| \Re[ \mb'_i (E) / \mb_i(E) ] \right| \leq C,
\eeq
so
\beq
|  \Re [ v^T |\mb|S \mb' |\mb| \mb^{-1} v] | \leq C.
\eeq
Therefore,
\beq
\left| \Im\left[ \tr\left(  ( 1-S \mb^2 )^{-1} S \mb' \mb \right)  \right] \right| \leq \frac{C}{\sqrt{ |E- \beta | }}.
\eeq
This finishes the second estimate.   We now turn to the case $ |z - \beta| = r$. When the argument of $\mb$, $\mu$, or $v$ is $z$, we will omit the argument and write $\mb = \mb(z)$, $\mu = \mu(z)$ and $v = v(z)$. We will include the argument when it is $\beta$, so that, for example, $\mb - \mb(\beta ) = \mb(z) - \mb (\beta)$.

Starting from \eqref{eqn:expand-bb} and using similar arguments as before, we have
\begin{align}
\tr\left(  ( 1-S \mb^2 )^{-1} S \mb' \mb \right) &= \frac{ v^T |\mb| S \mb' | \mb| \mb^{-1} v}{1- \mu + \mu v^T (U^*-1) v } + \O (r^{-1/2} ) \nonumber\\
&=  \frac{ v^T   \mb' \mb^{-1} v}{1- \mu + \mu v^T (U^*-1) v } + \O (r^{-1/2} ),
\end{align}
where we also used \eqref{eqn:est-aa1} and Lemma \ref{lem:mder-edge}.   In the second line we used $|\mb| S |\mb| = F$ and $\mu(z) =1  + \O (r^{1/2} )$,  the latter estimate following from \eqref{eqn:mu-bd-t1}. 
For $\mb'$ we have, from \eqref{eqn:m-der-exp},
\beq
\mb' = |\mb| v \frac{ v^T | \mb| }{ 1- \mu + \mu v^T (U^*-1) v }  + \O(1).
\eeq
Then
\begin{align}
\tr\left(  ( 1-S \mb^2 )^{-1} S \mb' \mb \right) &= \frac{ \langle v^3 |\mb| \mb^{-1} \rangle \langle v |\mb| \rangle}{ (1 - \mu + \mu v^T (U^*-1) v )^2} + \O(r^{-1/2} ) \nonumber\\
&= \frac{ \langle v(\beta)^3 \rangle \langle v (\beta) \mb ( \beta) \rangle }{(1 - \mu + \mu v^T (U^*-1) v )^2} + \O (r^{-1/2} ),
\end{align}
where we used Proposition \ref{prop:F-edge-pert} in the second line.   
For symmetric matrices $X$ and $Y$ with eigenvalues $\lambda_x$ and $\lambda_y$ with unit eigenvectors $v_x$ and $v_y$, we have the identity
\beq
\lambda_x - \lambda_y = v^T_x (X - Y) v_x + (v_x - v_y )^T (Y - \lambda_y ) (v_x - v_y).
\eeq
Therefore, by Proposition \ref{prop:F-edge-pert},
\beq
1 - \mu = v(\beta)^T (F(\beta) - F( z)) v(\beta) + \O (r).
\eeq
Now, using \eqref{eqn:m-perturb-edge}, 
\begin{align}
v(\beta)^T (F(\beta) - F( z)) v(\beta)  &= v(\beta)^T ( | \mb (\beta ) | S | \mb ( \beta )| - |\mb (z) | S | \mb(z) |) v(\beta) \nonumber\\
&= v(\beta)^T ( (| \mb (\beta)| - |\mb (z) | )S |\mb(\beta)| + | \mb(z) | S ( |\mb(\beta)| - \mb (z) ) ) v(\beta) \nonumber \\
&=v(\beta)^T ( (| \mb (\beta) |- |\mb(z) | )S |\mb(\beta)| + |\mb(\beta) | S (| \mb(\beta)| -| \mb(z)| ) ) v(\beta) \nonumber + \O(r) \\
&= 2 v(\beta)^T (|\mb(\beta)| - |\mb(z)|) |\mb(\beta)|^{-1} v(\beta) + \O(r).
\end{align}
By direct calculation,
\beq
|| ( |\mb(z) | - | \Re[ \mb (z) ]| ) ||_\infty \leq Cr
\eeq
so,
\beq
2 v(\beta)^T (|\mb(\beta)| - |\mb(z)|) |\mb(\beta)|^{-1} v(\beta) =2 v^T (\beta) ( \mb(\beta) - \Re[ \mb (z) ] ) \mb(\beta)^{-1} v (\beta) + \O(r).
\eeq
We have,
\begin{align}
\mu v^T (U^*-1) v &= v(\beta)^T (U^*-1) v (\beta) + \O (r) \nonumber \\
&= - 2 \i v(\beta)^T \Im[ \mb(z) ] ( \mb (\beta))^{-1} v (\beta) + \O (r)
\end{align}
and so,
\begin{align}
1 - \mu + \mu v^T (U^*-1) v  &= 2v(\beta)^T \frac{\mb(\beta) - \mb(z) }{ \mb (\beta) } v(\beta) + \O (r).
\end{align}
From Lemma 9.11 of \cite{qve} we have (note that in the notation there, the parameter $h'_x$ equals $h'_x = |m_x| f_x \langle |m| f \rangle^{1/2}$ - the proof contains a typo, incorrectly identifying $h'_x$ as $h_x' = |m_x| f_x \langle |m| f \rangle^{-1/2}$; compare with Lemma 7.13 of \cite{alt2018dyson}),
\beq
\Im[ \mb_i (\beta - E) ]= \left(  \frac{\langle | \mb (\beta ) | v (\beta) \rangle }{\langle v (\beta)^3 \rangle } \right)^{1/2} |E|^{1/2} |\mb_i( \beta) | v_i(\beta) + \O(|E|).
\eeq
In particular,
\beq \label{eqn:Im-m-msc}
\Im [ \mb_i ( \beta - E) ] =  \frac{ | \mb_i(\beta)| v_i(\beta ) }{ q} \Im[ \msc (2 - E) ] + \O (|E|)
\eeq
where we introduced the following for notational simplicity,
\beq
q^2 = \frac{\langle v (\beta)^3 \rangle } {\langle | \mb (\beta ) | v (\beta) \rangle },
\eeq
and $\msc$ denotes the Stietljes transform of the semicircle distribution,
\beq
\msc(z) := \int \frac{ \rhosc(x)}{ x -z } \d x = \frac{ -z + \sqrt{z^2 -4}}{2}.
\eeq
Integrating \eqref{eqn:Im-m-msc} we obtain the estimate,
\begin{align}
\mb_i (\beta + z) - \mb_i ( \beta) &=  \frac{ | \mb_i(\beta)| v_i(\beta ) }{ q} ( \msc (2 + z ) - \msc (2) ) + \O ( r \log ( \eta ) ) \nonumber\\
&= \frac{ | \mb_i(\beta)| v_i(\beta ) }{ q}  \sqrt{z} + \O ( r \log ( \eta ) ).
\end{align}
Hence,
\begin{align}
\tr\left(  ( 1-S \mb^2 )^{-1} S \mb' \mb \right) &= - \frac{1}{4  z} + \O ( r^{-1/2} \log ( \eta ) ).
\end{align}
This completes the proof. \qed

\bep \label{prop:expect-bdry}
Let $f$ be a $C^3$ function of compact support.  Let $\cc_r$ be the following domain:
\beq
\cc_r := \{ z \in \cc :  | \Im [z] |  \neq 0 , |z-\alpha | > r, |z - \beta| > r \}.
\eeq
Then
\begin{align} \label{eqn:bdry-eqn}
&\lim_{r \dto 0} \frac{1}{ \pi} \int_{\cc_r}( \bar{\del}_z \tilf (z) )  \tr (S \mb \mb' ) + \tr ( (1 - S \mb^2 )^{-1} S \mb'  \mb ) + \frac{1}{N^2} \sum_{j, a} \sfo_{aj} \mb_j' \mb_j \mb_a^2 \d x \d y \nonumber\\
= & \frac{1}{ \pi } \int_\alpha^\beta f(x) \left(   \frac{1}{2} \partial_x \Im\left[ \tr (  S \mb^2(x) )\right] + \frac{1}{4} \partial_x \Im \left[ \frac{1}{N^2} \sum_{i, j} s^{(4)}_{ij} ( \mb_i (x) \mb_j (x) )^2 \right] \right) \d x \nonumber\\
+ & \frac{1}{ \pi} \int_\alpha^\beta f(x) \Im\left[  \tr ( (1 - S \mb(x)^2 )^{-1} S \mb'(x)  \mb(x) ) \right] \d x +  \frac{1}{4 } f( \beta) + \frac{1}{4 } f( \alpha ),
\end{align}
where for real $E$, $\mb(E) $ denotes the boundary value $ \mb(E):= \mb (E + \i 0)$.  
We have the estimates
\beq \label{eqn:exp-kern-bd-1}
\left|  \partial_x \Im\left[ \tr (  S \mb^2(x) )\right] \right| + \left|  \partial_x \Im \left[ \frac{1}{N^2} \sum_{i, j} s^{(4)}_{ij} ( \mb_i (x) \mb_j (x) )^2 \right] \right| \leq \frac{C}{ \sqrt{ |E- \alpha| |E- \beta| }} \1_{ \{ \alpha < E < \beta \} }
\eeq
and
\beq \label{eqn:exp-kern-bd-2}
\Im\left[  \tr ( (1 - S \mb(x)^2 )^{-1} S \mb'(x)  \mb(x) ) \right] \leq \frac{C}{ \sqrt{ |E- \alpha| |E- \beta| }} \1_{ \{ \alpha < E < \beta \} }.
\eeq
\eep
\proof All of the functions  on the left side of \eqref{eqn:bdry-eqn} that multiply $\bar{\del}_z \tilf (z)$ are bounded in $\cc_r$, and have extensions from the upper half-plane to the real axis away from the edges of the support $\alpha, \beta$.  Moreover, the extension is real-valued outside the support of $\rho$.  Applying Green's theorem \eqref{eqn:green-thm}, we see from \eqref{eqn:m-der-edge-bd} that,
\begin{align}
&\lim_{r \dto 0} \frac{1}{ \pi} \int_{\cc_r}( \bar{\del}_z \tilf (z) )  \left( \tr (S \mb \mb' )+\frac{1}{N^2} \sum_{j, a} \sfo_{aj} \mb_j' \mb_j \mb_a^2  \right) \d x \d y \nonumber \\
= & \frac{1}{ \pi} \int f(x) \left(   \frac{1}{2} \partial_x \Im\left[ \tr (  S \mb^2(x+\i 0) )\right] + \frac{1}{4} \partial_x \Im \left[ \frac{1}{N^2} \sum_{i, j} s^{(4)}_{ij} ( \mb_i (x+ \i 0) \mb_j (x+\i0) )^2 \right] \right) \d x.
\end{align}
For the final term, we first apply Green's Theorem to obtain
\begin{align}
&\frac{1}{ \pi} \int_{ \cc_r}  ( \bar{\del}_z \tilf (z) ) \tr ( (1 - S \mb^2 )^{-1} S \mb'  \mb ) \nonumber \\
= &\frac{1}{ \pi} \int_{ \alpha+r}^{ \beta-r} f(x) \Im\left[  \tr ( (1 - S \mb(x+\i 0)^2 )^{-1} S \mb'(x+\i 0)  \mb(x+ \i 0) ) \right] \d x \nonumber\\
+ & \frac{\i}{ 2 \pi} \int_{ |z-\beta| = r} ( f(x) + \i y f'(x) ) \tr ( (1 - S \mb^2 )^{-1} S \mb'  \mb ) \d z \nonumber\\
+ & \frac{\i}{ 2 \pi}  \int_{ |z-\alpha| = r} ( f(x) + \i y f'(x) ) \tr ( (1 - S \mb^2 )^{-1} S \mb'  \mb ) \d z
\end{align}
By \eqref{eqn:bdry-1-S-a1}, we have
\beq
\lim_{r \dto 0 }   \frac{\i}{ 2 \pi} \int_{ |z-\beta| = r} ( f(x) + \i y f'(x) ) \tr ( (1 - S \mb^2 )^{-1} S \mb'  \mb ) \d z =  \frac{ f (\beta)}{4} .
\eeq
The claim follows from this and an analogous calculation for $z$ near $\alpha$. \qed

We have the following lemma.
\bel \label{lem:logdet-der}
For $E \in (\alpha, \beta)$ we have,
\beq
-\frac{1}{2} \frac{\d}{ \d E} \log \det ( 1 - S \mb^2 (E) ) = \tr ((1 - S \mb^2(E))^{-1} S \mb'(E) \mb(E) )),
\eeq
where we denote the boundary values by $\mb(E) := \mb(E + \i 0)$.
\eel
\proof Fix $\kappa >0$ and $\Ikap := ( \alpha + \kappa, \beta - \kappa )$ as usual.  The infimum over $E \in \Ikap$ of the distance of the spectrum of the matrix $1- S \mb(E)^2$ from the negative real axis is strictly positive.  Therefore, we can define,
\beq
\log(1 - S \mb^2 (E) ) = \frac{1}{ 2 \pi \i } \int_\Gamma \log (z) \frac{1}{ z - (1 - S \mb^2 (E) ) } \d z
\eeq
where the contour $\Gamma$ encloses the spectrum of the matrix  $1- S \mb^2 (E)$ and does not intersect the negative real axis.  Above, we use the principle branch of the logarithm.  Moreover, we can take a single contour which works for all $E \in \Ikap$.  Clearly,
\beq
\frac{\d }{ \d E} \tr \log ( 1 - S \mb^2 (E) ) =- \frac{2}{ 2 \pi \i } \int_\Gamma \log(z) \tr\left( \frac{1}{ (z - (1 - S \mb^2 (E) ) )^2} S \mb'(E) \mb(E) \right) \d z
\eeq
where we used the cyclicity of the trace.  A straightforward calculation using the Jordan canonical form (our matrix $S \mb^2 (E)$ is not necessarily diagonalizable) gives the general identity 
\beq
\frac{1}{ 2 \pi \i } \int_\Gamma F'(z) \frac{1}{ z - M } \d z = \frac{1}{ 2 \pi \i } \int_\Gamma F(z) \frac{1}{ ( z - M)^2 } \d z .
\eeq
for any contour $\Gamma$ enclosing the spectrum of $M$ (the identity may be proven using the explicit form of an inverse of a Jordan block and the Cauchy integral formula; it may also be obtained via a limiting procedure approximating non-diagonalizable $M$ by diagonalizable matrices).

Therefore,
\begin{align}
-\frac{1}{2} \frac{ \d }{ \d E} \tr \log (1 - S \mb^2 (E) ) &=  \frac{1}{ 2 \pi \i} \tr\left[ \left( \int_\Gamma \log(z) \frac{1}{ (z- ( 1 - S \mb^2 (E) ) )^2} \d z\right) S \mb'(E) \mb(E)\right]   \nonumber\\
 &=  \frac{1}{ 2 \pi \i } \tr\left[ \left( \int_\Gamma \frac{1}{z} \frac{1}{ (z- ( 1 - S \mb^2 (E) ) )} \d z\right) S \mb'(E) \mb(E)\right]   \nonumber\\
&= \tr \left( \frac{1}{  1 - S \mb^2 (E) }S \mb'(E) \mb (E) \right).
\end{align}
We used the analytic functional calculus in the last equality. 
On the other hand, an argument again using the Jordan canonical form gives,
\beq
\tr \log (1 - S \mb^2 (E) ) = \log \det ( 1 - S \mb^2 (E) )
\eeq
and the claim follows.
\qed 

We next calculate the various boundary integrals for the special case where the test function is close to an indicator function.

\bep
Let $E \in \Ikap$ for some $\kappa >0$.  Let $ t= N^{-\omega}$ for some $\omega >0$.  Let $f(x)$ be a function such that $f(x) = 0$ for $x \leq E-t$ and $f(x) =1 $ for $x \geq E + t$ and $|f(x)| \leq 1$ for all $x$.  Then, we have
\beq
 \frac{1}{ \pi } \int_\alpha^\beta f(x) \left(   \frac{1}{2} \partial_x \Im\left[ \tr (  S \mb^2(x) )\right] \right) \d x = - \frac{1}{ 2 \pi }\Im\left[ \tr (  S \mb^2(E) )\right] + \O (t) 
\eeq
and
\beq
\frac{1}{ \pi } \int_\alpha^\beta f(x) \left(  \frac{1}{4} \partial_x \Im \left[ \frac{1}{N^2} \sum_{i, j} s^{(4)}_{ij} ( \mb_i (x) \mb_j (x) )^2 \right] \right) \d x = - \frac{1}{ 4 \pi} \Im \left[ \frac{1}{N^2} \sum_{i, j} s^{(4)}_{ij} ( \mb_i (E) \mb_j (E) )^2 \right] + \O (t),
\eeq
as well as
\begin{align}
&\frac{1}{ \pi} \int_\alpha^\beta f(x) \Im\left[  \tr ( (1 - S \mb(x)^2 )^{-1} S \mb'(x)  \mb(x) ) \right] \d x \nonumber\\
= & -\frac{1}{4} + \frac{1}{ 2 \pi} \Im\left[ \log \det (1 - S \mb^2 (E) ) \right] + \O (t).
\end{align}
As above, we denote boundary values by $\mb(E) := \mb(E + \i 0)$. 
\eep
\proof The first two estimates are straightforward.  For the last estimate we have, using Lemma \ref{lem:logdet-der}, 
\begin{align}
&\frac{1}{ \pi} \int_\alpha^\beta f(x) \Im\left[  \tr ( (1 - S \mb(x)^2 )^{-1} S \mb'(x)  \mb(x) ) \right] \d x \nonumber \\
 =&  \frac{1}{2 \pi} \left(\Im \log \det (1 - S \mb^2 (E) ) - \Im \log \det (1 - S \mb^2 (\beta^-) ) \right) + \O(t).
\end{align}
We need to calculate the second term in the brackets (here, $\beta^-$ means $\lim_{x \to \beta^-}$).   Let $ x< \beta$ and consider the following identity, a consequence of the matrix determinant lemma,
\begin{align}
\det ( 1 - S \mb^2 (x) ) &= \det (U) \det (U^*  - F) \nonumber\\
&= \det(U) \det(U^* - A) \left( 1 - v^T (U^*-A)^{-1} v \right) \nonumber\\
&= \det(1 - AU)\left( 1 - v^T (U^*-A)^{-1} v \right) ,
\end{align}
where we have omitted the argument $x$ above. 
Since $\|AU\|_{ \ell^2 \to \ell^2 } \leq 1 -c$ all of the eigenvalues of $1-AU$ are within a disc of radius $1-c$ centered at $1$.  Therefore,
\beq
\lim_{ x \to \beta^-} \log \det (1 - AU ) = \log \det (1 - A(\beta) U(\beta) ) = \log \det (1 - A(\beta ))
\eeq
where we used that $U (\beta) = 1$.  The imaginary part of this vanishes.  Let $x = \beta - r$, for small $r>0$.  Then, using \eqref{eqn:U-bd},
\begin{align}
v^T (U^*-A)^{-1} v - 1 &= v^T ( (U^*-A)^{-1}- (1-A)^{-1} ) v = v^T(1-U^* ) v + \O (r) .
\end{align}
Now, by direct calculation,
\begin{align}
v^T(1-U^* ) v  = -2 \i v^T \Im [ \mb ] / | \mb | v + \O (r).
\end{align}
Hence, we have,
\beq
c r^{1/2} \leq \Im [1  -v^T (U^*-A)^{-1} v ] \leq C r^{1/2} 
\eeq
and 
\beq
| \Re[ 1 -v^T (U^*-A)^{-1} v ] |  \leq C r.
\eeq
Therefore, 
\beq
\lim_{ r \to 0 } \Im \log (1 - v^T (U^* - A)^{-1} v ) =   \pi /2.
\eeq
This yields the claim.
\qed

The following is a straightforward consequence of all of the above calculations as well as Lemma \ref{lem:lss-expect} (one must restore the remaining part of $\cc \backslash \Oma$ to the integral appearing on the RHS of \eqref{eqn:expect-expand}, but this is similar to other calculations appearing in Section \ref{sec:clt}, for example the proof of Lemma \ref{lem:var-lead}).
\bet \label{thm:expect-bdry}
Let $f(x)$ be any regular test function with data $( \|f''\|_1^{-1}, C, c)$.  Let $W$ be a matrix of general Wigner-type.  Assume that there is a $c_1 >0$ such that $\|f''\|_1 \leq N^{1-c_1}$.  
Then there is a $c_2 >0$ such that
\begin{align}
&\ee[ \tr f (W) ] - N \int f(x) \rho (x) \d x \nonumber\\
=&  \frac{1}{ \pi } \int f(x) \left(   \frac{1}{2} \partial_x \Im\left[ \tr (  S \mb^2(x) )\right] + \frac{1}{4} \partial_x \Im \left[ \frac{1}{N^2} \sum_{i, j} s^{(4)}_{ij} ( \mb_i (x) \mb_j (x) )^2 \right] \right) \d x \nonumber\\
+ & \frac{1}{ \pi} \int f(x) \Im\left[  \tr ( (1 - S \mb(x)^2 )^{-1} S \mb'(x)  \mb(x) ) \right] \d x \nonumber \\
+ & \frac{1}{4 } f( \beta)  + \frac{1}{4 } f( \alpha )  + \O (N^{-c_2} ).
\end{align}
Note that the estimates \eqref{eqn:exp-kern-bd-1} and \eqref{eqn:exp-kern-bd-2} hold for the functions appearing above. 
\eet

\noindent{\bf Proof of Theorem \ref{thm:sev-expect}}. The result for Gaussian divisible ensembles follows in a straightforward manner from Theorems \ref{thm:homog}, \ref{thm:expect-bdry}, and Theorem 1.4 of \cite{meso}.  The removal of the Gaussian component is straightforward, and can be done in a similar manner as the removal done in the proof of Theorem \ref{thm:main} given in Section \ref{sec:proof-of-main}. 
\qed

\appendix

\section{Auxilliary CLT proofs} \label{sec:aux-clt}

\subsection{Proof of Lemma \ref{52}} \label{sec:Txy-error-2m}

We expand,
\begin{align}
\ee \left| \frac{1}{N} \sum_{i \neq y } s_{xi} Q_i [ G_{iy} (z) G_{iy} (w) ] \right|^2  &= \frac{1}{N^2} \sum_{i \neq y } s_{xi}^2 \ee[ |Q_i [ G_{iy} (z) G_{iy} (w)] |^2] \nonumber \\
&+ \frac{2}{N^2} \sum^{(y)}_{i \neq j } s_{ix} s_{jx} \ee[ Q_i [G_{iy} (z) G_{iy} (w) ] Q_j [\bar{G}_{jy} (z) \bar{G}_{jy} (w) ] ].
\end{align}
The first line is $\O (N^{-1} (N \eta)^{-2} N^{\eps })$ since the off-diagonal resolvent entries are
\beq
G_{iy} =\O ( (N \eta)^{-1/2}  N^{\eps/2} )
\eeq
with overwhelming probability by \eqref{eqn:entry-ll}. Using the identity
\beq
G_{ab} = \Gc_{ab} + \frac{ G_{ac} G_{cb}}{G_{cc}}
\eeq
valid for $a, b \neq c$ we have (recall $i \neq j$ and $i, j \neq y$),
\begin{align}
&\ee[ Q_i [G_{iy} (z) G_{iy} (w) ] Q_j [\bar{G}_{jy} (z) \bar{G}_{jy} (w) ] ] \nonumber \\
=& \ee \bigg\{ Q_i \left[ \left( \Gj_{iy} (z) + \frac{ G_{ij} (z) G_{jy} (z) }{ G_{jj} (z) } \right)\left( \Gj_{iy} (w) + \frac{ G_{ij} (w) G_{jy} (w) }{ G_{jj} (w) } \right) \right] \nonumber \\
&\times Q_j \left[ \left( \bar{G}^{(i)}_{jy} (z) + \frac{ \bar{G}_{ji} (z) \bar{G}_{iy} (z) }{ \bar{G}_{ii} (z) } \right)\left( \bar{G}^{(i)}_{jy} (w) + \frac{ \bar{G}_{ji} (w) \bar{G}_{iy} (w) }{ \bar{G}_{ii} (w) } \right) \right] \bigg\}.
\end{align}
Every off-diagonal resolvent entry contributes $\O ((N \eta)^{-1/2})$.  Since the indices $i, j$ and $y$ are all distinct, all resolvent entries not appearing in the denominator are off-diagonal.  If we expand out the product on the right side, we only need to consider terms that have less than $6$ off-diagonal resolvent entries.  The term with four entries has expectation $0$ since 
\beq
\ee[ Q_i (A) Q_j (B) ] = 0
\eeq
 if either $A$ is independent of the $j$th row of $W$ or if $B$ is independent of the $i$th row of $W$.  Similarly, the term with the product of five off-diagonal entries is, generically,
\beq
\ee[ Q_i [  \Gj_{iy} (z)  \frac{ G_{ij} (w) G_{jy} (w) }{ G_{jj} (w) }] Q_j [ \bar{G}^{(i)}_{jy} (z)\bar{G}^{(i)}_{jy} (w)  ] ]=0.
\eeq
This yields the claim. \qed

\subsection{Proof of Lemma \ref{lem:h12-calc}} \label{sec:h12-calc}

Note that under our assumptions, $\|\phi'\|_1 \leq C K$.  Denote $R = [-r, r]^2$. Using the fundamental theorem of calculus to write 
\beq
\left(\phi(x) - \phi(y)\right)^2 = \int_x^y\int_x^y \phi'(u)\phi'(v)\, \d u \, \d v,
\eeq
using symmetry to reduce the integral to one over $x<u < v  < y$, and switching the limits of integration, we have
\begin{align}
\int_{ R} \left( \frac{ \phi (x) - \phi (y) }{x-y} \right)^2 \d x \d y = 4 \int_{u < v} \phi'(u) \phi'(v) \int_{R \cap \{x < u, y > v \} } \frac{1}{ (x-y)^2} \d x \d y \d u \d v.
\end{align}
As long as $tM < r/2$, the integrand is non-zero only if $|v| \leq r/2$ and $|u| \leq r/2$, so for such $u, v$, with $u < v$,
\beq
\int_{R \cap \{ x < u, y > v \} } \frac{1}{ (x-y)^2 } \d x \d y = | \log (u-v) | + \O (1).
\eeq
Hence,
\beq
\int_{u < v} \phi'(u) \phi'(v) \int_{R \cap \{x < u, y > v \} } \frac{1}{ (x-y)^2} = \int_{u < v } \phi'(u) \phi'(v) |\log (u-v) |+ \O (1).
\eeq
Fix a small $1/2 > \delta >0$, to be optimized over later.  We split the region above into three, where $\delta t < |u-v| < t /\delta$ and the other two complementary regions.  Recall the indefinite integral $\int \log x\, \d x = x \log x -x$.  
We have
\begin{multline}
\int_{ 0 < v-u < \delta t} |\phi'(u) \phi'(v)| |\log (u-v) | \d u \d v \\ \leq \frac{K}{t} \int | \phi'(v)| \int_{0 < v-u < \delta t} | \log (v-u ) |  \d u \d v \leq C K^2 \delta | \log ( \delta t ) |.
\end{multline}
Now consider the region $|u-v| > t / \delta$.  Then either $|u| > t / (2 \delta)$ or $|v| > t / ( 2 \delta)$.  Using in this region that $| \log(u-v)| \leq C( |\log(t/\delta)|+1)$, we have
\begin{align}
\int_{|u-v| > t / \delta} | \phi'(u) \phi'(v) |& |\log (u-v) | \d u \d v\nonumber \\  &\leq C (  |\log(t/\delta)| +1)\int_{ |u-v| >t /\delta, |u| >t / (2 \delta ) } | \phi'(u) \phi'(v) | \d u \d v  \\
&\leq C K^2( |\log (t / \delta ) | +1) \int_{ |u| >t / (2 \delta )} t/u^2 \d u \leq C \delta K^2(1+| \log (t / \delta ) | ).
\end{align}
 In the second inequality, we used $\int_{\mathbb{R}} | \phi'(v)|\, \d v \leq C K$, which follows from \eqref{e:phiderivbound}.
For $t \delta < |u-v| \leq t / \delta$ we have
\beq
| \log (u-v ) | = | \log (t) | + \O ( | \log \delta |),
\eeq
and so
\begin{align}\label{e:previousthree}
\int_{ t \delta < u - v < t / \delta} \phi'(u) \phi' (v) | \log (u-v) | \d u \d v = \log (t) \int_{ t \delta < u - v < t / \delta} \phi'(u) \phi' (v) \d u \d v + \O ( K^2| \log ( \delta ) |).
\end{align}
Now, 
\beq
\int_{ |u-v| < t \delta } | \phi' (u) \phi' (v)| \d u \d v \leq K^2 \delta
\eeq
by using $| \phi'(u) | \leq K /t$ from \eqref{e:phiderivbound}.  Similarly,
\beq
\int_{ |u-v| > t / \delta} | \phi'(u) \phi'(v) | \d u \d v \leq K^2 \delta
\eeq
using that either $|u|$ or $|v| $ is larger than $t / (2 \delta)$ and the estimate $|\phi'(u) | \leq K t / u^2$.  
 Note,
\begin{align}
\int_{u < v} \phi'(u) \phi'(v) \, \d u \, \d v= \frac{1}{2},
\end{align}
so putting the previous three equations in \eqref{e:previousthree}, we find
\begin{align}
\int_{ R} \left( \frac{ \phi (x) - \phi (y) }{x-y} \right)^2 \d x \d y &= 2 | \log (t) | + \O ( (K+1)^2 \delta(  | \log ( \delta t )| + | \log (t/\delta ) | ) + | \log \delta | ).
\end{align}
We choose $\delta = 1 / | \log (t) |$. \qed

\subsection{Proof of Lemma \ref{lem:lss-expect}} \label{sec:lss-expect}

We begin with the Helffer--Sj{\"o}strand formula (see \cite[Section 11.2]{erdos2017dynamical}), which gives us the estimate
\begin{align}
\tr f(W) - N \int f(E) \rho (E) \d E &= \frac{1}{ 2 \pi} \int_{\Oma} ( \i y \chi (y) + \i ( f(x) + \i f'(x) y ) \chi' (y) ) \sum_i (G_{ii}(z) - \bm_i (z)) \d x \d y \nonumber \\
&+ \O (N^{\mfa+\eps-1} \|f''\|_1 ), \label{eqn:expect-a2}
\end{align}
that holds with overwhelming probability.  
By the cumulant expansion,
\begin{align}
z \ee[ G_{ii} ] +1 &= \sum_{a} \ee[ G_{ia} W_{ai} ] \nonumber \\
&= - \frac{1}{N} \sum_{a} s_{ia} \ee[ G_{ia}^2 + G_{ii} G_{aa} ] \nonumber \\
&+ \frac{1}{N} s_{ii} \ee[ G_{ii}^2] \nonumber  \\
&+ \frac{1}{2 N^{3/2}} \sum_{a} \sth_{ia} \ee[ \partial_{ia}^2 G_{ia} ] + \frac{1}{6 N^2} \sum_{a} \sfo_{ia} \ee[ \partial_{ia}^3 G_{ia} ] + \O(N^{-3/2+\eps} ).
\end{align}
We start with the last line.  First, for $i \neq a$ we see that by \eqref{eqn:entry-ll} we have with overwhelming probability
\beq
\partial_{ia}^2 G_{ia} = 6 G_{ia} \mb_i \mb_a + \O (N^{\eps} (N \eta)^{-1} ),
\eeq
so
\begin{align}
\frac{1}{2 N^{3/2}} \sum_{a} \sth_{ia} \ee[ \partial_{ia}^2 G_{ia} ]  &= \frac{3}{N^{3/2}} \sum_{a\neq i} \sth_{ia} \mb_i \mb_a \ee[G_{ia} ] + \O ( N^{\eps-3/2} \eta^{-1} ) = \O ( N^{\eps-3/2} \eta^{-1} ) 
\end{align}
using the isotropic local law \eqref{eqn:iso} in the final estimate. Similarly, for $i \neq a$, we have with overwhelming probability
\beq
\partial_{ia}^3 G_{ia} =- 6 \mb_i^2 \mb_a^2 + \O(N^{\eps} (N \eta)^{-1/2} ),
\eeq
so
\beq
\frac{1}{6 N^2} \sum_{a} \sfo_{ia} \ee[ \partial_{ia}^3 G_{ia} ] =- \frac{1}{N^2} \sum_{a} \sfo_{ia} \mb_i^2 \mb_a^2 + \O( N^{\eps-1} (N \eta)^{-1/2} ).
\eeq
Clearly, by \eqref{eqn:entry-ll},
\beq
 \frac{1}{N} s_{ii} \ee[ G_{ii}^2] = \frac{1}{N} s_{ii} \mb_i^2 + \O (N^{\eps-1} ( N \eta)^{-1/2} ).
\eeq
Next, using \eqref{eqn:tr-ll},
\begin{align}
- \frac{1}{N} \sum_{a} s_{ia} \ee[ G_{ii} G_{aa} ] =& - \frac{1}{N} \sum_a s_{ia} \ee[ (G_{ii} - \mb_i )] \mb_a - \frac{1}{N} \sum_{a} s_{ia} \mb_i \ee[ (G_{aa} - \mb_a ) \mb_i ] \\
- &\frac{1}{N} \sum_a s_{ia} \mb_i \mb_a + \O (N^{\eps-1} N^{-1/2} \eta^{-3/2} ).
\end{align}
We therefore see we have proven
\begin{align}
- \frac{1}{ \mb_i} \ee[ G_{ii} - \mb_i ] &= - \frac{1}{N} \sum_a \mb_i s_{ia} \ee[ G_{aa} - \mb_a ] + \frac{1}{N} s_{ii} \mb_i^2 \nonumber \\
&- \frac{1}{N^2} \sum_a \sfo_{ia} \mb_i^2 \mb_a^2 - \ee[ T_{ii} (z, z) ] + \O (N^{\eps-3/2} \eta^{-3/2} ).
\end{align}
For $T(z, z)$ we see from Proposition \ref{prop:T-self} and Lemma \ref{lem:stab-bulk} that, 
\beq
\ee[ T_{ii} (z, z) ] = [ (1- S \mb^2 )^{-1} S \mb^2 ]_{ii} + \O(N^{\eps-3/2} \eta^{-3/2} ).
\eeq
Therefore,
\begin{align}
\ee[ G_{ii} ] - \mb_i (z) &=- \sum_j (1 - \mb^2 S)^{-1}_{ij} (S \mb^3)_{jj} + \sum_j (1- \mb^2 S)^{-1}_{ij} [ \mb (1 - S \mb^2 )^{-1} S \mb^2 ]_{jj} \nonumber \\
+& \frac{1}{N^2} \sum_{j,a} (1 - \mb^2 S)^{-1}_{ij} \sfo_{ja} \mb_j^3 \mb_a^2 + \O (N^{\eps-3/2} \eta^{-3/2} ). \label{eqn:expect-a1}
\end{align}
Recalling
\beq
\sum_{i} (1 - \mb^2 S)^{-1}_{ij} = \frac{ \mb_j'(z)}{ \mb^2_j (z) },
\eeq
and we now embark on the proof of the lemma.

\noindent{\bf Proof of Lemma \ref{lem:lss-expect}}.  The first estimate is a consequence of Lemma \ref{lem:H}, and \eqref{eqn:expect-a1} and \eqref{eqn:expect-a2}.  For the second, observe that by Proposition \ref{prop:basic-m}, Proposition \ref{prop:stab-elements}, and Lemma \ref{lem:stab-bulk}, the function on the second line of \eqref{eqn:expect-expand} is a bounded holomorphic function where $\bar{\del}_z \tilf (z) \neq 0$.  The contributions from $f'(x)$ and $f(x)$ are clearly bounded.  For the contribution from $f''(x)$, we argue by integration by parts as in the proof of Lemma \ref{lem:var-sublead}.  Let $H(z)$ be the function on the second line of \eqref{eqn:expect-expand}.  Then,
\begin{align}
\left| \int_{\Oma} f''(x) y \chi (y) H(z) \d x \d y \right| = \left| \int_{\Oma} f'(x) y \chi (y) H'(z) \d x \d y \right| \leq C
\end{align}
as $|H'(z) y | \leq C$.  We conclude the desired estimate. \qed

\section{Free convolution} \label{a:free-1}

\subsection{Perturbations of variance matrix \texorpdfstring{$S$}{S}}
We require the following result about the behavior of solutions of two different quadratic vector equations with different $S$ matrices.  It is a consequence of Proposition 10.1(c) of  \cite{alt2018dyson}. 
\bep \label{prop:stab-1}
Let $S$ satisfy Assumptions \ref{it:ass-1} and \ref{it:ass-2}.  Let $\hatS$ be another symmetric matrix with positive entries.   Define
\beq
\eps := \max_{i, j} N | S_{ij} - \hatS_{ij} |.
\eeq
Denote the solutions of the associated quadratic vector equations by $\bm$ and $\hatmb$.  There exists $\eps_* >0$ such that the following holds. If $\eps \leq \eps_*$, then
\beq
\sup_{\substack{z \in \hh\\ i \in \unn{1}{N}}} | \bm_i (z) - \hatmb_i (z) | \leq C \eps^{1/2}.
\eeq
\eep

As a consequence, we have the following.
\bep \label{prop:stab-a1}
Let $S$ be a variance matrix satisfying Assumptions \ref{it:ass-1} and \ref{it:ass-2}.  There is an $\eps_* >0$ such that for any $\hatS$ satisfying 
\beq
\eps = \max_{i, j} N |S_{ij} - \hatS_{ij} |
\eeq
with $\eps < \eps_*$ we have that $\hatS$ also satisfies Assumptions \ref{it:ass-1} and \ref{it:ass-2}.  Moreover, if the support of the spectral measure associated with $\hatS$ is denoted by $[\hat{\alpha}, \hat{\beta} ]$ then $|\hat{\alpha} - \alpha| + | \hat{\beta} - \beta| \leq C \eps$.
\eep
\proof 
From Proposition \ref{prop:stab-1} we see that $\hatmb$ satisfies Assumption \ref{it:ass-1}.  Hence, the characterization of the associated density of states of  \cite[Theorem 4.1]{qve} applies.  In particular, around the respective extremal edges of $\rho$ and $\hat{\rho}$, they both have constant order neighborhoods in which they have square-root behavior \cite[(4.5d)]{qve}.  Since $\| \hatmb - \mb \|_\infty \leq C \eps^{1/2}$ we see that if $\eps >0$ is small enough, the extremal edges are within $\O ( \eps)$ of each other.  Then, since $\rho (E) \geq c$ for $E$ away from the edges, we get the same for $\hat{\rho}$ if $\eps $ is small enough, which verifies Assumption~\ref{it:ass-2}. \qed

\subsection{Comparison of free convolution and variance matrix flow} \label{a:free-2}

Let $W$  a matrix of general Wigner-type, with density of states $\rho(E)$ and variance matrix $S$.  Consider the two flows
\beq
S_t := S + t e e^T + (t/N) \1, \qquad \hat{S}_t := S + t e e^T,
\eeq
where $e$ is the constant vector normalized so $\|e\|_2 =1 $.  Then, $S_t$ is the matrix of variances of
\beq
W_t := W + \sqrt{t} G,
\eeq
where $G$ is a GOE matrix independent of $W$.  On the other hand, let $\hatp_t$ and $\hatm_t$ be  
the spectral measure and Stieltjes transform associated with the matrix $\hat{S}_t$.  
Differentiating the QVE for $\hat{S}_t$, we find
\beq\label{b3}
\frac{\partial_t \hatm_{t, i} }{ \hatm_{t,i}^2} = (\hat{S}_t \del_t \hatm_t )_i + \hatm_t.
\eeq
Rearranging \eqref{b3} gives
\beq
\del_{t} \hatm_{t, i} = \hatm_{t} \left( \frac{1}{ 1 - \hatm_t^2 \hat{S}_t } \hatm_t^2 \right)_i = \hatm_{t} \del_z \hatm_{t, i},
\eeq
and so 
\beq
\del_t \hatm_{t} = \hatm_{t} \del_z \hatm_t.
\eeq
Hence $\hatm_{t}$ is also the free convolution of $\hatp$ with the semicircle distribution at time $t$.  The following lemma compares the two measures for short times $t$.

\bel \label{lem:diag-S}
Consider two variance matrices $S$ and $\hat{S}$.  Suppose that $S$ obeys Assumptions \ref{it:ass-1} and \ref{it:ass-2}.  Suppose that $S_{ij} = \hat{S}_{ij}$ for $i \neq j$, and
\beq
\max_i | \hat{S}_{ii} - S_{ii} | \leq \frac{ \eps}{N}.
\eeq
Then there is $\eps_* >0$ such that if $\eps < \eps_*$ we have that the following holds.   The spectral edges are within $\O ( \eps /N)$ of each other, and $\hatS$ satisfies Assumptions \ref{it:ass-1} and \ref{it:ass-2}.  
The densities satisfy, in a neighbourhood of their support,
\beq \label{eqn:a-dens}
| \rho (E) - \hatp (E) | \leq \frac{C \eps}{N} \left( \frac{1}{ \min\{ |E- \alpha|, |E- \beta | \} + \eps N^{-1} } \right)^{1/2}.
\eeq
and so
\beq
\left| \int f(x) \rho (E) - \int f(x) \hatp (E) \right| \leq \|f\|_\infty C \frac{\eps}{N},
\eeq
as well as
\beq
\left| \gamma_i - \hat{\gamma}_i \right| \leq \frac{C \eps}{N}
\eeq
for the $N$-quantiles.  Further,
\beq
 \|\bm (z)- \hatbm (z) \|_\infty \leq C \frac{\eps}{N \eta}.
\eeq
\eel
\proof By Propositions \ref{prop:stab-1} and \ref{prop:stab-a1}, $\hat{S}$ also obeys Assumptions \ref{it:ass-1} and \ref{it:ass-2}.  Therefore, $\hatbm$ is bounded and we can apply Proposition \ref{prop:mstability} with $\bd(z)$ satisfying
\beq
\sup_z \| \bd (z)\|_\infty \leq \frac{ C \eps}{N},
\eeq
and $\bg (z) = \hatbm (z)$.  Note that again by Proposition \ref{prop:stab-1} that $\| \hatbm(z) - \bm(z) \|_\infty \leq (\eps N^{-1} )^{1/2}$ so we can ignore the indicator functions in the statements of Proposition \ref{prop:mstability}.  It follows immediately that in the spectral bulk, $\| \hatbm(z) - \bm(z) \|_\infty \leq C \eps N^{-1}$.   Near the edges, say near $\beta$, we get for $\eta $ such that $ |E- \beta| + \eta \geq C \eps N^{-1}$ that
\beq
\| \hatbm (z) - \bm (z) \|_\infty \leq \frac{C \eps}{ N ( \sqrt{ |E- \beta| + \eta } )}.
\eeq
For $|E- \beta| + \eta \leq C \eps N^{-1}$ we have
\beq
\| \hatbm (z) - \bm (z) \|_\infty \leq \sqrt{ \frac{ \eps}{N} } \leq \frac{C \eps}{ N ( \sqrt{ |E- \beta| + \eta } )}.
\eeq
We see the spectral edges are within $\O ( \eps /N)$ of each other, and conclude the estimate \eqref{eqn:a-dens}.  Integrating the densities of states from the left spectral edges, we see that the eigenvalue counting functions satisfy
\beq
\left| n(E) - \hat{n}(E) \right| \leq C \frac{\eps}{N} \left( \frac{\eps}{N} + |E-\alpha| \right)^{1/2},
\eeq
and the estimates for the quantiles with $i \leq N/2$ follow.  The estimates for $i \geq N/2$ are proven via integration from the right spectral edge.  \qed

Recall also the definition of Dyson Brownian motion.  Let $\{ \lambda_i (0) \}_{i=1}^N$ be some eigenvalue distribution and for Brownian motions independent of $\{ \lambda_i (0) \}_{i=1}^N$ define
\beq \label{eqn:a-dbm}
\d \lambda_i (t) = \sqrt{ \frac{2}{N} } \d B_i (t) + \frac{1}{N} \sum_{j \neq i } \frac{1}{ \lambda_i (t) - \lambda_j (t) } \d t.
\eeq
Note that the particles $\{ \lambda_i (t) \}_{i=1}^N$ have the same distribution as the eigenvalues of $W + B_t$ where $B_t$ is a symmetric matrix of Brownian motions (so that $B_1$ has the same distribution as the GOE).  
We can establish the following basic proposition.
\bep
Let $W$ be a matrix of general Wigner-type satisfying Assumptions \ref{it:ass-1} and \ref{it:ass-2}, with variance matrix $S$ and density of states $\rho$.  Denote the free convolution of $\rho$ with the semicircle distribution at time $t$ by $\rho_t$, which is the density of states associated with $S_t := S + t e e^T$ where $e$ is the constant vector with $\|e\|_2 =1$.  Consider the measures associated with the matrix $\hat{S}_t = S + te e^T + t/N$, denoted by $\hatp_t$.  There is an $\eps_* > 0$ such that if $t < \eps_*$, the matrices $S_t$ and $\hat{S}_t$ both satisfy Assumptions \ref{it:ass-1} and \ref{it:ass-2}.  The eigenvalues of the matrices
\beq
W_t := W + \sqrt{t} G
\eeq
satisfy the local laws and optimal rigidity estimates as in Theorems \ref{thm:local-law} and \ref{thm:rigidity}.  Furthermore, the particles in \eqref{eqn:a-dbm} obey
\beq \label{eqn:dbm-quant}
\sup_{0 \leq t \leq \eps_*  } | \lambda_i (t) - \gamma_i (t) | \leq \frac{N^{\eps}}{N^{2/3} ( \min\{ i, (N+1-i ) \} )^{1/3} }
\eeq
for either choice of the quantiles $\gamma_i(t)$ or $\hat{\gamma}_i (t)$, with overwhelming probability.
\eep
\proof All of the statements are clear from Lemma \ref{lem:diag-S} and Theorems \ref{thm:local-law} and \ref{thm:rigidity} applied to $W_t$, except for the fact that the estimate \eqref{eqn:dbm-quant} which holds uniformly in $t$.  From Theorems \ref{thm:local-law} and \ref{thm:rigidity}, the estimates hold over a mesh of $t_n$ with $|t_n -t_{n-1} | \leq N^{-100}$ with overwhelming probability.  The extension from this mesh to all $t$ can be proven either as in Appendix B of \cite{fixed}, or using Weyl's inequality together with the fact that the eigenvalues have the same distribution as $W + B_t$ for $B_t$ a matrix-valued Brownian motion; that is, 
\beq
\sup_{t_i \leq t \leq t_{i+1}} |\lambda_i(t) - \lambda_i (t_i) | \leq \sup_{t_i \leq t \leq t_{i+1}} \|B_{t} - B_{t_i} \|_{\ell^2 \to \ell^2}.
\eeq  The  latter norm can be controlled via large deviations bounds for Brownian motion. \qed

\subsection{Auxiliary stability}\label{a:stab}

The following stability result is a consequence of Theorem 4.2 and Proposition 4.3 of \cite{univ-wig-type}. 
\bep\label{prop:mstability}
Let $S$ satisfy Assumptions \ref{it:ass-1} and \ref{it:ass-2}.  There exist constants $C ,c, \delta_*, \lambda_*  >0$ and a vector-valued function $\bs (z)$ such that the following holds.  Let $\bg : \hh \to \hh^N$ and $\bd : \hh \to \cc^N$ satisfy
\beq
- \frac{1}{ \bg_i (z) } = z + \sum_{j=1}^N S_{ij} \bg_j (z) + \bd_i (z).
\eeq
Define
\beq
\Theta(z) =\left| \frac{1}{N} \sum_i \bar{ \bs_i} (z) (\bg_i (z) -  \bm_i (z) ) \right|.
\eeq
Then
\beq \label{gm}
\|\bg - \bm \|_\infty \1 ( \| \bg  - \bm \|_\infty \leq \lambda_* ) \leq C ( \Theta + \|\bd\|_\infty)
\eeq
and
\beq\label{cubic}
\left| \Theta^3 + \pi_2 \Theta^2 + \pi_1 \Theta \right| \1 ( \| \bg  - \bm \|_\infty \leq \lambda_* ) \leq C\left( \|\bd\|_\infty + \|\bd\|_\infty^2\right)
\eeq
hold for some coefficients $\pi_1(z)$ and $\pi_2 (z)$ that may depend on $S$ and $\bg$.  

In any compact subset of $\cc$,
\beq
| \pi_2(E + \i \eta) | \asymp 1 \text{  if  } E \in [\alpha - \delta_*, \beta+\delta_*],
\eeq
and $| \pi_2(z) | \leq C $ otherwise.
We also have
\beq
|\pi_1(E + \i \eta)| \asymp1 \text{  if  } E \notin [\alpha - \delta_* , \alpha + \delta_*] \cup [ \beta - \delta_*, \beta + \delta_*].
\eeq
Further, if $|E- \beta| \leq \delta_*$, then
\beq
| \pi_1 (z) | \asymp \sqrt{| E-\beta | + \eta },
\eeq
with a similar estimate for $|E-\alpha| \leq \delta_*$. We also have $\| \bs \|_\infty \leq C$.
\eep

\section{Green's function comparison for linear spectral statistics} \label{a:gfct-lss}

Let $f$ be a regular test function with data $(\eta_*, C', c')$ and $W$ a matrix of general Wigner-type satisfying Assumptions \ref{it:ass-1} and \ref{it:ass-2}.  Assume $\eta_* = N^{\delta_*-1}$.  Fix a $\delta_* > \mfb >0$ and quasi-analytic extension $\tilf (z)$ with $\chi (y)$ satisfying
\beq
\chi (y) = \begin{cases} 1  & |y| < 1/2 \\ 0 & |y| >1 \end{cases}.
\eeq
We have the Helffer--Sj{\"o}strand formula
\beq
 \tr f (W)  - N\int f (x) \rho (x) \d x = \frac{1}{ \pi} \int_{ |y| > N^{-\mfb} \eta_* } ( \bar{ \del}_z \tilf (z) ) N ( m_N (z) - m (z) ) \d x \d y + \O (N^{\eps-\mfb} ),
\eeq
with overwhelming probability, for any $\eps >0$.  Denote
\beq
X :=  \frac{1}{ \pi} \int_{ |y| > N^{-\mfb} \eta_* } ( \bar{ \del}_z \tilf (z) ) N ( m_N (z) - m (z) ) \d x \d y.
\eeq
Let $\Adel$ be the following set of matrices:
\beq
\Adel := \{ W : |G_{ij} (z) | \leq C, \mbox{ for all } \eta > N^{\delta-1} \}.
\eeq
The following is a straightforward consequence of the Ward identity.
\bel
For $\eta > N^{\delta-1}$, we have for any $k$ that
\beq
\left| \del_{ab}^k \tr \frac{1}{ W-z} \right| \leq \frac{C_k}{ \eta}
\eeq
for any $ W \in \Adel$. 
\eel
\proof We only prove $k=1$, higher $k$ being similar.  We have
\beq
(1+\delta_{ab} )\del_{ab} \sum_i G_{ii} (z) = 2 \sum_i G_{ia} (z) G_{ib} (z).
\eeq
By the Ward identity,
\beq
\sum_i |G_{ia} (z) |^2 = \frac{\Im[ G_{aa} (z) ]}{ \eta}.
\eeq
This proves the claim. \qed

From the above, the following is easily proven.
\bep
For $W \in \Adel$ with $\delta < \delta_* - \mfb$ we have,
\beq
\left| \del_{ab}^k X \right| \leq C_k \log(N),
\eeq
for any $k$.
\eep
\proof Clearly,
\begin{align}
\left| \del_{ab}^k X \right| \leq \left| \int_{y > N^{-\mfb} \eta_* } f''(x) \chi(y) y \del_{ab}^k m_N ( x + \i y ) \d x \d y\right|  + C_k.
\end{align}
  By the Cauchy integral formula,
\beq
\left| \partial_z \del_{ab}^k m_N ( x + \i y ) \right| \leq \frac{C_k}{ y^2}.
\eeq
Hence,
\begin{align}
&\left| \int_{ y > N^{-\mfb} \eta_* } f''(x) \chi(y) y \del_{ab}^k m_N ( x + \i y ) \d x \d y\right| \nonumber\\
=& \left| \int_{ y > N^{-\mfb} \eta_* } f' (x) \chi (y)y \del_z \del_{ab}^k m_N ( x + \i y ) \d x \d y \right|  \nonumber\\
\leq & C_k \int_{ y > N^{-\mfb} \eta_* } \frac{ |f'(x) | }{ |y| } \d x \d y \leq C_k \log(N).
\end{align}
This yields the claim. \qed

With the above estimates, one can prove the following in a standard way using the four moment method. See for example, the argument given in Section 16 of \cite{erdos2017dynamical} or the argument we gave in Section \ref{sec:proof-of-main}. 
\bet \label{thm:lss-gfct}
Let $f$ be a regular test function and $W$ and $X$ two matrices of general Wigner-type obeying Assumptions \ref{it:ass-1} and \ref{it:ass-2} such that,
\beq
\ee[ W^k_{ij} ] = \ee[ X^k_{ij} ] , \qquad k=1, 2, 3
\eeq
and 
\beq
\left| \ee[ W^4_{ij} ] - \ee[ X^4_{ij} ] \right| \leq \frac{ C t}{N^2}.
\eeq
Then for any $F \in C^5$ and $\eps >0$ we have,
\beq
\left| \ee[ F ( \tr f (W) )] - \ee[ F ( \tr f (X) ) ] \right| \leq  ||F||_{C^5} N^{\eps} t.
\eeq
\eet

\subsection{Proofs of Theorem \ref{thm:meso-clt-gen} and \ref{thm:global-clt-gen}.}  \label{a:clts}
Consider first the case $f = g (N^{\omst } (x- E_0 ))$ with $g$ of compact support and $W = X + \sqrt{t_0} G$ with $t_0 = N^{-\omega_0}$ for $\omega_0 < \omst$.  For such Gaussian divisible ensembles, the result follows from Corollary 4.3 of \cite{huang2019rigidity}.  The result for general matrices is then a consequence of Theorem \ref{thm:lss-gfct}.  We now consider the case where $f = h (N^{\omst} ( x -E_0))$ where $h$ is as in the statement of theorem.  The result can be deduced for Gaussian divisible ensembles with sufficiently large Gaussian component in the same manner as Theorem \ref{thm:final-clt-homog}.  Again, the result for general matrices follows from Theorem \ref{thm:lss-gfct}. Theorem \ref{thm:global-clt-gen} is a direct consequence of Proposition \ref{prop:char-func}.

{\footnotesize

\bibliography{mybib}{}

\begin{thebibliography}{10}

\bibitem{adhikari2019linear}
K.~Adhikari, I.~Jana, and K.~Saha.
\newblock Linear eigenvalue statistics of random matrices with a variance
  profile.
\newblock {\em Preprint, arXiv:1901.09404}, 2019.

\bibitem{univ-wig-type}
O.~Ajanki, L.~Erd{\H{o}}s, and T.~Kr{\"u}ger.
\newblock Universality for general {W}igner-type matrices.
\newblock {\em Probab. Theory Related Fields}, 169(3):667--727, 2017.

\bibitem{qve}
O.~Ajanki, L.~Erd{\H{o}}s, and T.~Kr{\"u}ger.
\newblock Quadratic vector equations on complex upper half-plane.
\newblock {\em Mem. Amer. Math. Soc.}, 261(1261), 2019.

\bibitem{sing}
O.~Ajanki, T.~Kr{\"u}ger, and L.~Erd{\H{o}}s.
\newblock Singularities of solutions to quadratic vector equations on the
  complex upper half-plane.
\newblock {\em Comm. Pure Appl. Math}, 70(9):1672--1705, 2017.

\bibitem{alt2018dyson}
J.~Alt, L.~Erdos, and T.~Kr{\"u}ger.
\newblock The {D}yson equation with linear self-energy: spectral bands, edges
  and cusps.
\newblock {\em Doc. Math.}, 25:1421--1539, 2020.

\bibitem{az}
G.~W. Anderson and O.~Zeitouni.
\newblock A {CLT} for a band matrix model.
\newblock {\em Probab. Theory Related Fields}, 134(2):283--338, 2006.

\bibitem{bao2021quantitative}
Z.~Bao and Y.~He.
\newblock Quantitative {CLT} for linear eigenvalue statistics of {W}igner
  matrices.
\newblock {\em Preprint, arXiv:2103.05402}, 2021.

\bibitem{benaych2014central}
F.~Benaych-Georges, A.~Guionnet, and C.~Male.
\newblock Central limit theorems for linear statistics of heavy tailed random
  matrices.
\newblock {\em Comm. Math. Phys.}, 329(2):641--686, 2014.

\bibitem{semi-lec}
F.~Benaych-Georges and A.~Knowles.
\newblock Lectures on the local semicircle law for {W}igner matrices.
\newblock {\em Panoramas et Synth{\`e}ses}, 53, 2016.

\bibitem{benaych2016fluctuations}
F.~Benaych-Georges and A.~Maltsev.
\newblock Fluctuations of linear statistics of half-heavy-tailed random
  matrices.
\newblock {\em Stochastic Process. Appl.}, 126(11):3331--3352, 2016.

\bibitem{biane}
P.~Biane.
\newblock On the free convolution with a semi-circular distribution.
\newblock {\em Indiana Univ. Math. J.}, pages 705--718, 1997.

\bibitem{bourgade2018extreme}
P.~Bourgade.
\newblock Extreme gaps between eigenvalues of {W}igner matrices.
\newblock {\em J. Eur. Math. Soc., to appear}, 2021.

\bibitem{fixed-wig}
P.~Bourgade, L.~Erd{\H{o}}s, H.-T. Yau, and J.~Yin.
\newblock Fixed energy universality for generalized {W}igner matrices.
\newblock {\em Comm. Pure Appl. Math.}, 69(10):1815--1881, 2016.

\bibitem{bourgade2019gaussian}
P.~Bourgade and K.~Mody.
\newblock Gaussian fluctuations of the determinant of {W}igner matrices.
\newblock {\em Electron. J. Probab.}, 24, 2019.

\bibitem{costin1995gaussian}
O.~Costin and J.~L. Lebowitz.
\newblock Gaussian fluctuation in random matrices.
\newblock {\em Phys. Rev. Lett.}, 75(1):69, 1995.

\bibitem{dallaporta2011note}
S.~Dallaporta and V.~Vu.
\newblock A note on the central limit theorem for the eigenvalue counting
  function of {W}igner matrices.
\newblock {\em Electron. Comm. Probab.}, 16:214--322, 2011.

\bibitem{dyson1}
F.~Dyson.
\newblock Statistical theory of the energy levels of complex systems, {I},
  {II}, and {III}.
\newblock {\em J. Math. Phys.}, 3(1):140--156, 157--165, 166--175, 1962.

\bibitem{dyson2}
F.~Dyson.
\newblock Correlations between eigenvalues of a random matrix.
\newblock {\em Comm. Math. Phys.}, 19(3):235--250, 1970.

\bibitem{mde-notes}
L.~Erd\H{o}s.
\newblock The matrix {D}yson equation and its applications for random matrices.
\newblock {\em Preprint, arXiv:1903.10060}, 2019.

\bibitem{erdos2020functional}
L.~Erd{\H{o}}s, G.~Cipolloni, and D.~Schr{\"o}der.
\newblock Functional central limit theorems for {W}igner matrices.
\newblock {\em Preprint, arXiv:2012.13218}, 2020.

\bibitem{diffusion}
L.~Erd{\H{o}}s, A.~Knowles, H.-T. Yau, and J.~Yin.
\newblock Delocalization and diffusion profile for random band matrices.
\newblock {\em Comm. Math. Phys.}, 323(1):367--416, 2013.

\bibitem{erdos2010bulk}
L.~Erd{\H{o}}s, S.~P{\'e}ch{\'e}, J.~A. Ramirez, B.~Schlein, and H.-T. Yau.
\newblock Bulk universality for {W}igner matrices.
\newblock {\em Comm. Pure Appl. Math.}, 63(7):895--925, 2010.

\bibitem{local-relax}
L.~Erd{\H{o}}s, B.~Schlein, and H.-T. Yau.
\newblock Universality of random matrices and local relaxation flow.
\newblock {\em Invent. Math.}, 185(1):75--119, 2011.

\bibitem{localrelaxation}
L.~Erd{\H{o}}s, B.~Schlein, H.-T. Yau, and J.~Yin.
\newblock The local relaxation flow approach to universality of the local
  statistics for random matrices.
\newblock {\em Ann. Inst. Henri Poincar{\'e} Probab. Stat.}, 48(1):1--46, 2012.

\bibitem{es}
L.~Erd{\H{o}}s and K.~Schnelli.
\newblock Universality for random matrix flows with time-dependent density.
\newblock {\em Ann. Inst. Henri Poincar{\'e} Probab. Stat.}, 53(4):1606--1656,
  2017.

\bibitem{erdos2017dynamical}
L.~Erd\"os and H.-T. Yau.
\newblock A dynamical approach to random matrix theory.
\newblock {\em Courant Lecture Notes in Mathematics}, 28, 2017.

\bibitem{yaubernoulli}
L.~Erdos, H.-T. Yau, and J.~Yin.
\newblock Universality for generalized {W}igner matrices with {B}ernoulli
  distribution.
\newblock {\em J. Comb.}, 2(1):15--81, 2010.

\bibitem{erdos2012rigidity}
L.~Erd{\H{o}}s, H.-T. Yau, and J.~Yin.
\newblock Rigidity of eigenvalues of generalized {W}igner matrices.
\newblock {\em Adv. Math.}, 229(3):1435--1515, 2012.

\bibitem{FR}
P.~J. Forrester and E.~M. Rains.
\newblock Interrelationships between orthogonal, unitary and symplectic matrix
  ensembles.
\newblock {\em Random matrix models and their applications}, 40:171--207, 2001.

\bibitem{gaudin1}
M.~Gaudin.
\newblock Sur la loi limite de l'espacement des valeurs propres d'une matrice
  al{\'e}atoire.
\newblock {\em Nucl. Phys.}, 25:447--458, 1961.

\bibitem{guionnet}
A.~Guionnet.
\newblock Large deviations upper bounds and central limit theorems for
  non-commutative functionals of {G}aussian large random matrices.
\newblock {\em Ann. Inst. Henri Poincar{\'e} Probab. Stat.}, 38(3):341--384,
  2002.

\bibitem{gust}
J.~Gustavsson.
\newblock Gaussian fluctuations of eigenvalues in the {GUE}.
\newblock {\em Ann. Inst. Henri Poincar{\'e} Probab. Stat.}, 41(2):151--178,
  2005.

\bibitem{he2020fluctuations}
Y.~He.
\newblock Bulk eigenvalue fluctuations of sparse random matrices.
\newblock {\em Ann. Appl. Probab.}, 30(6):2846--2879, 2020.

\bibitem{heknowles}
Y.~He and A.~Knowles.
\newblock Mesoscopic eigenvalue statistics of {W}igner matrices.
\newblock {\em Ann. Appl. Probab.}, 27(3):1510--1550, 2017.

\bibitem{huang2019rigidity}
J.~Huang and B.~Landon.
\newblock Rigidity and a mesoscopic central limit theorem for {D}yson
  {B}rownian motion for general $\beta$ and potentials.
\newblock {\em Probab. Theory Related Fields}, 175(1):209--253, 2019.

\bibitem{kato}
T.~Kato.
\newblock {\em Perturbation theory for linear operators}, volume 132.
\newblock Springer Science \& Business Media, 2013.

\bibitem{KM}
A.~Kuijlaars and K.~McLaughlin.
\newblock Generic behavior of the density of states in random matrix theory and
  equilibrium problems in the presence of real analytic external fields.
\newblock {\em Comm. Pure. Appl. Math.}, 53(6):736--785, 2000.

\bibitem{comparison}
B.~Landon, P.~Lopatto, and J.~Marcinek.
\newblock Comparison theorem for some extremal eigenvalue statistics.
\newblock {\em Ann. Probab.}, 48(6):2894--2919, 2020.

\bibitem{meso}
B.~Landon and P.~Sosoe.
\newblock Applications of mesoscopic {CLT}s in random matrix theory.
\newblock {\em Ann. Appl. Probab.}, 30(6):2769--2795, 2020.

\bibitem{fixed}
B.~Landon, P.~Sosoe, and H.-T. Yau.
\newblock Fixed energy universality of {D}yson {B}rownian motion.
\newblock {\em Adv. Math.}, 346:1137--1332, 2019.

\bibitem{convergence}
B.~Landon and H.-T. Yau.
\newblock Convergence of local statistics of {D}yson {B}rownian motion.
\newblock {\em Comm. Math. Phys.}, 355(3):949--1000, 2017.

\bibitem{landonedge}
B.~Landon and H.-T. Yau.
\newblock Edge statistics of dyson brownian motion.
\newblock {\em Preprint, arXiv:1712.03881}, 2017.

\bibitem{li-xu}
Y.~Li and Y.~Xu.
\newblock On fluctuations of global and mesoscopic linear eigenvalue statistics
  of generalized {W}igner matrices.
\newblock {\em preprint, arXiv:2001.08725}, 2020.

\bibitem{lodhia2020covariance}
A.~Lodhia and A.~Maltsev.
\newblock Covariance kernel of linear spectral statistics for half-heavy tailed
  {W}igner matrices.
\newblock {\em Preprint, arXiv:2010.04219}, 2020.

\bibitem{lodhiasimm}
A.~Lodhia and N.~J. Simm.
\newblock Mesoscopic linear statistics of {W}igner matrices.
\newblock {\em preprint, arXiv:1503.03533}, 2015.

\bibitem{lytovapastur}
A.~Lytova and L.~Pastur.
\newblock Central limit theorem for linear eigenvalue statistics of random
  matrices with independent entries.
\newblock {\em Ann. Probab.}, 37(5):1778--1840, 2009.

\bibitem{mehta1}
M.~Mehta.
\newblock A note on correlations between eigenvalues of a random matrix.
\newblock {\em Comm. Math. Phys.}, 20(3):245--250, 1971.

\bibitem{mehta}
M.~Mehta.
\newblock {\em Random Matrices}.
\newblock Academic Press, 2004.

\bibitem{gm1}
M.~L. Mehta and M.~Gaudin.
\newblock On the density of eigenvalues of a random matrix.
\newblock {\em Nucl. Phys.}, 18:420--427, 1960.

\bibitem{orourke}
S.~O'Rourke.
\newblock Gaussian fluctuations of eigenvalues in {W}igner random matrices.
\newblock {\em J. Stat. Phys.}, 138(6):1045--1066, 2010.

\bibitem{pastur}
L.~A. Pastur.
\newblock The spectrum of random matrices.
\newblock {\em Teoret. Mat. Fiz.}, 10(1):102--112, 1972.

\bibitem{potters2020first}
M.~Potters and J.-P. Bouchaud.
\newblock {\em A First Course in Random Matrix Theory: For Physicists,
  Engineers and Data Scientists}.
\newblock Cambridge University Press, 2020.

\bibitem{rogershi1993}
L.~Rogers and Z.~Shi.
\newblock Interacting brownian particles and the {W}igner law.
\newblock {\em Probab. Th. Rel. Fields}, 95:555--570, 1993.

\bibitem{shcherbina}
M.~Shcherbina.
\newblock Central limit theorem for linear eigenvalue statistics of the
  {W}igner and sample covariance random matrices.
\newblock {\em Zh. Mat. Fiz. Anal. Geom.}, 7(2), 2011.

\bibitem{taovu-conj}
T.~Tao and V.~Vu.
\newblock Random matrices: Localization of the eigenvalues and the necessity of
  four moments.
\newblock {\em Acta Math. Vietnam.}, 36(2), 2010.

\bibitem{tao2010random}
T.~Tao and V.~Vu.
\newblock Random matrices: Universality of local eigenvalue statistics up to
  the edge.
\newblock {\em Comm. in Math. Phys.}, 298(2):549--572, 2010.

\bibitem{tao2011random}
T.~Tao and V.~Vu.
\newblock Random matrices: Universality of local eigenvalue statistics.
\newblock {\em Acta Math.}, 206(1):127--204, 2011.

\bibitem{wig2}
E.~Wigner.
\newblock Characteristic vectors of bordered matrices infinite dimensions.
\newblock {\em Ann. of Math}, 62(3):548--564, 1955.

\bibitem{wig1}
E.~Wigner.
\newblock On the distribution of the roots of certain symmetric matrices.
\newblock {\em Ann. of Math}, 67(2):325--327, 1958.

\end{thebibliography}


\begin{thebibliography}{9999}
\bibitem[EKYY]{EKYY} Erdos-Knowles-Yau-Yin.  Semicircle for general class




\end{thebibliography}

\bibliographystyle{abbrv}}

\end{document}